\documentclass[a4paper,fleqn,usenatbib]{mnras}

\usepackage{newtxtext,newtxmath}

\usepackage[T1]{fontenc}
\usepackage{ae,aecompl}

\usepackage{graphicx}	
\usepackage{amsmath}	

\def\chisq{\hbox{$\chi^2$}}

\def\chisqr{\hbox{$\chi^2_{\rm r}$}}
\def\msun{\hbox{${\rm M}_{\odot}$}}

\def\rsun{\hbox{${\rm R}_{\odot}$}}

\def\mstar{\hbox{$M_{\star}$}}
\def\rstar{\hbox{$R_{\star}$}}

\def\teff{\hbox{$T_{\rm eff}$}}
\def\logg{\hbox{$\log g$}}

\def\kms{\hbox{km\,s$^{-1}$}}

\def\vsini{\hbox{$v \sin i$}}

\def\mic{\hbox{$\mu$m}}

\def\Bl{\hbox{$B_{\rm \ell}$}}

\def\Prot{\hbox{$P_{\rm rot}$}}

\def\emr{} 

\title[Magnetic field \& rotation periods of M dwarfs]{Magnetic fields \& rotation periods of M dwarfs from SPIRou spectra} 
\author[J.-F.~Donati et al.]{J.-F.~Donati$^{1}$\thanks{E-mail: jean-francois.donati@irap.omp.eu},
           L.T.~Lehmann$^{1}$, P.I.~Cristofari$^{1,2}$, P.~Fouqu\'e$^{1}$, C.~Moutou$^1$, 
\newauthor P.~Charpentier$^{1}$, M.~Ould-Elhkim$^{1}$, A.~Carmona$^{3}$, X.~Delfosse$^{3}$, E.~Artigau$^{4}$, 
\newauthor S.H.P.~Alencar$^{5}$, C.~Cadieux$^{4}$, L.~Arnold$^{6}$, P.~Petit$^{1}$, J.~Morin$^{7}$, T.~Forveille$^{3}$, 
\newauthor R.~Cloutier$^{8}$, R.~Doyon$^{4}$, G.~H\'ebrard$^{9}$, and the SLS collaboration 
\vspace{1mm}
\\ 
$^1$ Univ.\ de Toulouse, CNRS, IRAP, 14 avenue Belin, 31400 Toulouse, France \\ 
$^2$ Center for Astrophysics, Harvard \& Smithsonian, 60 Garden street, Cambridge, MA 02138, United States \\ 
$^3$ Univ.\ Grenoble Alpes, CNRS, IPAG, 38000 Grenoble, France \\ 
$^4$ Universit\'e de Montr\'eal, D\'epartement de Physique, IREX, Montr\'eal, QC H3C 3J7, Canada \\ 
$^5$ Departamento de F\'{\i}sica -- ICEx -- UFMG, Av. Ant\^onio Carlos, 6627, 30270-901 Belo Horizonte, MG, Brazil\\  
$^6$ Canada-France-Hawaii Telescope, 65-1238 Mamalahoa Hwy., Kamuela, HI 96743, USA \\ 
$^7$ LUPM, Univ.\ de Montpellier, CNRS, F-34095 Montpellier, France \\
$^8$ Dept.\ of Physics \& Astronomy, McMaster University, Hamilton, ON L8S 4L8, Canada \\ 
$^9$ Institut d'Astrophysique de Paris, CNRS, Sorbonne Univ., 98 bis bd Arago, 75014 Paris, France \\ 
}

\date{Submitted 2023 June 06 -- Accepted 2023 July 25 } 

\pubyear{2023}

\begin{document}

\label{firstpage}
\pagerange{\pageref{firstpage}--\pageref{lastpage}}
\maketitle

\begin{abstract}
We present near-infrared spectropolarimetric observations of a sample of 43 weakly- to moderately-active M dwarfs, carried with SPIRou at the 
Canada-France-Hawaii Telescope in the framework of the SPIRou Legacy Survey from early 2019 to mid 2022.  We use the 6700 circularly polarised spectra collected 
for this sample 
to investigate the longitudinal magnetic field and its temporal variations for all sample stars, from which we diagnose, through quasi-periodic Gaussian 
process regression, the periodic modulation and longer-term fluctuations of the longitudinal field.  We detect the large-scale field for 40 of our 43 sample stars, 
and infer a reliable or tentative rotation period for 38 of them, using a Bayesian framework to diagnose the confidence level at which each rotation period 
is detected.  We find rotation periods ranging from 14 to over 60~d for the early-M dwarfs, and from 70~d to 200~d for most mid- and late-M dwarfs (potentially up to 430~d for one of them).  
We also find that the strength of the detected large-scale fields does not decrease with increasing period or Rossby number for the slowly rotating dwarfs of our sample as it does for higher-mass, 
more active stars, suggesting that these magnetic fields may be generated through a different dynamo regime than those of more rapidly rotating stars.  
We also show that the large-scale fields of most sample stars 
evolve on long timescales, with some of them globally switching sign as stars progress on their putative magnetic cycles.  
\end{abstract}

\begin{keywords}
stars: magnetic fields --
stars: low-mass stars --
stars: rotation --
techniques: polarimetric
\end{keywords}



\section{Introduction}
\label{sec:int}

Magnetic fields of M dwarfs have triggered sustained interest since the first detection of a strong field at surface of the M3 dwarf AD~Leo \citep{Saar85}.  
First diagnosed through the Zeeman broadening of unpolarized spectral lines, giving access to the small-scale field at the surface of the star, magnetic fields of M dwarfs 
were then detected through polarized Zeeman signatures in spectral lines \citep{Donati06a}, yielding information on the large-scale field topology.  Both quantities are now 
routinely monitored on a large sample of partly- to fully-convective M dwarfs, outlining how their fields change with stellar parameters such as mass, rotation period and age 
\citep{Donati09, Reiners12, Kochukhov21, Reiners22}, and what it implies in terms of the underlying dynamo processes amplifying and sustaining them in the convective envelopes 
or interiors of these stars \citep[e.g.,][]{Shulyak15}.  

With the detection of many planets and planetary systems around nearby M dwarfs over the last two decades \citep[e.g.,][]{Bonfils13, Gaidos16}, there is even more interest in investigating the 
magnetic fields of our low-mass stellar neighbours.  Not only do these fields trigger all sorts of activity phenomena such as flares or surface brightness inhomogeneities and thereby induce 
different kinds of radial velocity (RV) perturbations \citep[e.g.,][]{Reiners10,Hebrard14}, but they can also generate star / planet interactions for close-in planets and potentially affect 
their orbital parameters \citep[e.g.,][]{Strugarek15}, or even impact the habitability of rocky planets located in the habitable zones of their host stars \citep{Vidotto13}.  

Using spectropolarimetric observations, one can measure the longitudinal magnetic field of stars, i.e., the line-of-sight-projected component of the magnetic field vector averaged 
over the visible hemisphere of the star, noted \Bl, and its modulation with time if stars are monitored over a given time frame.  For stars whose large-scale field is not symmetric with respect 
to the rotation axis, doing so gives access to the rotation period of stars, as first discovered in the context of chemically peculiar (Cp) stars \citep{Babcock49, Stibbs50} then extensively 
used for all classes of magnetic stars \citep{Landstreet92,Donati09}.  Time series of phase-resolved polarized Zeeman signatures of spectral lines can then be 
analysed with Principal Component Analysis \citep{Lehmann22}, or inverted into maps of the large-scale magnetic field using tomographic techniques inspired from medical imaging 
\citep[e.g.,][]{Semel89,Donati06b,Kochukhov21}.  

Whereas rotation periods of active M dwarfs are often known from their photometric variability \citep{Kiraga07}, this is far less the case for weakly-active stars with presumably long 
rotation periods.  Yet, estimating these rotation periods is essential, for instance to avoid confusing planetary RV signatures from those induced by activity.  Carrying out 
velocimetric observations in the near-infrared (nIR) where the RV impact of activity is smaller obviously helps in this respect \citep[e.g.,][]{Carmona23}.  This is the same with 
spectropolarimetry as Zeeman signatures are comparatively larger in the nIR than in the optical for a given field topology and spectral line depth, making it especially interesting for 
studying magnetic fields and rotation periods of M dwarfs that are brightest in this spectral window.  

In this paper, we concentrate on series of \Bl\ measurements collected with SPIRou, the nIR spectropolarimeter \citep{Donati20} mounted at the Cassegrain focus of the 
Canada-France-Hawaii Telescope (CFHT) atop Maunakea, for a sample of 43 M dwarfs monitored in the context of the SPIRou Legacy Survey (SLS) over a time frame of 7 semesters (from 2019a
to 2022a).  This sample and the corresponding SPIRou raw frames are identical to those analysed by \citet{Fouque23} to find out the rotation periods of the sample stars from \Bl\ measurements.  
In this new study however, everything else is different, from the 
data reduction to the modeling of the extracted spectra and Zeeman signatures.  Being carried out with the same reference tools used to process and analyse extensive sets of ESPaDonS and 
NARVAL optical spectropolarimetric data \citep[e.g.,][]{Morin08b,Hebrard16}, our study can thereby serve as a comparison point to double check the consistency of new results based on 
SPIRou data vs older ones derived from ESPaDOnS and NARVAL data, and to assess the agreement between results of various studies based on the same SPIRou data but reduced and analysed 
with a different set of tools \citep[e.g.,][]{Fouque23}.  

After briefly outlining what our stellar sample and observations consist of (Sec.~\ref{sec:obs}), 
we describe the \Bl\ values we retrieve and whether the field is detected and variable with time (Sec.~\ref{sec:lsb}).  We then detail the modeling of these time series to investigate whether 
and how reliably we detect the rotation periods of all sample stars (Sec.~\ref{sec:per}) and clarify the particular cases of a number of individual stars when needed (Sec.~\ref{sec:ist}).  
We finally summarize and discuss in Sec.~\ref{sec:dis} the interest of our new results for our understanding of large-scale magnetic fields and dynamo action in slowly rotating M dwarfs.  

\begin{table*}
\caption[]{Stellar sample studied in this paper.  For each star, columns 2 to 6 list the effective temperature \teff, the logarithmic gravity \logg, the metallicity ${\rm [M/H]}$, the mass 
\mstar\ and radius \rstar\ \citep[from][]{Cristofari22b}, whereas columns 7 to 10 give the number of successful visits $n$ and the number of rejected spectra $r$, the LSD mask used (M0 or M3), 
{\emr the number of 2~\kms\ pixels on which the LSD profiles were integrated to derive \Bl\ values (see Eq.~\ref{eq:blf})}, and the average error bar on \Bl.   Stars are ordered by decreasing \teff.  } 
\begin{tabular}{lccccccccc}
\hline
Star   & \teff        & \logg & ${\rm [M/H]}$  & \mstar        & \rstar          & $n / r$ & mask & width & $\sigma_B$ \\ 
       & (K)          &       &                & (\msun)       & (\rsun)         &         &      & (pix) &   (G)      \\ 
\hline 
Gl 338B & $ 3952\pm30$ & $4.71\pm0.05$ & $-0.08\pm0.10$ & $0.58\pm0.02$ & $0.609\pm0.012$ & 50 / 0 & M0 & 13 & 1.4 \\
Gl 410 & $ 3842\pm31$ & $4.87\pm0.05$ & $0.05\pm0.10$ & $0.55\pm0.02$ & $0.543\pm0.009$ & 132 / 2 & M0 & 19 & 4.4 \\
Gl 846 & $ 3833\pm31$ & $4.69\pm0.05$ & $0.07\pm0.10$ & $0.57\pm0.02$ & $0.568\pm0.009$ & 201 / 1 & M0 & 13 & 2.1 \\
Gl 205 & $ 3771\pm31$ & $4.70\pm0.05$ & $0.43\pm0.10$ & $0.58\pm0.02$ & $0.588\pm0.010$ & 156 / 4 & M0 & 13 & 1.0 \\
Gl 880 & $ 3702\pm31$ & $4.72\pm0.05$ & $0.26\pm0.10$ & $0.55\pm0.02$ & $0.563\pm0.009$ & 168 / 0 & M0 & 13 & 1.7 \\
Gl 514 & $ 3699\pm31$ & $4.74\pm0.05$ & $-0.07\pm0.10$ & $0.50\pm0.02$ & $0.497\pm0.008$ & 167 / 10 & M0 & 13 & 2.9 \\
Gl 382 & $ 3644\pm31$ & $4.75\pm0.05$ & $0.15\pm0.10$ & $0.51\pm0.02$ & $0.511\pm0.009$ & 114 / 4 & M0 & 13 & 2.4 \\
Gl 412A & $ 3620\pm31$ & $4.79\pm0.05$ & $-0.42\pm0.10$ & $0.39\pm0.02$ & $0.391\pm0.007$ & 165 / 8 & M0 & 13 & 4.1 \\
Gl 15A & $ 3611\pm31$ & $4.80\pm0.05$ & $-0.33\pm0.10$ & $0.39\pm0.02$ & $0.345\pm0.015$ & 235 / 7 & M0 & 13 & 2.7 \\
Gl 411 & $ 3589\pm31$ & $4.74\pm0.05$ & $-0.38\pm0.10$ & $0.39\pm0.02$ & $0.383\pm0.008$ & 166 / 2 & M3 & 13 & 2.0 \\
Gl 752A & $ 3558\pm31$ & $4.69\pm0.05$ & $0.11\pm0.10$ & $0.47\pm0.02$ & $0.469\pm0.008$ & 128 / 2 & M3 & 13 & 2.4 \\
Gl 48 & $ 3529\pm31$ & $4.68\pm0.05$ & $0.08\pm0.10$ & $0.46\pm0.02$ & $0.469\pm0.008$ & 190 / 3 & M3 & 13 & 3.8 \\
Gl 617B & $ 3525\pm31$ & $4.84\pm0.06$ & $0.20\pm0.10$ & $0.45\pm0.02$ & $0.460\pm0.008$ & 144 / 6 & M3 & 17 & 5.3 \\
Gl 480 & $ 3509\pm31$ & $4.88\pm0.06$ & $0.26\pm0.10$ & $0.45\pm0.02$ & $0.449\pm0.008$ & 107 / 1 & M3 & 13 & 3.9 \\
Gl 436 & $ 3508\pm31$ & $4.75\pm0.05$ & $0.03\pm0.10$ & $0.42\pm0.02$ & $0.425\pm0.008$ & 92 / 0 & M3 & 13 & 3.3 \\
Gl 849 & $ 3502\pm31$ & $4.88\pm0.06$ & $0.35\pm0.10$ & $0.46\pm0.02$ & $0.458\pm0.008$ & 205 / 1 & M3 & 13 & 3.6 \\
Gl 408 & $ 3487\pm31$ & $4.79\pm0.05$ & $-0.09\pm0.10$ & $0.38\pm0.02$ & $0.390\pm0.007$ & 157 / 6 & M3 & 21 & 7.8 \\
Gl 687 & $ 3475\pm31$ & $4.71\pm0.05$ & $0.01\pm0.10$ & $0.39\pm0.02$ & $0.414\pm0.007$ & 212 / 8 & M3 & 13 & 2.8 \\
Gl 725A & $ 3470\pm31$ & $4.77\pm0.06$ & $-0.26\pm0.10$ & $0.33\pm0.02$ & $0.345\pm0.006$ & 211 / 4 & M3 & 15 & 4.2 \\
Gl 317 & $ 3421\pm31$ & $4.71\pm0.06$ & $0.23\pm0.10$ & $0.42\pm0.02$ & $0.423\pm0.008$ & 77 / 2 & M3 & 13 & 4.6 \\
Gl 251 & $ 3420\pm31$ & $4.71\pm0.06$ & $-0.01\pm0.10$ & $0.35\pm0.02$ & $0.365\pm0.007$ & 178 / 3 & M3 & 15 & 5.1 \\
GJ 4063 & $ 3419\pm31$ & $4.77\pm0.06$ & $0.42\pm0.10$ & $0.42\pm0.02$ & $0.422\pm0.008$ & 219 / 3 & M3 & 13 & 4.4 \\
Gl 725B & $ 3379\pm31$ & $4.82\pm0.06$ & $-0.28\pm0.10$ & $0.25\pm0.02$ & $0.280\pm0.005$ & 208 / 3 & M3 & 13 & 5.0 \\
PM J09553-2715 & $ 3366\pm31$ & $4.76\pm0.06$ & $-0.03\pm0.10$ & $0.29\pm0.02$ & $0.302\pm0.006$ & 76 / 3 & M3 & 13 & 5.9 \\
Gl 876 & $ 3366\pm31$ & $4.80\pm0.06$ & $0.15\pm0.10$ & $0.33\pm0.02$ & $0.333\pm0.006$ & 91 / 1 & M3 & 13 & 3.5 \\
GJ 1012 & $ 3363\pm31$ & $4.66\pm0.06$ & $0.07\pm0.10$ & $0.35\pm0.02$ & $0.367\pm0.007$ & 137 / 7 & M3 & 13 & 6.3 \\
GJ 4333 & $ 3362\pm31$ & $4.72\pm0.06$ & $0.25\pm0.10$ & $0.37\pm0.02$ & $0.386\pm0.008$ & 186 / 4 & M3 & 13 & 5.2 \\
Gl 445 & $ 3356\pm31$ & $4.85\pm0.06$ & $-0.24\pm0.10$ & $0.24\pm0.02$ & $0.266\pm0.005$ & 91 / 5 & M3 & 13 & 7.2 \\
GJ 1148 & $ 3354\pm31$ & $4.70\pm0.06$ & $0.11\pm0.10$ & $0.34\pm0.02$ & $0.365\pm0.007$ & 98 / 5 & M3 & 13 & 6.1 \\
PM J08402+3127 & $ 3347\pm31$ & $4.76\pm0.06$ & $-0.08\pm0.10$ & $0.28\pm0.02$ & $0.299\pm0.006$ & 130 / 5 & M3 & 13 & 6.9 \\
GJ 3378 & $ 3326\pm31$ & $4.81\pm0.06$ & $-0.05\pm0.10$ & $0.26\pm0.02$ & $0.279\pm0.005$ & 178 / 3 & M3 & 13 & 6.2 \\
GJ 1105 & $ 3324\pm31$ & $4.63\pm0.07$ & $-0.04\pm0.10$ & $0.27\pm0.02$ & $0.283\pm0.005$ & 167 / 2 & M3 & 13 & 6.0 \\
Gl 699 & $ 3311\pm31$ & $5.11\pm0.06$ & $-0.37\pm0.10$ & $0.16\pm0.02$ & $0.185\pm0.004$ & 247 / 3 & M3 & 13 & 4.6 \\
Gl 169.1A & $ 3307\pm31$ & $4.71\pm0.06$ & $0.13\pm0.10$ & $0.28\pm0.02$ & $0.292\pm0.006$ & 172 / 7 & M3 & 13 & 5.7 \\
PM J21463+3813 & $ 3305\pm33$ & $5.06\pm0.08$ & $-0.38\pm0.10$ & $0.18\pm0.02$ & $0.208\pm0.004$ & 185 / 4 & M3 & 13 & 10.1 \\
Gl 15B & $ 3272\pm31$ & $4.89\pm0.06$ & $-0.42\pm0.10$ & $0.16\pm0.02$ & $0.182\pm0.004$ & 179 / 6 & M3 & 13 & 8.3 \\
GJ 1289 & $ 3238\pm32$ & $5.00\pm0.07$ & $0.05\pm0.10$ & $0.21\pm0.02$ & $0.233\pm0.005$ & 204 / 9 & M3 & 23 & 15.3 \\
Gl 447 & $ 3198\pm31$ & $4.82\pm0.06$ & $-0.13\pm0.10$ & $0.18\pm0.02$ & $0.201\pm0.004$ & 57 / 0 & M3 & 13 & 5.9 \\
GJ 1151 & $ 3178\pm31$ & $4.71\pm0.06$ & $-0.16\pm0.10$ & $0.17\pm0.02$ & $0.193\pm0.004$ & 158 / 4 & M3 & 13 & 7.9 \\
GJ 1103 & $ 3170\pm31$ & $4.67\pm0.06$ & $-0.03\pm0.10$ & $0.19\pm0.02$ & $0.224\pm0.005$ & 60 / 9 & M3 & 13 & 8.7 \\
Gl 905 & $ 3069\pm31$ & $4.78\pm0.08$ & $0.05\pm0.11$ & $0.15\pm0.02$ & $0.165\pm0.004$ & 219 / 3 & M3 & 13 & 6.8 \\
GJ 1002 & $ 2980\pm33$ & $4.70\pm0.08$ & $-0.33\pm0.11$ & $0.12\pm0.02$ & $0.139\pm0.003$ & 140 / 6 & M3 & 13 & 9.3 \\
GJ 1286 & $ 2961\pm33$ & $4.55\pm0.12$ & $-0.23\pm0.10$ & $0.12\pm0.02$ & $0.142\pm0.004$ & 104 / 10 & M3 & 13 & 10.0 \\
\hline
\end{tabular}
\label{tab:str}
\end{table*}

\section{SPIRou observations}
\label{sec:obs}
In this paper, we focus on the 43 stars listed in Table~\ref{tab:str}, the same as in \citet{Fouque23}, observed with SPIRou more than 50 times over the 
duration of the SLS \citep[for a description of the whole sample, see][]{Moutou23}.  
We recall that SPIRou collects nIR spectra of stars at a resolving power of 70\,000, spanning a spectral window ranging from 0.95 to 
2.50~\mic\ \citep[$Y$$J$$H$$K$ bands, with a small 2-nm gap at 2.438~\mic,][]{Donati20}.  SPIRou is also a spectropolarimeter, capable of measuring polarization in 
spectral lines.  It does so by carrying out, at each visit, a sequence of 4 sub-exposures in the pre-defined positions of the polarimeter quarter-wave 
Fresnel-rhomb retarders that can yield the requested polarization state with minimal errors.  In this study, all stars were observed in circular polarization, leading to 
one Stokes $I$ (unpolarized) and one Stokes $V$ (circular polarization) spectrum per star and per visit;  from the same data, we also compute a null 
polarization check called $N$, expected to yield a null signature when the polarimeter and reduction pipeline behave nominally\footnote{\emr We however caution that the
opposite is not necessarily true and that $N=0$ is not definite evidence that the polarimeter and pipeline are working well.} \citep{Donati97b}.  

As opposed to \citet{Fouque23} where the nominal SPIRou reduction package \texttt{APERO} \citep[optimized for RV precision,][]{Cook22} was used, all data in 
this study were processed with the alternate package \texttt{Libre-ESpRIT}, i.e., the nominal ESPaDOnS reduction pipeline \citep{Donati97b} adapted for 
SPIRou data \citep{Donati20}.  Optimized for polarimetry and checked against magnetic standard (Cp) stars, \texttt{Libre-ESpRIT} can be considered as a reference in this 
respect \citep{Donati97b,Donati20}.  {\emr The main differences between the two pipelines are the way the spectra of both science channels (i.e., both orthogonal polarization states) 
are extracted from the raw frames of each sub-exposure on the one hand, and how Stokes $V$ spectra are derived from those of both science channels and all sub-exposures 
on the other.  We do not expect significant differences between both pipelines for these steps, that were cross-checked on spectra of a few reference stars (e.g., AD~Leo), 
except on how error bars are propagated from the raw frames to the Stokes $V$ spectra.} 

We then applied our version of Least-Squares Deconvolution \citep[LSD,][]{Donati97b} to all reduced Stokes $I$, $V$, and $N$ spectra 
of all stars, in contrast to \citet{Fouque23} who used a different LSD implementation.  In our case, we carried out LSD using 2 main line masks, an M0 
mask for stars whose effective temperature \teff\ is larger than 3600~K and an M3 mask for all others.  In these masks, that we constructed using VALD-3 
\citep{Ryabchikova15} assuming a logarithmic gravity of $\logg=5$ 
and a solar metallicity ${\rm [M/H]}$, we only use atomic lines whose relative depth with respect to the continuum (including microturbulence only) is larger 
than 10 per cent, and with known magnetic sensitivity (Land\'e factor).  We end up exploiting about 800 atomic lines for the M0 mask, and 575 lines for the M3 
mask.  We also tried incorporating weaker lines in the mask (down to relative depths of 7, 5 or 3 percent), but found that the 10 per cent threshold in relative 
line depth gives the best results in terms of reliability of the Zeeman detections.  For each star, the few epochs for which LSD Stokes $I$ profiles were 
of much lower quality than the typical one (e.g., as a result of poor weather) were rejected from our data set.  
{\emr This second step is also different between the two studies, with \citet{Fouque23} using not only a different implementation of LSD, but also different line 
masks and different thresholds on line strengths.  To our knowledge, the alternate LSD tools used by \citet{Fouque23} have not been extensively cross-checked with 
ours, and may generate potential differences between the two approaches.  }  

For each set of LSD Stokes $I$, $V$ and $N$ profiles associated with one visit, we compute \Bl, its $N$ equivalent and the corresponding error bars following 
\citet{Donati97b}, i.e., from the first moment of the $V(v)$ (and $N(v)$) profile normalised by the equivalent with of the $I(v)$ profile, 
(both integrated over velocity $v$): 
\begin{eqnarray}
\Bl = -2.14\times10^{11} \frac{\int v V(v) \mathrm{d}v}{ \lambda g c \int [1-I(v)] \mathrm{d}v}
\label{eq:blf}
\end{eqnarray}
with \Bl\ in G, $v$ and $c$ (the speed of light) in \kms, and where $g$ and $\lambda$ (in nm) refer to the equivalent Land\'e factor and wavelength of the resulting 
Stokes $I$, $V$ (and $N$) LSD profiles.  
Error bars are derived analytically from Eq.~\ref{eq:blf} and the noise in LSD profiles, themselves computed by propagating the photon noise in the $I$, $V$ and $N$ spectra 
\citep{Donati97b}.  
The integration is carried out over a window centred on the median stellar RV, of width $\pm13$~\kms\ about the center for most stars, {\emr except for a few whose Zeeman signatures are 
stronger and / or wider than average and for which this interval was widened to $\pm15$~\kms\ (Gl~725A and Gl~251), $\pm17$~\kms\ (Gl~617B), $\pm19$~\kms\ (Gl~410), $\pm21$~\kms\ 
(Gl~408) and up to $\pm23$~\kms\ (GJ~1289).  Keeping this window as narrow as possible is indeed key for minimizing the error bar on \Bl\ (denoted $\sigma_B$) while retaining all 
information about the longitudinal field.  
In practice, we looked at the time averaged absolute value of the Stokes $V$ profiles for each star, and verified that no signal is detected above the continuum photon noise outside 
of the selected interval;  we also checked that the standard deviation of Stokes $V$ profiles for each star shows no signal outside of this window.   
We find that the average $\sigma_B$ per star varies from 1.0 to 15.3~G (depending on the stellar magnitude and spectral type) in our observations, with a median of 4.4~G (see 
Table~\ref{tab:str}).   This third step is also slightly different in \citet{Fouque23}, where integration is carried out on a significantly wider window.  However, 
we do not expect major discrepancies from such differences, mostly larger error bars on \Bl\ values as a result of the wider integration window.}  

\begin{table*}
\caption[]{Statistics on the \Bl\ values from the LSD Stokes $V$ and $N$ profiles, for each star of our sample.  Columns 2 to 5, computed from LSD Stokes $V$ profiles, 
respectively list the weighted-average and standard deviation of \Bl, the average error bar $\sigma_B$, the reduced chi square \chisqr\ of \Bl\ and \Bl$-$<\Bl> (where <> notes the 
weighted average), and the detection status (DS) of  \Bl\ and \Bl$-$<\Bl> (with DD, MD and ND standing for Definite Detection, Marginal Detection and No Detection, see text for how these cases 
are defined).  Columns 6 to 9 give the same quantities, but derived from the LSD $N$ profiles. The last column recalls the number of visits $n$ for each star.  } 
\begin{tabular}{lccccccccccc}
\hline
       & \multicolumn{4}{c}{Stokes $V$}                     && \multicolumn{4}{c}{$N$} \\ 
Star   & <\Bl> / RMS & $\sigma_B$ & \chisqr            & DS && <\Bl> / RMS & $\sigma_B$ & \chisqr             & DS && $n$ \\ 
       & (G)         & (G)        & \Bl\ / \Bl$-$<\Bl> &    && (G)         & (G)        & \Bl\ / \Bl$-$<\Bl>  &    &&     \\ 
\hline 
Gl 338B & -3.8 / 4.3 & 1.4 & 9.93 / 2.13 & DD / DD && -0.1 / 1.6 & 1.4 & 1.30 / 1.30 & ND / ND && 50 \\
Gl 410 & -0.1 / 16.7 & 4.4 & 14.20 / 14.20 & DD / DD && -0.3 / 4.2 & 4.5 & 0.89 / 0.89 & ND / ND && 132 \\
Gl 846 & 0.0 / 3.4 & 2.1 & 2.70 / 2.70 & DD / DD && 0.3 / 2.2 & 2.1 & 1.07 / 1.05 & ND / ND && 201 \\
Gl 205 & 1.5 / 3.2 & 1.0 & 10.60 / 8.19 & DD / DD && -0.1 / 1.0 & 1.0 & 1.07 / 1.05 & ND / ND && 156 \\
Gl 880 & 0.3 / 4.2 & 1.7 & 6.25 / 6.21 & DD / DD && -0.1 / 1.9 & 1.7 & 1.20 / 1.19 & ND / ND && 168 \\
Gl 514 & -2.5 / 5.3 & 2.9 & 3.31 / 2.57 & DD / DD && 0.0 / 2.9 & 2.9 & 0.96 / 0.96 & ND / ND && 167 \\
Gl 382 & -1.3 / 4.8 & 2.4 & 3.90 / 3.63 & DD / DD && -0.1 / 2.5 & 2.4 & 1.04 / 1.04 & ND / ND && 114 \\
Gl 412A & 7.0 / 8.9 & 4.1 & 4.77 / 1.80 & DD / DD && -0.1 / 4.0 & 4.1 & 0.92 / 0.92 & ND / ND && 165 \\
Gl 15A & 0.1 / 3.4 & 2.7 & 1.59 / 1.59 & DD / DD && -0.1 / 3.0 & 2.8 & 1.20 / 1.20 & ND / ND && 235 \\
Gl 411 & 2.2 / 3.3 & 2.0 & 2.69 / 1.49 & DD / MD && 0.2 / 2.4 & 2.1 & 1.30 / 1.29 & ND / ND && 166 \\
Gl 752A & 0.2 / 3.7 & 2.4 & 2.31 / 2.30 & DD / DD && 0.0 / 2.4 & 2.5 & 0.98 / 0.98 & ND / ND && 128 \\
Gl 48 & -2.7 / 5.9 & 3.8 & 2.42 / 1.90 & DD / DD && 0.3 / 3.7 & 3.9 & 0.93 / 0.93 & ND / ND && 190 \\
Gl 617B & 12.6 / 14.4 & 5.3 & 7.47 / 1.85 & DD / DD && 0.4 / 5.8 & 5.4 & 1.18 / 1.17 & ND / ND && 144 \\
Gl 480 & 1.0 / 5.8 & 3.9 & 2.23 / 2.16 & DD / DD && 0.1 / 3.7 & 3.9 & 0.89 / 0.89 & ND / ND && 107 \\
Gl 436 & -5.1 / 6.5 & 3.3 & 4.03 / 1.63 & DD / MD && 0.3 / 2.7 & 3.3 & 0.66 / 0.65 & ND / ND && 92 \\
Gl 849 & 2.1 / 5.5 & 3.6 & 2.33 / 1.98 & DD / DD && 0.6 / 3.4 & 3.7 & 0.86 / 0.83 & ND / ND && 205 \\
Gl 408 & -29.9 / 29.0 & 7.8 & 15.94 / 1.37 & DD / ND && 0.7 / 7.8 & 7.9 & 0.97 / 0.96 & ND / ND && 157 \\
Gl 687 & 0.7 / 4.4 & 2.8 & 2.47 / 2.41 & DD / DD && -0.0 / 2.8 & 2.8 & 0.97 / 0.97 & ND / ND && 212 \\
Gl 725A & -7.6 / 9.2 & 4.2 & 4.74 / 1.47 & DD / MD && -0.1 / 4.5 & 4.3 & 1.11 / 1.11 & ND / ND && 211 \\
Gl 317 & -1.9 / 6.2 & 4.6 & 1.81 / 1.65 & MD / MD && 0.6 / 4.6 & 4.7 & 0.99 / 0.97 & ND / ND && 77 \\
Gl 251 & 9.3 / 11.0 & 5.1 & 4.65 / 1.39 & DD / MD && -0.3 / 4.9 & 5.2 & 0.91 / 0.90 & ND / ND && 178 \\
GJ 4063 & 7.3 / 8.6 & 4.4 & 3.92 / 1.17 & DD / ND && 0.4 / 4.2 & 4.4 & 0.91 / 0.90 & ND / ND && 219 \\
Gl 725B & -2.5 / 6.0 & 5.0 & 1.49 / 1.24 & DD / ND && 0.6 / 5.0 & 5.0 & 0.99 / 0.98 & ND / ND && 208 \\
PM J09553-2715 & 8.2 / 10.3 & 5.9 & 3.52 / 1.57 & DD / ND && -0.3 / 5.5 & 5.9 & 0.86 / 0.86 & ND / ND && 76 \\
Gl 876 & 1.7 / 7.3 & 3.5 & 4.37 / 4.13 & DD / DD && 0.8 / 3.6 & 3.5 & 1.03 / 0.97 & ND / ND && 91 \\
GJ 1012 & 1.9 / 6.7 & 6.3 & 1.14 / 1.04 & ND / ND && 0.3 / 6.1 & 6.3 & 0.94 / 0.94 & ND / ND && 137 \\
GJ 4333 & 2.0 / 8.0 & 5.2 & 2.36 / 2.21 & DD / DD && 0.3 / 5.1 & 5.3 & 0.96 / 0.95 & ND / ND && 186 \\
Gl 445 & -0.1 / 7.3 & 7.2 & 1.02 / 1.02 & ND / ND && -0.0 / 7.4 & 7.3 & 1.04 / 1.04 & ND / ND && 91 \\
GJ 1148 & -5.5 / 8.4 & 6.1 & 2.01 / 1.19 & DD / ND && -0.7 / 6.0 & 6.1 & 0.98 / 0.96 & ND / ND && 98 \\
PM J08402+3127 & 12.9 / 17.5 & 6.9 & 6.93 / 3.41 & DD / DD && -0.3 / 7.6 & 6.9 & 1.19 / 1.19 & ND / ND && 130 \\
GJ 3378 & 5.3 / 9.2 & 6.2 & 2.26 / 1.52 & DD / MD && 0.8 / 5.8 & 6.2 & 0.87 / 0.85 & ND / ND && 178 \\
GJ 1105 & -0.2 / 6.1 & 6.0 & 1.01 / 1.01 & ND / ND && 0.6 / 6.1 & 6.1 & 0.99 / 0.98 & ND / ND && 167 \\
Gl 699 & 1.6 / 6.9 & 4.6 & 2.28 / 2.16 & DD / DD && 0.6 / 5.1 & 4.7 & 1.18 / 1.17 & ND / ND && 247 \\
Gl 169.1A & -1.8 / 7.0 & 5.7 & 1.49 / 1.40 & MD / MD && 0.9 / 5.9 & 5.8 & 1.05 / 1.02 & ND / ND && 172 \\
PM J21463+3813 & 0.7 / 11.1 & 10.1 & 1.22 / 1.22 & ND / ND && -0.0 / 10.5 & 10.2 & 1.07 / 1.07 & ND / ND && 185 \\
Gl 15B & -0.4 / 10.8 & 8.3 & 1.70 / 1.69 & DD / DD && 0.6 / 8.0 & 8.3 & 0.93 / 0.92 & ND / ND && 179 \\
GJ 1289 & 47.4 / 40.6 & 15.3 & 19.68 / 10.12 & DD / DD && 0.4 / 14.8 & 15.4 & 0.92 / 0.92 & ND / ND && 204 \\
Gl 447 & 10.1 / 12.0 & 5.9 & 5.20 / 2.31 & DD / DD && -1.0 / 6.0 & 6.0 & 1.01 / 0.99 & ND / ND && 57 \\
GJ 1151 & 1.7 / 16.8 & 7.9 & 4.60 / 4.55 & DD / DD && 1.0 / 7.8 & 7.9 & 0.97 / 0.96 & ND / ND && 158 \\
GJ 1103 & 4.0 / 10.2 & 8.7 & 1.72 / 1.50 & MD / ND && -0.2 / 8.7 & 8.8 & 0.98 / 0.98 & ND / ND && 60 \\
Gl 905 & -6.7 / 15.6 & 6.8 & 5.37 / 4.38 & DD / DD && -0.1 / 7.1 & 6.8 & 1.07 / 1.07 & ND / ND && 219 \\
GJ 1002 & -0.5 / 10.0 & 9.3 & 1.15 / 1.14 & ND / ND && 0.1 / 9.2 & 9.4 & 0.95 / 0.95 & ND / ND && 140 \\
GJ 1286 & 15.4 / 18.2 & 10.0 & 5.53 / 3.14 & DD / DD && -1.7 / 8.7 & 10.1 & 0.76 / 0.74 & ND / ND && 104 \\
\hline
\end{tabular}
\label{tab:sta}
\end{table*}

\section{Longitudinal field and its temporal variations}
\label{sec:lsb}

We start by carrying out a statistical analysis on the series of \Bl\ values for each star, computed from both LSD Stokes $V$ and $N$ profiles.  
We compute in particular the reduced chi squares \chisqr\ of both \Bl\ and \Bl$-$<\Bl> series (where <> notes the weighted average over all 
epochs).  Whereas the former indicates whether the longitudinal field is significantly different from zero, i.e., is detected, the latter informs 
on whether \Bl\ fluctuates about its mean value by more than what the estimated error bars allow, i.e., that \Bl\ is variable with time.  
Following \citet{Donati97b}, we use the \chisq\ probability function \citep{Press92} to diagnose a detection, taking $n$ as the number 
of degrees of freedom.  We consider that we can claim a definite detection (DD) if the false-alarm probability (FAP) is $<10^{-5}$, a marginal 
detection (MD) if the FAP is $<10^{-3}$, and that we have no detection (ND) otherwise.  All \chisqr\ values are listed in Table~\ref{tab:sta}, 
along with the corresponding Detection Status (DS).  We also include in Table~\ref{tab:sta} the weighted average and standard deviations of \Bl\ 
values, as well as the average error bar $\sigma_B$.  

For all stars but 5 (namely GJ~1012, GJ~1105, Gl~445, PM~J21463+3813 and GJ~1002), we obtain either a clear detection (35 stars) or a marginal 
detection (3 stars, Gl~317, Gl~169.1A and GJ~1103) of the longitudinal field.  For the 5 stars with no detection, one can check in particular that 
<\Bl> is close to zero (within $\sigma_B$) and that the standard deviation of \Bl\ is similar to $\sigma_B$.  As detailed in Sec.~\ref{sec:per}, 
\Bl\ seems to be detected for the last 2 of these 5 stars (PM~J21463+3813 and GJ~1002) even though our statistical test 
conclude to a ND.  It suggests that our error bars on \Bl\ are slightly overestimated for a few of our targets, rendering our detection criterion 
a bit too stringent in such cases.  

Out of the 38 stars where \Bl\ is either clearly or marginally detected, 5 of them (GJ~4063, GJ~1148, PM~J09553-2715, Gl~725B, GJ~1103) are 
diagnosed as showing no temporal variations of \Bl, whereas 5 (Gl~411, Gl~436, Gl~317, GJ~3378, Gl~169.1A) are identified as 
having a marginally variable \Bl.  Once again, we will see later in the paper (Sec.~\ref{sec:per}) that 4 of the 5 stars whose \Bl\ is listed as 
non variable are in fact found to show either clear or probable periodic \Bl\ fluctuations, presumably for the same reason as that mentioned previously, 
i.e., that our detection criterion is too stringent at times.  Only GJ~1148 ends up showing no \Bl\ variations despite the large-scale field being clearly 
detected, possibly because the magnetic topology is almost perfectly axisymmetric or the star is seen almost exactly pole-on.  
Most of our targets thus show clear \Bl\ detections and temporal variations, in particular GJ~1289, Gl~410, Gl~205, Gl~880, GJ~1151, Gl~905 
and Gl~876 for which the \chisqr\ of \Bl$-$<\Bl> is larger than 4 and up to 16.6.  

We finally note that \Bl\ values derived from LSD $N$ profiles (columns 6 to 9 of Table~\ref{tab:sta}) are all consistent with 0, with 
standard deviations equal to $\sigma_B$ and \chisqr\ equal to $1.00\pm0.12$ on average over the sample of 43 stars.  We note that 
\chisqr\ ranges from 0.7 to 1.3, reflecting mostly statistical photon noise fluctuations;  moreover, the weighted-average <\Bl> is always very close 
to 0, causing the \chisqr\ associated with \Bl\ and \Bl$-$<\Bl> to be almost identical for all stars.  This is what we expect if the SPIRou 
polarimeter and the \texttt{Libre-ESpRIT} reduction package are working nominally regarding polarimetry, with no spurious signatures showing up 
in the null polarization check down to the photon noise level.  

By comparing <\Bl> and $\sigma_B$ values in Table~\ref{tab:sta} to their equivalents in \citet{Fouque23}, we see that both our \Bl\ values and error bars 
are smaller than those of \citet{Fouque23} by a factor of 2$-$3.  This is most obvious for stars where |<\Bl>| is larger than 10~G, but it is also the case 
for other stars with weaker fields.  As Stokes $I$ and $V$ profiles spectra derived by \texttt{Libre-ESpRIT} and \texttt{APERO} apparently yield, 
for a few stars, consistent LSD profiles when the same LSD code (the one used here) is applied to both, {\emr we suspect that the discrepancy mentioned above 
mostly reflects differences in the LSD implementation used by \citet{Fouque23}, i.e., in the step referred to  as step 2 of our description of how SPIRou data 
were analyzed (see Sec.~\ref{sec:obs}), and to a lesser extent in the way their \Bl\ values were derived from LSD profiles (step 3).  } 
The origin of this difference is currently being investigated.

\section{Periodicity of the \Bl\ variations}
\label{sec:per}

To investigate whether the temporal variability of \Bl\ is periodic, a standard Fourier analysis is not ideal for these mildly to weakly active M dwarfs, 
where \Bl\ is expected to evolve on a timescale of the same order of magnitude as the rotation period itself.  We therefore use instead Gaussian process 
regression (GPR) to the \Bl\ data, with a quasi-periodic (QP) kernel whose covariance function $c(t,t')$ is given by 
\begin{eqnarray}
c(t,t') = \theta_1^2 \exp \left( -\frac{(t-t')^2}{2 \theta_3^2} -\frac{\sin^2 \left( \frac{\pi (t-t')}{\theta_2} \right)}{2 \theta_4^2} \right) 
\label{eq:covar}
\end{eqnarray}
where $\theta_1$ is the amplitude (in G) of the Gaussian Process (GP), $\theta_2$ its recurrence period (i.e., \Prot, in d),
$\theta_3$ the timescale (in d) on which \Bl\ evolves, and $\theta_4$ a smoothing parameter setting the level of harmonic complexity.  
Although our error bars on \Bl\ are slightly overestimated (see Sec.~\ref{sec:lsb}), we nonetheless introduce a fifth 
hyper-parameter $\theta_5$ setting the amount of additional white noise potentially required by GPR (e.g., as a result of short-term intrinsic variability) 
to achieve the fit to the \Bl\ values (denoted $y$) that maximizes likelihood $\mathcal{L}$, defined as:
\begin{eqnarray}
2 \log \mathcal{L} = -n \log(2\pi) - \log|C+\Sigma+S| - y^T (C+\Sigma+S)^{-1} y
\label{eq:llik}
\end{eqnarray}
where $C$ is the covariance matrix for our observation epochs, $\Sigma$ the diagonal variance matrix associated with \Bl, and $S=\theta_5^2 I$ the 
contribution of the additional white noise ($I$ being the identity matrix).  Coupled to a MCMC run to explore the parameter domain, one can find 
out the optimal set of hyper parameters and corresponding posterior distributions.  

\begin{figure*}
\centerline{\includegraphics[scale=0.58,angle=-90]{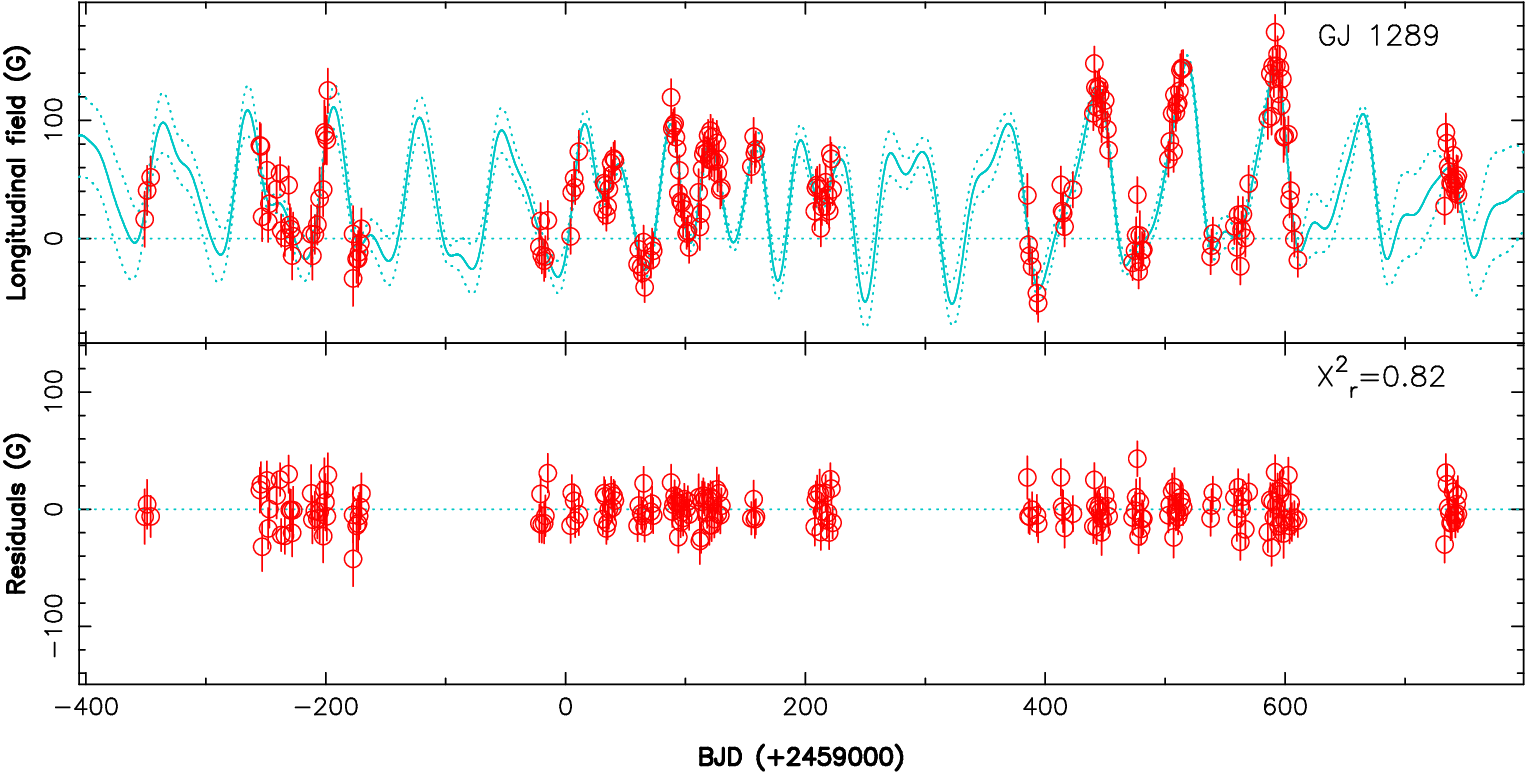} \vspace{2mm}} 
\centerline{\includegraphics[scale=0.58,angle=-90]{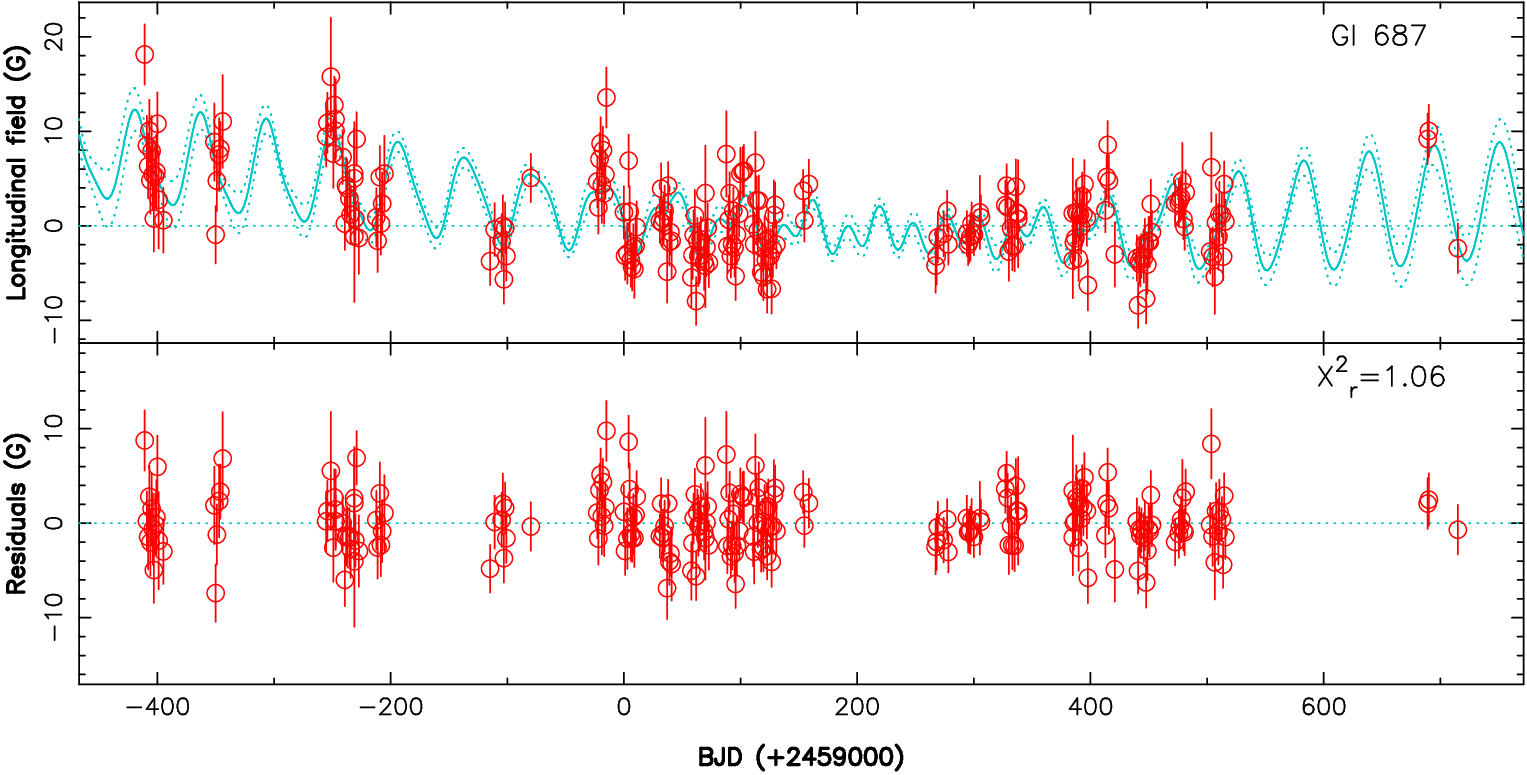} \vspace{2mm}} 
\centerline{\includegraphics[scale=0.58,angle=-90]{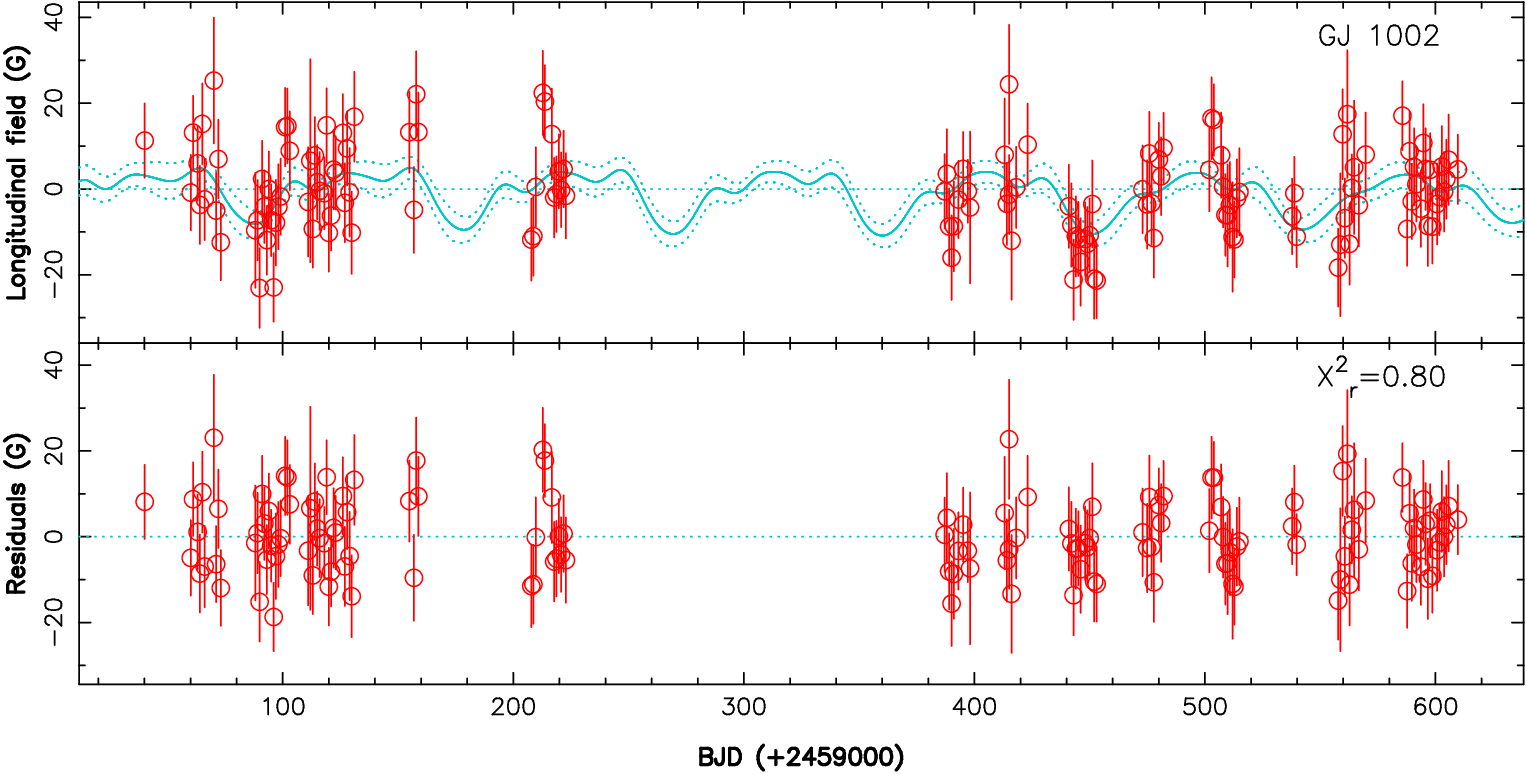}} 
\caption[]{QP GPR fit and error bars (full and dotted cyan curves) of the \Bl\ data (red open symbols with error bars) for GJ~1289 (top), Gl~687 (middle) 
and GJ~1002 (bottom).  The corresponding $\Delta \log \mathcal{L}_M$ are respectively equal to 168.5 (the highest of the whole sample), 34.6 and 7.2, indicating a clear 
detection of the QP \Bl\ modulation in the first 2 cases and a marginal detection in the third case (see Table~\ref{tab:gpr}).  } 
\label{fig:gjs}
\end{figure*}

\renewcommand{\arraystretch}{1.2}
\begin{table*}
\caption[]{Results of the QP GPR applied to the \Bl\ time series of our sample stars.  Columns 2 lists the recovered period whenever detected, whereas 
columns 3 to 6 give the values of the 4 other hyper-parameters (with some fixed in a few cases).  The achieved \chisqr, RMS, $\Delta \log \mathcal{L}_M$ 
with respect to a model with no modulation, and the detection status of the periodic modulation are mentioned in columns 7 to 10.  In column~7, we also mention 
in parenthesis the \chisqr\ for the GPR fit of the long-term \Bl\ variations only, to emphasize how much it changes with respect to that of the main GPR fit.  }  
\begin{tabular}{lccccccccc}
\hline
Star   & \Prot\  & GP ampl.\ & Evol.\ timescale & Smoothing & White noise & \chisqr & RMS & $\Delta \log \mathcal{L}_M$ & DS \\ 
       & $\theta_2$ (d) & $\theta_1$ (G) & $\theta_3$ (d) & $\theta_4$ & $\theta_5$ (G) &         & (G) &   &  \\ 
\hline 
Gl 338B & $42.2\pm4.1$ & $2.1^{+1.0}_{-0.7}$ & $120$ & $0.99\pm0.59$ & $0.6^{+0.4}_{-0.2}$ & 0.79 (1.97) & 1.2 & 7.2 & MD \\
Gl 410 & $13.91\pm0.09$ & $17.2^{+2.8}_{-2.4}$ & $59^{+9}_{-8}$ & $0.48\pm0.06$ & $2.2^{+1.3}_{-0.8}$ & 0.71 (14.20) & 3.7 & 85.0 & DD \\
Gl 846 & $21.84\pm0.14$ & $3.0^{+0.4}_{-0.4}$ & $70^{+15}_{-12}$ & $0.31\pm0.04$ & $0.4^{+0.3}_{-0.2}$ & 0.60 (2.70) & 1.6 & 58.6 & DD \\
Gl 205 & $34.58\pm0.46$ & $2.9^{+0.4}_{-0.4}$ & $53^{+11}_{-9}$ & $0.37\pm0.05$ & $0.3^{+0.2}_{-0.1}$ & 0.69 (6.94) & 0.8 & 85.5 & DD \\
Gl 880 & $37.21\pm0.30$ & $4.4^{+0.8}_{-0.6}$ & $113^{+21}_{-18}$ & $0.41\pm0.06$ & $0.4^{+0.3}_{-0.2}$ & 0.67 (5.51) & 1.4 & 94.8 & DD \\
Gl 514 & $30.32\pm0.21$ & $4.0^{+1.4}_{-1.0}$ & $300$ & $0.60\pm0.23$ & $1.5^{+0.5}_{-0.4}$ & 1.14 (2.05) & 3.1 & 29.0 & DD \\
Gl 382 & $21.91\pm0.16$ & $4.2^{+0.8}_{-0.7}$ & $131^{+50}_{-36}$ & $0.35\pm0.07$ & $0.8^{+0.6}_{-0.3}$ & 0.69 (3.63) & 2.0 & 37.5 & DD \\
Gl 412A & $36.9\pm2.5$ & $4.5^{+1.2}_{-1.0}$ & $78^{+28}_{-21}$ & $1.22\pm0.58$ & $1.2^{+0.8}_{-0.5}$ & 0.84 (1.33) & 3.7 & 14.6 & DD \\
Gl 15A & $43.26\pm0.36$ & $3.6^{+1.3}_{-1.0}$ & $300$ & $0.78\pm0.26$ & $1.0^{+0.5}_{-0.3}$ & 1.02 (1.41) & 2.7 & 22.1 & DD \\
Gl 411 & $427\pm34$ & $1.3^{+0.4}_{-0.3}$ & $300$ & $0.24\pm0.14$ & $0.7^{+0.4}_{-0.3}$ & 0.95 (1.21) & 2.0 & 11.9 & DD \\
Gl 752A & $45.0\pm4.2$ & $2.8^{+0.7}_{-0.5}$ & $63^{+27}_{-19}$ & $0.51\pm0.16$ & $0.8^{+0.6}_{-0.3}$ & 0.90 (2.31) & 2.3 & 19.0 & DD \\
Gl 48 & $52.1\pm1.9$ & $4.4^{+1.0}_{-0.8}$ & $61^{+20}_{-15}$ & $0.47\pm0.17$ & $1.0^{+0.7}_{-0.4}$ & 0.80 (1.90) & 3.4 & 32.6 & DD \\
Gl 617B & $40.4\pm3.0$ & $5.4^{+1.3}_{-1.0}$ & $69^{+35}_{-23}$ & $0.60\pm0.22$ & $1.7^{+1.1}_{-0.7}$ & 0.86 (1.56) & 4.9 & 17.7 & DD \\
Gl 480 & $25.00\pm0.24$ & $6.6^{+3.5}_{-2.3}$ & $300$ & $1.16\pm0.57$ & $2.5^{+0.8}_{-0.6}$ & 1.26 (1.75) & 4.4 & 9.7 & MD \\
Gl 436 & $48\pm13$ & $1.7^{+1.6}_{-0.8}$ & $149^{+106}_{-62}$ & $0.82\pm0.78$ & $1.3^{+1.0}_{-0.6}$ & 0.90 (1.63) & 3.1 & 1.6 & ND \\
Gl 849 & $41.76\pm0.61$ & $4.8^{+1.5}_{-1.1}$ & $209^{+58}_{-45}$ & $0.47\pm0.13$ & $1.0^{+0.7}_{-0.4}$ & 0.89 (1.97) & 3.4 & 39.0 & DD \\
Gl 408 & $171.0\pm8.4$ & $6.3^{+1.5}_{-1.2}$ & $200$ & $0.21\pm0.10$ & $1.5^{+1.2}_{-0.7}$ & 0.66 (1.16) & 6.3 & 16.5 & DD \\
Gl 687 & $56.69\pm0.56$ & $5.9^{+2.0}_{-1.5}$ & $300$ & $0.78\pm0.25$ & $1.2^{+0.5}_{-0.3}$ & 1.06 (1.83) & 2.9 & 34.6 & DD \\
Gl 725A & $103.1\pm6.1$ & $3.4^{+0.7}_{-0.6}$ & $86^{+51}_{-32}$ & $0.32\pm0.10$ & $0.8^{+0.7}_{-0.4}$ & 0.78 (1.47) & 3.7 & 24.1 & DD \\
Gl 317 & $39.0\pm3.8$ & $5.6^{+2.4}_{-1.7}$ & $107^{+70}_{-42}$ & $1.68\pm0.70$ & $1.3^{+1.0}_{-0.6}$ & 0.75 (1.04) & 4.0 & 7.2 & MD \\
Gl 251 & $108.0\pm2.2$ & $4.0^{+1.2}_{-0.9}$ & $300$ & $0.30\pm0.12$ & $1.2^{+0.9}_{-0.5}$ & 0.83 (1.39) & 4.7 & 15.3 & DD \\
GJ 4063 & $40.7\pm3.5$ & $2.7^{+1.0}_{-0.7}$ & $93^{+42}_{-29}$ & $1.17\pm0.69$ & $1.0^{+0.7}_{-0.4}$ & 0.84 (1.03) & 4.0 & 7.9 & MD \\
Gl 725B & $135\pm15$ & $3.2^{+0.9}_{-0.7}$ & $122^{+81}_{-49}$ & $0.52\pm0.25$ & $1.0^{+0.8}_{-0.4}$ & 0.89 (1.20) & 4.7 & 12.9 & DD \\
PM J09553-2715 & $73.0\pm3.5$ & $7.4^{+5.1}_{-3.0}$ & $300$ & $0.81\pm0.60$ & $2.0^{+1.3}_{-0.8}$ & 0.85 (1.57) & 5.4 & 6.8 & MD \\
Gl 876 & $83.7\pm2.9$ & $6.9^{+2.0}_{-1.6}$ & $201^{+89}_{-62}$ & $0.40\pm0.13$ & $1.1^{+0.9}_{-0.5}$ & 0.70 (3.15) & 2.9 & 36.7 & DD \\
GJ 1012 &        & $2.3^{+2.6}_{-1.2}$ & $300$ & $1.00$ & $1.6^{+1.2}_{-0.7}$ & 0.92 (1.04) & 6.0 & -4.0 & ND \\
GJ 4333 & $71.0\pm1.5$ & $6.6^{+1.4}_{-1.2}$ & $150^{+52}_{-39}$ & $0.37\pm0.08$ & $1.1^{+0.9}_{-0.5}$ & 0.77 (2.21) & 4.6 & 49.8 & DD \\
Gl 445 &        & $1.4^{+1.5}_{-0.7}$ & $300$ & $1.00$ & $2.2^{+1.5}_{-0.9}$ & 1.02 (1.02) & 7.3 & -2.2 & ND \\
GJ 1148 &        & $2.7^{+1.7}_{-1.0}$ & $300$ & $1.00$ & $1.9^{+1.4}_{-0.8}$ & 1.04 (1.11) & 6.2 & 1.8 & ND \\
PM J08402+3127 & $89.5\pm8.0$ & $14.0^{+4.2}_{-3.2}$ & $219^{+58}_{-46}$ & $1.50$ & $1.7^{+1.3}_{-0.7}$ & 0.87 (1.13) & 6.4 & 7.6 & MD \\
GJ 3378 & $95.1\pm2.3$ & $4.9^{+1.6}_{-1.2}$ & $300$ & $0.44\pm0.21$ & $1.7^{+1.3}_{-0.7}$ & 0.92 (1.39) & 5.9 & 22.8 & DD \\
GJ 1105 &        & $2.1^{+2.1}_{-1.1}$ & $300$ & $1.00$ & $1.4^{+1.1}_{-0.6}$ & 0.93 (1.01) & 5.8 & -3.9 & ND \\
Gl 699 & $136\pm13$ & $5.6^{+0.8}_{-0.7}$ & $100$ & $0.40$ & $1.5^{+0.8}_{-0.5}$ & 0.95 (1.65) & 4.5 & 44.5 & DD \\
Gl 169.1A & $92.3\pm3.6$ & $5.6^{+2.6}_{-1.8}$ & $300$ & $1.08\pm0.53$ & $1.6^{+1.1}_{-0.6}$ & 0.91 (1.20) & 5.5 & 12.7 & DD \\
PM J21463+3813 & $93.9\pm3.4$ & $7.7^{+2.8}_{-2.0}$ & $300$ & $0.51\pm0.37$ & $2.1^{+1.7}_{-0.9}$ & 0.82 (1.14) & 9.1 & 14.2 & DD \\
Gl 15B & $113.3\pm4.3$ & $10.5^{+5.0}_{-3.4}$ & $250^{+83}_{-62}$ & $0.79\pm0.39$ & $2.4^{+1.6}_{-1.0}$ & 0.92 (1.50) & 8.0 & 23.8 & DD \\
GJ 1289 & $73.66\pm0.92$ & $53.2^{+12.4}_{-10.1}$ & $152^{+32}_{-27}$ & $0.48\pm0.09$ & $4.2^{+2.7}_{-1.6}$ & 0.82 (9.19) & 13.9 & 168.5 & DD \\
Gl 447 & $24.1\pm3.7$ & $11.1^{+5.8}_{-3.8}$ & $74^{+48}_{-29}$ & $1.42\pm0.69$ & $2.3^{+1.5}_{-0.9}$ & 0.85 (1.31) & 5.4 & 7.5 & MD \\
GJ 1151 & $175.6\pm4.9$ & $14.9^{+4.2}_{-3.3}$ & $300$ & $0.43\pm0.14$ & $1.6^{+1.3}_{-0.7}$ & 0.72 (1.85) & 6.7 & 75.2 & DD \\
GJ 1103 & $142.6\pm9.6$ & $8.3^{+4.1}_{-2.7}$ & $300$ & $0.51\pm0.29$ & $2.6^{+2.1}_{-1.2}$ & 0.79 (1.50) & 7.7 & 9.1 & MD \\
Gl 905 & $114.3\pm2.8$ & $13.3^{+2.5}_{-2.1}$ & $129^{+25}_{-21}$ & $0.43\pm0.09$ & $1.7^{+1.2}_{-0.7}$ & 0.84 (2.86) & 6.2 & 94.5 & DD \\
GJ 1002 & $89.8\pm2.8$ & $8.3^{+5.0}_{-3.1}$ & $300$ & $0.96\pm0.66$ & $2.1^{+1.7}_{-0.9}$ & 0.80 (1.14) & 8.3 & 7.2 & MD \\
GJ 1286 & $178\pm15$ & $16.7^{+4.6}_{-3.6}$ & $300$ & $0.29\pm0.09$ & $4.6^{+2.6}_{-1.6}$ & 1.02 (2.86) & 10.1 & 28.2 & DD \\
\hline
\end{tabular}
\label{tab:gpr}
\end{table*}

As in \citet{Donati23}, we use modified Jeffreys priors for $\theta_1$ (GP amplitude) and $\theta_5$ (white noise), with a knee set to $\sigma_B$, 
and a uniform prior for $\theta_4$ (smoothing parameter) in the range [0,3].  For $\theta_3$ (evolution timescale), we start with a log Gaussian prior centred on 150~d and 
a standard deviation of a factor of 3, then recenter it on the derived value in a second step (keeping the same standard deviation).  Finally, for the rotation period 
($\theta_2$, handled in linear space), we start with a uniform prior in the range [10,500]~d, then change it to a Gaussian prior on each of the (potentially multiple) regions 
of maximum likelihood (with a standard deviation equal to 25\% of the most probable local period), and ultimately select the period featuring the highest likelihood.  

We start with a first MCMC run where all parameters are free to vary.  In a few cases where the temporal variations of \Bl\ are weak, we choose to fix $\theta_4$ to 1.0 
or 1.5, to obtain a smooth fit to the \Bl\ data.  If $\theta_3$ reaches 300~d or more, suggesting that \Bl\ evolves slowly from one year to the next, we fix it to this 
value.  In one case where the data are sparse (Gl~338B), we fix $\theta_3$ to its optimal value (120~d).  In another one where the \Bl\ fluctuations are complex and 
varying rapidly (Gl~699), we have to fix $\theta_4$ to its optimal value (0.4) and $\theta_3$ to 100~d to ensure the latter does not get much smaller than the rotation 
period (which would prevent QP GPR to safely identify periodicity).  

To estimate the confidence level in the derived rotation period, we compare the results of this MCMC run with with those of 
another model where only the amplitude of the long-term \Bl\ variations is adjusted (by arbitrarily imposing $\theta_2=1500$~d, $\theta_3=300$~d and $\theta_4=1$).  We then 
compute the variation in marginal log likelihood $\Delta \log \mathcal{L}_M$ between the two solutions to assess whether and how reliably the QP term is detected and 
characterized.  If $\Delta \log \mathcal{L}_M>10$, 
we can claim a definite detection (DD) and a clear period;  if it falls in the range [5,10], we only have a marginal detection (MD) and a probable period;   
otherwise we have no detection (ND) of a quasi-periodic behaviour.  We also store the achieved \chisqr\ when fitting the long-term \Bl\ variations only, as a starting point reference.  
A few examples are shown in Fig.~\ref{fig:gjs} in the case of GJ~1289, Gl~687 and GJ~1002, where the 
detection of the QP modulation is obvious ($\Delta \log \mathcal{L}_M=168.5$), clear ($\Delta \log \mathcal{L}_M=34.6$) and marginal ($\Delta \log \mathcal{L}_M=7.2$), respectively.  

Table~\ref{tab:gpr} summarizes the results obtained for the 43 stars of our sample.  We obtain definite detections of the QP \Bl\ fluctuations for 29 stars, and marginal detections 
for 9.  Altogether, this is 11 more stars with either definite or marginal detections of the rotation period than in \citet{Fouque23}, as a likely result of the different data 
reduction and analysis.  Note that even in the case of definite detections, there is still a small chance that the derived period is off, e.g., when GPR confused the true period with 
an harmonic or an alias.  For instance, the period we derive for Gl~846 is twice larger than that quoted in \citet{Fouque23};  in this case, we made sure that $\Delta \log \mathcal{L}_M$ 
is significantly larger for our period than for that of \citet{Fouque23}.  For all other stars for which both studies derived rotation periods, estimates are consistent within error bars (see Sec.~\ref{sec:dis}).  
GPR plots for all stars with definite or marginal detections of the QP \Bl\ fluctuations (other than those of Fig.~\ref{fig:gjs}) plus Gl~436 (see Sec.~\ref{sec:ist}) are shown in Figs.~\ref{fig:a1}-\ref{fig:a12}.  

We stress that detecting the rotation period critically depends on whether \Bl\ fluctuations are large enough, and therefore on the epochs at which stars are observed as 
these fluctuations significantly evolve with time.  GJ~1151 is one obvious example (see Fig.~\ref{fig:a12} top panel), with the modulation barely visible for the first 2 seasons and 
getting much larger in 2022.  We speculate that this is likely why no periodic modulation is detected yet for a few of our sample stars, including Gl~436 whose rotation period is 
already known \citep{Bourrier18}.  More spectropolarimetric observations of these targets are thus expected to end up revealing a clear periodicity at some point.  

We note that the derived \chisqr\ of the GPR fit is on average smaller than 1 (<\chisqr>$=0.86\pm0.13$) and that $\theta_5$ is in most cases consistent with 0, confirming that our error 
bars are indeed slightly overestimated (by about 8\% on average), as anticipated in Sec.~\ref{sec:lsb}.  This is presumably why we sometimes obtain marginal (or even clear) 
detections of the \Bl\ modulation for stars in which the field or its variation was listed as not formally detected in Table~\ref{tab:sta}, like PM~J21463+3813 and GJ~1002 (see 
Figs.~\ref{fig:a10} middle panel and \ref{fig:a12} bottom panel).  

We finally point out that the derived rotation periods, when coupled to the stellar radii listed in Table~\ref{tab:str}, yield equatorial rotation velocities lower than 2~\kms\ for all 
stars, with a median value of 0.2~\kms, implying in practice undetectable rotational broadening of line profiles in SPIRou spectra.

\section{Comments on individual stars}
\label{sec:ist}

In this section, we discuss various specific points for about half the stars in our sample, focussing in particular on those for which clarifications are helpful.  
We do not discuss much the stars for which the periodicity of \Bl\ data is detected very clearly (e.g., GJ~1289, Gl~880, Gl~905, Gl~205, Gl~410), apart from 
mentioning that the \Bl\ curves we derive for GJ~1289, Gl~205 and Gl~410 (featuring peak-to-peak amplitudes in the range 150--200~G, 5--15~G and 40--90~G respectively) are 
consistent with those reported in previous studies based on optical ESPaDOnS and NARVAL data \citep{Moutou17, Hebrard16, Donati08c}.  We do not discuss either the 3 stars 
(GJ~1012, Gl~445 and GJ~1105) where neither \Bl\ nor its temporal variations and periodicity are detected.  

\subsection{Gl~338B}

Gl~338B is one of the stars for which we have the fewest data points (50), hence why we fixed $\theta_3=120$~d, which is already on the long side for an early-M star.   It is also 
one of the stars for which \citet{Fouque23} obtain no period from their \Bl\ analysis.  Our estimate ($42.2\pm4.1$~d, see Fig.~\ref{fig:a1} top panel) is marginally reliable, featuring
$\Delta \log \mathcal{L}_M=7.2$ with 
respect to the GPR fit including long-term modulation only.  One other local maximum at $110\pm17$~d shows up in the corner plot when we broaden the prior on $\theta_2$, 
with a slightly higher $\log \mathcal{L}_M$ (+1.8) than that of the main one we identified (at $42.2\pm4.1$~d).  Since the latter better matches the peak of the Lomb-Scargle periodogram 
\citep{Press92} of the \Bl\ curve (located at 44~d), 
we kept it as the most likely despite the slightly smaller $\log \mathcal{L}_M$.  We also note local maxima at 18 and 25~d \citep[the first one close to the tentative period reported from 
activity indices by][]{Sabotta21}, but with $\log \mathcal{L}_M$ values that are significantly lower than the 2 mentioned previously.  

\subsection{Gl~846}

Gl~846 is one star for which the period we derive is different from (and twice larger than) that found by previous spectropolarimetric studies \citep[][]{Hebrard16, Fouque23}.  
There is indeed a local maximum near 11~d 
in the MCMC corner plot, which also corresponds to the main periodogram peak (at 11d);  however, this local maximum period is much less probable than the main one we derive, with a likelihood 
contrast between the two of $\Delta \log \mathcal{L}_M=-27.3$.  We thus confirm that the rotation period of Gl~846 is $21.84\pm0.14$~d, which makes sense given the shape of the recorded \Bl\ 
curve, showing only weak power in the first harmonics at some epochs (e.g., around BJD~2458800, see Fig.~\ref{fig:a1} bottom panel).  The peak-to-peak amplitude of the \Bl\ modulation that we 
derive, ranging from 10 to 20~G, is fully consistent with that found by \citet[][]{Hebrard16} from optical data.  

\subsection{Gl~412A}

Gl~412A was listed in \citet{Fouque23} as exhibiting no periodic variations of \Bl\ whereas we find a clear QP behaviour with a period of $36.9\pm2.5$~d, significantly larger than where the main periodogram 
peak is located (28~d) in the period range 10--200~d.  It reflects that \Bl\ and its fluctuations evolve rapidly with time (see Fig.~\ref{fig:a3} middle panel), with a timescale $\theta_3\simeq80$~d.  We note 
in particular that the large-scale field of Gl~412A underwent rather drastic variations in our last observing season (2022).  We also note that the period we derive does not agree with the current literature 
estimate \citep[i.e., $100.9\pm0.3$~d,][]{Suarez18}, where our GPR analysis finds no local maximum. 

\subsection{Gl~411}

We find that Gl~411 exhibits a \Bl\ modulation with a period of $427\pm34$~d (see Fig.~\ref{fig:a4} top panel), consistent with that derived by \citet{Fouque23} and with the main periodogram peak 
(at 478~d).  It is the longest period found in our sample, with no other peak at shorter periods, including harmonics.  
Although the \Bl\ modulation is classified as being only marginally detected (see Sec.~\ref{sec:lsb}), the periodicity we detect is apparently reliable, with a $\Delta \log \mathcal{L}_M=11.9$ 
with respect to the GPR fit including long-term modulation only.  If this period is indeed the rotation period, it would imply that Gl~411 is an unusually slow rotator for a mid-M dwarf;  if the 
rotation period is in fact shorter \citep[e.g., the one suggested by][i.e., $56.15\pm0.17$~d]{Diaz19}, it would mean that the large-scale field of Gl~411 is nearly axisymmetric and  
generates undetectable rotational modulation of \Bl, but exhibits long-term QP intensity fluctuations with time, possibly as part of a much longer activity cycle.  

\subsection{Gl~617B}

Gl~617B is another of the sample stars for which \citet{Fouque23} finds no rotation period, and with no published value suggested in the literature.  We find clear periodicity in our \Bl\ data at 
a period of $40.4\pm3.0$~d (see Fig.~\ref{fig:a5} top panel), which coincides with the main peak of the periodogram (at 41~d).  We note that the periodogram also shows significant power at about twice this period;  
besides, the GPR fit finds a local maximum at $86\pm14$~d that is almost as likely as the main one ($\Delta \log \mathcal{L}_M=-1.5$) and could well be the true rotation period.  The shorter period being more 
in line with those of our other samples stars of similar spectral type (except Gl~411, see above), we select it as the main one.  

\subsection{Gl~480}

Like Gl~617B, Gl~480 has no published estimate of its rotation period, and \citet{Fouque23} did not succeed in deriving one.  We find a fairly 
precise estimate of $25.00\pm0.36$~d (see Fig.~\ref{fig:a5} middle panel), detected at almost the $\Delta \log \mathcal{L}_M=10$ threshold, and that exactly coincides with the maximum 
peak of the \Bl\ periodogram.  
We note that another local maximum is identified by GPR at a much longer period (of $170\pm14$~d), but whose significance is lower than the main one ($\Delta \log \mathcal{L}_M=-3.0$).  
We also find a third local maximum at about twice the rotation period we determined \citep[i.e., $50.9\pm1.3$~d and consistent with the period of the activity signal reported by][]{Feng20}, 
but again with a significance that is even lower than the previous one ($\Delta \log \mathcal{L}_M=-4.0$). 

\subsection{Gl~436}

This is again a star for which \citet{Fouque23} finds no period, and whose \Bl\ is no more than marginally fluctuating at the time of our observations (see Sec.~\ref{sec:lsb}).  There is however 
a fairly accurate period quoted in the literature \citep[$44.09\pm0.08$~d,][]{Bourrier18}.  Like in \citet{Fouque23}, our GPR analysis concludes that the periodicity is not detected in our SPIRou data, 
even marginally.  The period we derive, $48\pm13$~d, is nonetheless consistent with the literature value (albeit with a much larger error bar), hence why 
we mention it in Table~\ref{tab:gpr}.  We suspect that our non detection reflects that the \Bl\ modulation was weak at the time of our observations (see Fig.~\ref{fig:a5} bottom panel), leading 
to no more than an insignificant difference in $\log \mathcal{L}_M$ between the GPR fits.  

\subsection{Gl~408}

Gl~408 is another star for which our GPR analysis indicates clear periodicity of the \Bl\ fluctuations (see Fig.~\ref{fig:a6} middle panel), and for which no literature estimate 
\citep[including][]{Fouque23} exists. The period we derive, equal to $171.0\pm8.4$~d, is reliable (with a $\Delta \log \mathcal{L}_M$ contrast of 16.5 with respect to the GPR fit to 
long-term \Bl\ variations only) and coincides well with the main periodogram peak (169~d).  We fixed $\theta_3$ to a typical value of 200~d to help GPR converge given the small 
amplitude of the \Bl\ variations. 

\subsection{Gl~317}

Gl~317 is in a situation similar to Gl~480, with no known rotation period and none derived by \citet{Fouque23}.  We detect periodicity at a marginal level, with a period of $39.0\pm3.8$~d, i.e., 
3$\sigma$ lower than that given by the main peak of the periodogram (51~d) in the period range 10--200~d.  The low number of data points (77) and moderate sampling both contribute at keeping this 
detection marginal (see Fig.~\ref{fig:a7} top panel).   

\subsection{GJ~4063}

This is one of the few stars that features a \Bl\ curve with low amplitude fluctuations, listed as non detected in Sec.~\ref{sec:lsb}.  The GPR analysis nonetheless marginally detects periodicity 
in the data (at a level of $\Delta \log \mathcal{L}_M=7.9$, see Fig.~\ref{fig:a7} bottom panel) with a period of $40.7\pm3.5$~d that coincides both with the main periodogram peak (at 40~d) in the 10-100~d range, 
and is consistent with the literature estimate \citep{DiezAlonso19}.  This gives us confidence that the period we find is likely the true one, despite being only marginally detected.  

\subsection{Gl~725B}

With Gl~617B and Gl~408, Gl~725B is among the stars where we detect clear QP \Bl\ variations (see Fig.~\ref{fig:a8} top panel), and for which no estimate of the rotation period is available in the 
literature \citep[including][]{Fouque23}.  The GPR analysis yields a period of $135\pm15$~d, consistent within 1.3$\sigma$ with the main periodogram peak at 115~d, and 
features $\Delta \log \mathcal{L}_M$=12.9, i.e., a high enough contrast for the detection to be diagnosed as reliable.  

\subsection{PM~J09553-2715}  

Like GJ~4063, PM~J09553-2715 features low-amplitude \Bl\ fluctuations in addition to limited sampling (see Fig.~\ref{fig:a8} middle panel), but the GPR analysis is able to retrieve a 
period \citep[$73.0\pm3.5$~d, consistent with that of][]{Fouque23}, with a marginal confidence level ($\Delta \log \mathcal{L}_M=6.8$).  

\subsection{GJ~1148}

GJ~1148 is another star where \Bl\ is detected, but not its variations according to our statistical test (Sec.~\ref{sec:lsb}).  GPR is not able either to identify periodicity, even marginally, 
in the \Bl\ data.  A tentative periodicity shows up at $415\pm54$~d, i.e., almost as long as that found for Gl~411, but only at a very low confidence level ($\Delta \log \mathcal{L}_M=1.8$).  
We note that \citet{Fouque23} find a similar long-period signal in their \Bl\ data, although not reported explicitly in their paper.  

\subsection{PM~J08402+3127}  

PM~J08402+3127 features obvious long-term \Bl\ variations, including a clear sign switch between our first and second observing season (see Fig.~\ref{fig:a9} middle panel).  
The \Bl\ shorter-term modulation is weaker but nonetheless marginally detected with our QP GPR analysis, yielding a rotation period of $89.5\pm8.0$~d with a moderate confidence 
level ($\Delta \log \mathcal{L}_M=7.6$, more than 2$\sigma$ away from 
the main periodogram peak at 70~d.  The existing literature value \citep[$118\pm14$~d][]{DiezAlonso19}, is at best marginally consistent with our new estimate.  

\subsection{Gl~699}

Gl~699 is the star for which we have the largest number of visits (247), with both \Bl\ and its time fluctuations clearly detected.  Yet, the \Bl\ curve is complex and evolving rapidly 
with time (see Fig.~\ref{fig:a10} top panel).  As a result, the GPR modeling struggles to unambiguously determine the periodicity in the \Bl\ curve, and to converge to the known rotation period 
\citep[of $\simeq$140~d, e.g.,][]{ToledoPadron19, Fouque23}.  The main peak in the periodogram, located at 69~d, indicates that the \Bl\ modulation is dominated by the first harmonic.  Coupled to 
the uneven sampling, it leads GPR into reducing the evolution timescale $\theta_3$ to values much smaller than the rotation period, and lowering the smoothing parameter $\theta_4$ as well, making it 
hard to pinpoint periodicity.  We therefore fixed $\theta_3$ and $\theta_4$ to typical values, 100~d and 0.4 respectively, to help GPR converge.  The period we find, $136\pm13$~d, is consistent 
with the literature value, although with a rather large error bar.  

\subsection{PM~J21463+3813}

Although both \Bl\ and its fluctuations are listed as non detected per our statistical test (Sec.~\ref{sec:lsb}), the GPR analysis is nonetheless able to find a periodicity (see Fig.~\ref{fig:a10} 
bottom panel), with a confidence level high enough to claim a definite detection ($\Delta \log \mathcal{L}_M=14.2$).  The period we find, equal to $93.9\pm3.4$~d, is consistent with the main peak in the 
periodogram (88d).  No rotation period is mentioned in the literature for this star, including \citet{Fouque23} whose analysis did not succeed in detecting it.  

\subsection{Gl~447}

Despite the sparse and very unevenly sampled data set, GPR is able to identify a tentative period from the clearly detected \Bl\ values and temporal fluctuations (see Fig.~\ref{fig:a11} middle panel), 
equal to $24.1\pm3.7$~d and in agreement with the main peak in the periodogram (at 23~d).  The detection is only marginal, with $\Delta \log \mathcal{L}_M=7.5$, and may in fact be an artifact of the 
poor sampling (although the window function shows no peak in the corresponding period range).  We note that this period is much shorter than that reported by \citet{Suarez16}, and in fact suspiciously 
short for an inactive late-M dwarf like Gl~447.  No period was reported in \citet{Fouque23}.  

\subsection{GJ~1151}

GJ~1151 is a perfect demonstration of the critical need of long-term monitoring for detecting periodicities, and more generally the large-scale magnetic field and its temporal evolution.  As shown in 
Fig.~\ref{fig:a11} (bottom panel), \Bl\ was no more than marginally detected in the first 2 seasons, making it ambiguous to determine a rotation period.  In 2022 however, \Bl\ started to exhibit a much 
larger modulation, from which GPR is able to safely retrieve a rotation period of $175.6\pm4.9$~d, with one of the highest confidence level of the whole sample (($\Delta \log \mathcal{L}_M=75.2$), and 
located not far from the main peak in the periodogram (160~d).  We stress that this period is consistent with the one derived by \citet{Fouque23}, but different from the other existing literature 
estimates \citep[e.g.,][]{Irwin11, DiezAlonso19}.  

\subsection{GJ~1103}

GJ~1103 is another star on which the temporal fluctuations of \Bl\ are listed as non detected in Table~\ref{tab:sta} (Sec.~\ref{sec:lsb}), and for which the GPR analysis is able to identify a period 
(see Fig.~\ref{fig:a12} top panel).  In fact, given the relative sparseness of the data, 2 periods show up, one being the first harmonic of the second.  We find that the longer one, $142.6\pm9.6$~d is 
slightly more likely than the shorter one, but only by a small amount ($\Delta \log \mathcal{L}_M=0.6$), with the main peak in the periodogram (at 76~d) coinciding with the first harmonic.  
This period is only marginally detected ($\Delta \log \mathcal{L}_M=9.1$) and is consistent with that derived by \citet{Fouque23}.  

\subsection{GJ~1002}

As for PM~J21463+3813, GJ~1002 is listed in Table~\ref{tab:sta} as being non detected for both \Bl\ and its temporal fluctuations, whereas the GPR analysis succeeds in digging out a periodicity (see 
Fig.~\ref{fig:gjs} bottom panel), though at a marginal level ($\Delta \log \mathcal{L}_M=7.2$).  The period we find, $89.8\pm2.8$~d supports that of \citet{Fouque23} but not the recently published one of 
\citet{Suarez23}.

\section{Summary and discussion}
\label{sec:dis}

We scrutinized the complete set of SPIRou observations for the sample of 43 M dwarfs studied by \citet{Fouque23}, but reducing and analyzing the data with \texttt{Libre-ESpRIT} 
\citep[the nominal ESPaDOnS reduction package optimized for spectropolarimetry and adapted for SPIRou,][]{Donati97b, Donati20} and with LSD (with VALD-3 M0 and M3 masks using lines 
deeper than 10\% of the continuum) to reliably diagnose (with \chisq\ tests) whether \Bl\ and temporal fluctuations are detected, and if \Bl\ QP fluctuations are observed 
(using GPR and MCMC in a Bayesian framework).  Our reduction tools are different from those used in \citet{Fouque23}, where data were processed with \texttt{APERO}  
\citep[the nominal SPIRou reduction package optimized for RV precision,][]{Cook22} and analyzed with a different implementation of LSD.  

We find that the \Bl\ values and error bars derived with our reference reduction tools, consistent with those from previous studies based on optical ESPaDOnS and NARVAL data 
for stars observed in both domains \citep[e.g., Gl~205, Gl~410, Gl~846, GJ~1289,][]{Donati08c,Hebrard16,Moutou17}, are on average 2$-$3$\times$ smaller than those obtained by 
\citet{Fouque23}.  This issue, {\emr likely attributable to the alternate LSD implementation used in \citet{Fouque23} and to a lesser extent in the way \Bl\ values are derived 
from LSD profiles}, is currently being investigated by the team, but should not affect their conclusions regarding rotation periods.  

\begin{figure}
\centerline{\includegraphics[scale=0.50,angle=-90]{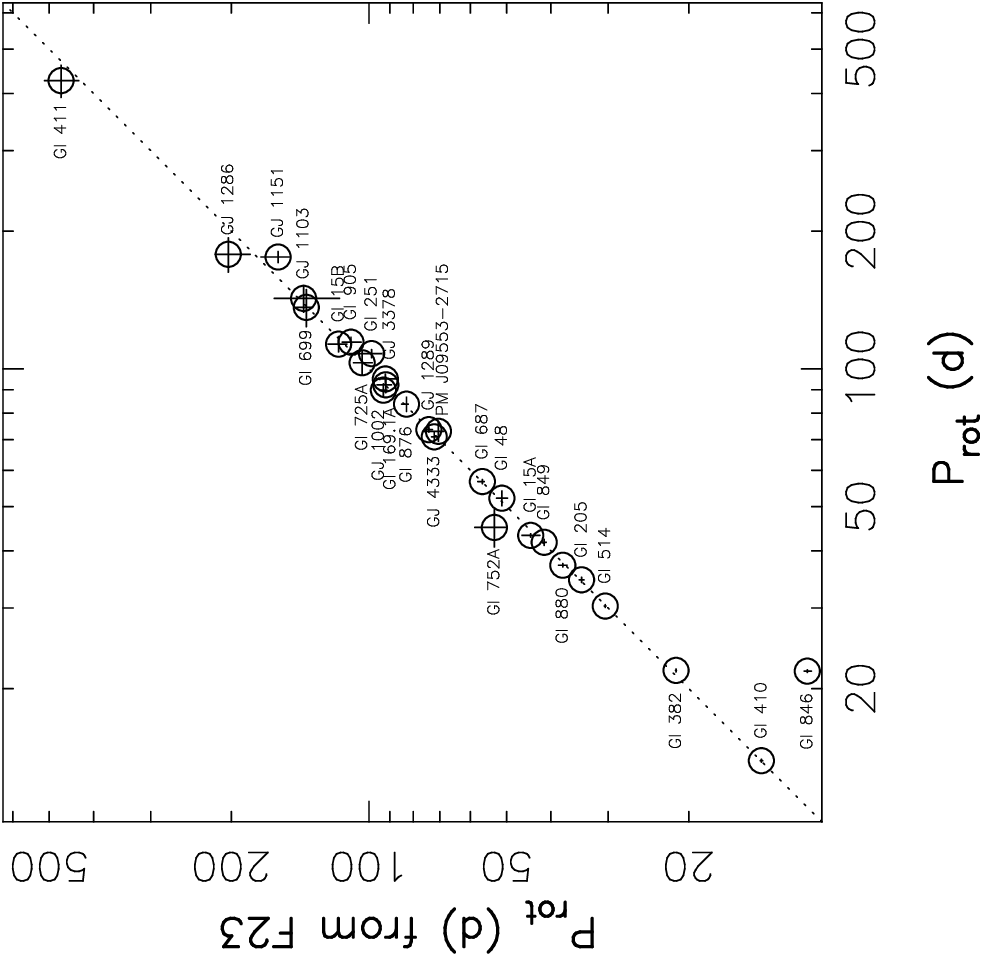}} 
\caption[]{\emr Comparison between rotation periods and error bars derived in this paper (on the horizontal axis) with those of \citet[][on the vertical axis]{Fouque23} for the 27 stars for 
which  \citet{Fouque23} measured a period (the dotted line depicting equality).  Except for Gl~846 where \citet{Fouque23} retrieves the first harmonic, both studies are consistent within better 
than 2$\sigma$.} 
\label{fig:prot}
\end{figure}

Altogether, we find that only 3 stars in the whole sample (GJ~1012, Gl~445 and GJ~1105) show no \Bl\ detection at all.  In the remaining set of 40 stars, one shows no \Bl\ modulation 
(GJ~1148, exhibiting no more than a very marginal low-amplitude variation with a long period of over 400~d), another one (Gl~436) features a low-amplitude modulation at the expected period, 
a third one (Gl~447) is so poorly sampled that the derived period of its weak modulation is suspicious.  For seven others, we are able to measure a period with marginal confidence that we 
tentatively identify as the stellar rotation period.  This is the first such measurement in the case of Gl~480 and Gl~317, whereas the period we derive is different from the literature value 
for Gl~338B and PM~J08402+3127, and consistent with the most recent estimate for GJ~4063, PM~J09553-2715 and GJ~1002 \citep[the latter two in agreement with][]{Fouque23}.  
Last but not least, we obtain definite detections of the periodic \Bl\ variations for 30 stars of our sample, for the first time 
in the case of Gl~617B, Gl~408, Gl~725B and PM~J21463+3813, contradicting the existing literature estimate for Gl~846 and Gl~412A, and in agreement 
with \citet{Fouque23} for the remaining 24 stars.  
Out of the 27 stars for which \citet{Fouque23} derived a rotation period, we find a consistent estimate for 26 of them, and disagree for only one (Gl~846) for which the period we 
infer is twice longer (i.e., the fundamental vs the first harmonic for \citealt{Fouque23}).  {\emr The comparison between rotation periods from both papers is shown in Fig.~\ref{fig:prot}.} 

On average, the periods we measure are shorter for early-M stars than for mid- to late-M stars.  For an unbiased sample of M dwarfs, the known trend is the opposite, with late-M 
dwarfs rotating statistically faster and being more active than early-M ones \citep[e.g.,][]{Delfosse98, Browning10, West15, Newton16}.  
Our result actually reflects that the SLS mainly focussed on inactive dwarfs to minimize the activity jitter in their RV curves, and therefore ended up discarding from the sample most 
late-M dwarfs with rotation periods shorter than a few tens of days \citep{Moutou17}.  Activity, known to grow with decreasing Rossby number Ro
\citep[defined as the rotation period normalised by the convective turnover time $\tau$, with $\tau$ ranging from 30 to 150~d from early- to late-type M dwarfs, e.g.,][]{Wright18},  
is indeed stronger for late-M dwarfs than for early-M dwarfs at a given rotation period.  One exception to our biased trend is Gl~411 for which we derive an ultra long period of $427\pm34$~d, the 
longest of our whole sample.  If it is indeed the rotation period, it may indicate that the evolution of Gl~411 was different from the bulk of our sample.  However, the abundance analysis of 
\citet{Cristofari22b} does not suggest that this is the case, with Gl~411 being apparently a rather standard member of the thick galactic disc (along with Gl~699) with low [M/H] and high [$\alpha$/Fe] 
(the abundance of $\alpha$ elements with respect to Fe).  Another option is that the rotation period of Gl~411 is much shorter \citep[e.g., the one suggested by][i.e., $56.15\pm0.17$~d]{Diaz19} and 
went undetected because of a perfectly axisymmetric large-scale field (generating no rotational modulation of \Bl) over the full timescale of our observations.  
Besides, we find that large-scale magnetic fields tend to evolve faster (i.e., lower $\theta_3$) for the higher-mass stars of our sample \citep[in agreement with][]{Fouque23}, which may simply 
reflect that these higher-mass stars are on average more active than the other sample stars.  

\begin{figure*}
\centerline{\includegraphics[scale=0.70,angle=-90]{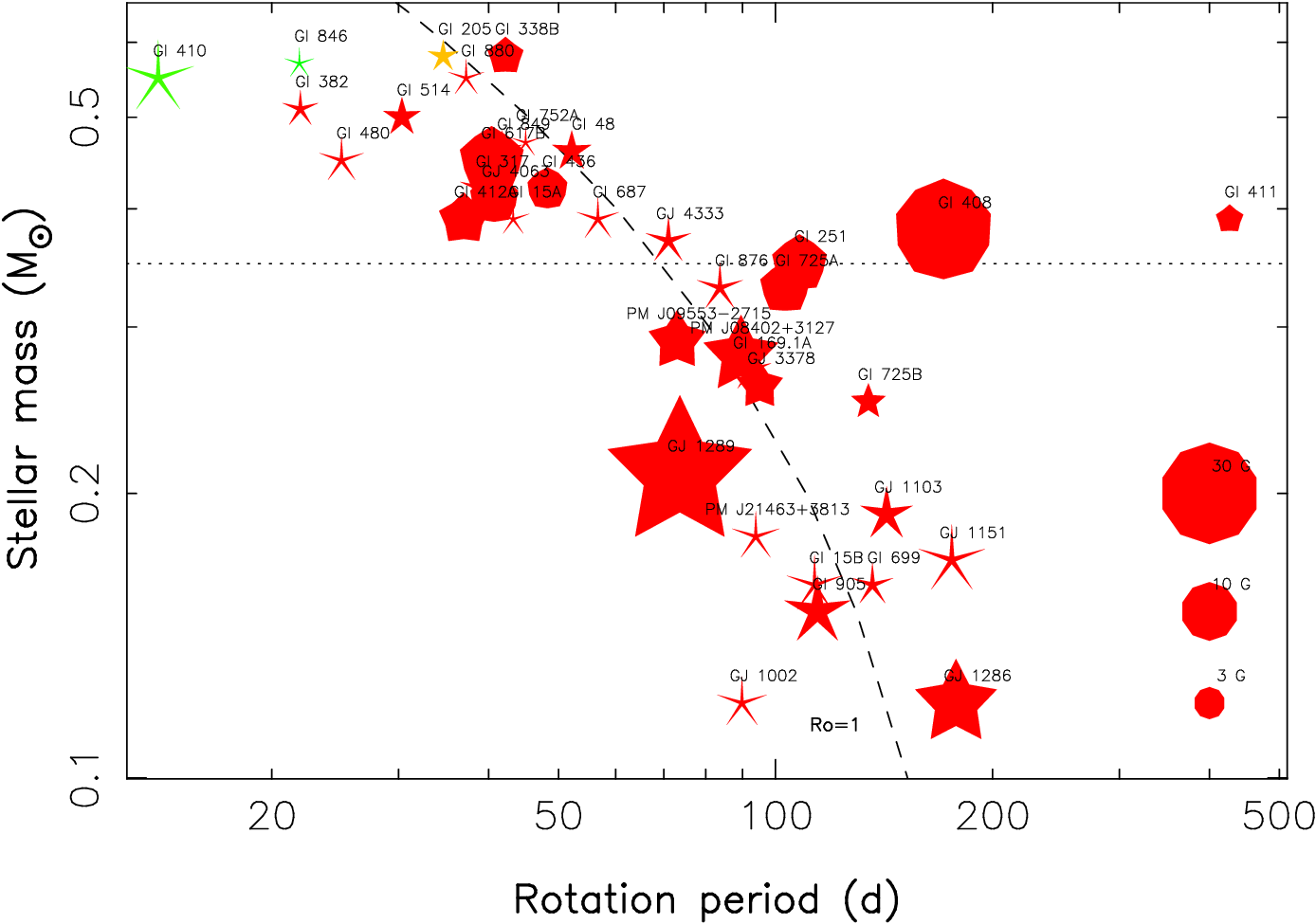}} 
\caption[]{Properties of the large-scale field for the 38 stars of our sample for which a rotation period was measured, leaving out Gl~447 for which the recovered period is suspicious as a result of 
the sparse data set and poor sampling (see text).  Symbol size depicts the strength of the large-scale field, whereas symbol shape describes the degree of axisymmetry (decagons for fully axisymmetric 
fields and stars for fully non-axisymmetric fields).  Symbol color tentatively 
illustrates the field topology (red to blue for purely poloidal to purely toroidal fields) for the few stars in which the toroidal field is detected.  The dashed line traces where Ro equals 1 
\citep[using convective turnover times from][]{Wright18}, whereas the dotted line marks the mass below which M dwarfs become fully convective.  The 3 red decagons in the bottom right corner 
indicate the symbol size for 3 typical values of the average longitudinal field.  } 
\label{fig:cfg}
\end{figure*}

Our results can be used to study how the large-scale magnetic fields of moderately- to weakly-active M dwarfs change with stellar parameters such as mass and rotation period, as in 
\citet{Donati09}.  
To estimate in a simple way the average amount of magnetic energy in the large-scale field of each sample star, we quadratically sum the average \Bl\ (second column of Table~\ref{tab:sta}) 
and the semi-amplitude of the QP GP $\theta_1$ (third column of Table~\ref{tab:gpr}) fitted to the \Bl\ curve, which can be respectively interpreted at first order as the amount of magnetic 
energy stored in the axisymmetric and non-axisymmetric components of the large-scale field.  The latter two components can also be used to estimate the degree of axisymmetry of the large-scale 
field;  for instance, a semi-amplitude $\theta_1$ much weaker than |<\Bl>| means a field that is mostly axisymmetric, whereas the opposite implies a field that is mostly non-axisymmetric.  

Estimating the relative amount of magnetic energy in the poloidal and toroidal components \citep[as in][]{Donati09} is more tricky, as the toroidal component is in fact hardly detectable in 
these low \vsini\ stars, except for the highest mass and fastest rotating targets of our sample \citep{Lehmann22} whose \vsini\ reaches up to $\simeq$1.5~\kms.  By using Zeeman-Doppler 
Imaging \citep[ZDI,][]{Donati06b} to carry out a preliminary analysis of our sample stars, we confirm that we are only able to detect a significant toroidal component (storing from 20 up to 
50\% of the magnetic energy) on 3 sample stars, namely Gl~410, Gl~846 and Gl~205, in agreement with previous studies based on optical \citep{Donati08c, Hebrard16, Moutou17} and nIR 
\citep{Cortes23} data.  For these 3 stars, we thus take the average amounts of toroidal energy derived with ZDI, whereas for all other stars, we assume that the field is fully poloidal.  
Note that it does not mean that the slowly rotating and / or less massive M dwarfs do not host significant toroidal fields, but rather that we are not able to detect them.  

We show in Fig.~\ref{fig:cfg} how our sample stars behave with respect to one another in a mass versus rotation period diagram, where the symbol size, shape and color illustrate the main 
characteristics of the large-scale field.  As opposed to previous results on more massive and more rapidly rotating stars \citep{Donati09,Vidotto14b,See15,Folsom18} that all clearly 
demonstrate that their large-scale fields get weaker with increasing Ro, we do not see the same trend in our sample, with large-scale field being more or less constant in strength or even 
increasing with increasing Ro.  We can see in particular stars rotating with periods of almost 200~d (like Gl~408, GJ~1151 or GJ~1286) hosting fields that are as strong or stronger than faster 
rotators (like GJ~1002, PM~J21463+3813 or Gl~251).  One trend that seems to emerge is that the lower mass dwarfs of our sample host fields that are less likely to 
be axisymmetric than the higher mass ones.  All 14 stars with $\mstar<0.3$~\msun\ and whose longitudinal field is detected indeed exhibit mostly non-asisymmetric large-scale fields, whereas 9 
of the remaining 24 higher mass dwarfs (10 out of 25 if we also include GJ~1148) feature mostly axisymmetric fields.  Besides, we see no obvious sign that the bistable magnetic behaviour 
reported for rapidly rotating late-M dwarfs \citep{Morin10} also applies for the slowly-rotating ones.  
A more detailed study will require every single star of the sample to be studied with ZDI, the first of such papers concentrating on 6 of them 
(Gl~617B, Gl~408, GJ~1289, GJ~1151, Gl~905 and GJ~1286) being ready for publication (Lehmann et al., in prep).  In parallel, a study of the small-scale fields of all sample stars will be 
carried out from the measurement of Zeeman broadening following \citet{Cristofari23}, which will allow us to diagnose 
how the large-scale and small-scale fields correlate with each other (Cristofari et al., in prep), as recently done for the young active M dwarf AU~Mic \citep{Donati23}.  

Another interesting feature that our data reveal is that most slowly rotating M dwarfs, including fully convective ones, undergo obvious large-scale field variations, with some switching 
polarity during our monitoring like Gl~876 (Fig.~\ref{fig:a8} bottom panel, see also \citealt{Moutou23}), PM~J08402+3127 (Fig.~\ref{fig:a9} middle panel) and Gl~169.1A 
(Fig.~\ref{fig:a10} middle panel).  Some other stars seem to succeed in amplifying their fields after a few years of relative magnetic quiescence like GJ~1151 (Fig.~\ref{fig:a11} bottom 
panel), or to achieve the opposite like Gl~905 (Fig.~\ref{fig:a12} middle panel).  Ideally, one would like to pursue such monitoring on a timescale of at least a decade to investigate 
whether M dwarfs, and in particular fully-convective ones, undergo activity cycles as claimed by \citet{Route16} from radio observations, and to study how the properties of the large-scale 
fields evolve as stars progress along their cycle \citep{Lehmann21}.  This observational approach would give the opportunity of scrutinizing for the first time the physical processes at 
work in magnetic cycles of fully-convective stars that lack a tachocline, an ingredient whose role in generating solar-like magnetic cycles is still debated \citep{Brun17}.  

Spectropolarimetric observations with SPIRou at CFHT are already being pursued for some of our sample stars, both in the framework of the SPICE Large Programme (a follow-up of the SLS 
carried out from mid-2022 until mid-2024, and focussed mainly on the lowest mass dwarfs) and within a multi-semester PI program (PI: A.\ Carmona) targeting the most promising SLS targets 
in terms of planet detection and characterization.  Altogether, additional observations will be collected for about 20 stars of our sample, that will be used to update our results in a 
couple of years and further confirm the rotation periods derived in \citet{Fouque23} and our paper.

\section*{Acknowledgements}
We thank an anonymous referee for valuable comments on a previous version of the manuscript.  
Our study is based on data obtained at the CFHT, operated by the CNRC (Canada), INSU/CNRS (France) and the University of Hawaii.  
This project received funding from the European Research Council (ERC) under the H2020 research \& innovation program (grant agreements \#740651 NewWorlds), 
from the French National Research Agency in the framework of the Investissements d'Avenir program (ANR-15-IDEX-02), and from the ``Origin of Life'' project 
of the Grenoble-Alpes University. 
The authors wish to recognise and acknowledge the very significant cultural role and reverence that the summit of Maunakea has always had 
within the indigenous Hawaiian community.  We are most fortunate to have the opportunity to conduct observations from this mountain.
This work also benefited from the SIMBAD CDS database at URL {\tt http://simbad.u-strasbg.fr/simbad} and the ADS system at URL {\tt https://ui.adsabs.harvard.edu}.

\section*{Data availability}  All data underlying this paper are part of the SLS, and will be publicly available from the Canadian Astronomy Data
Center by February 2024.

\bibliography{mdwspi}
\bibliographystyle{mnras}

\appendix

\section{QP GPR fits to \Bl\ data}
\label{sec:appA}

In this appendix, we show the QP GPR fit to the \Bl\ data for all stars of our sample exhibiting either clear or marginal periodicity (except those already shown in 
Fig.~\ref{fig:gjs}) plus that of Gl~436 (see main text in Sec.~\ref{sec:ist}).  

\begin{figure*}
\centerline{\includegraphics[scale=0.58,angle=-90]{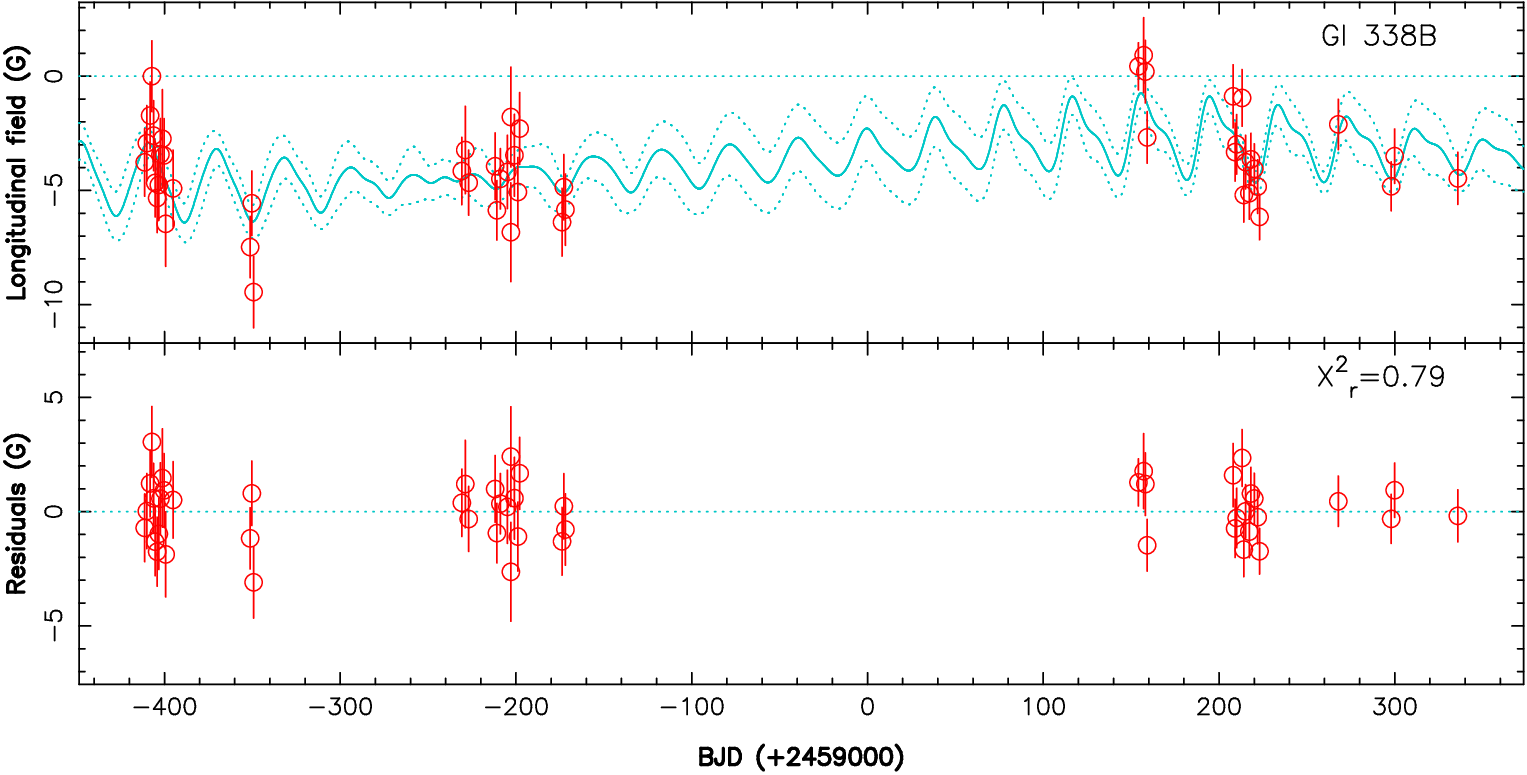} \vspace{2mm}} 
\centerline{\includegraphics[scale=0.58,angle=-90]{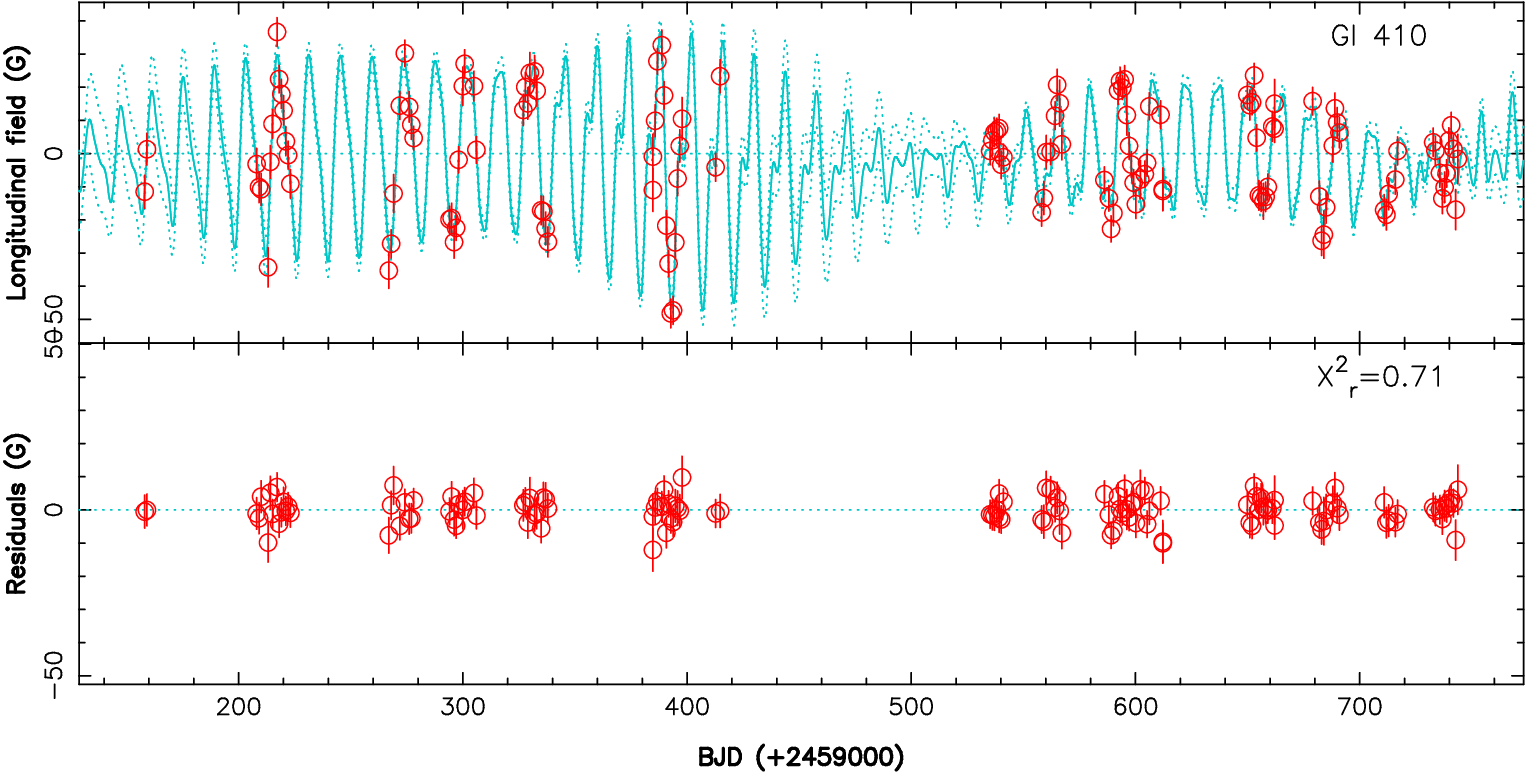} \vspace{2mm}} 
\centerline{\includegraphics[scale=0.58,angle=-90]{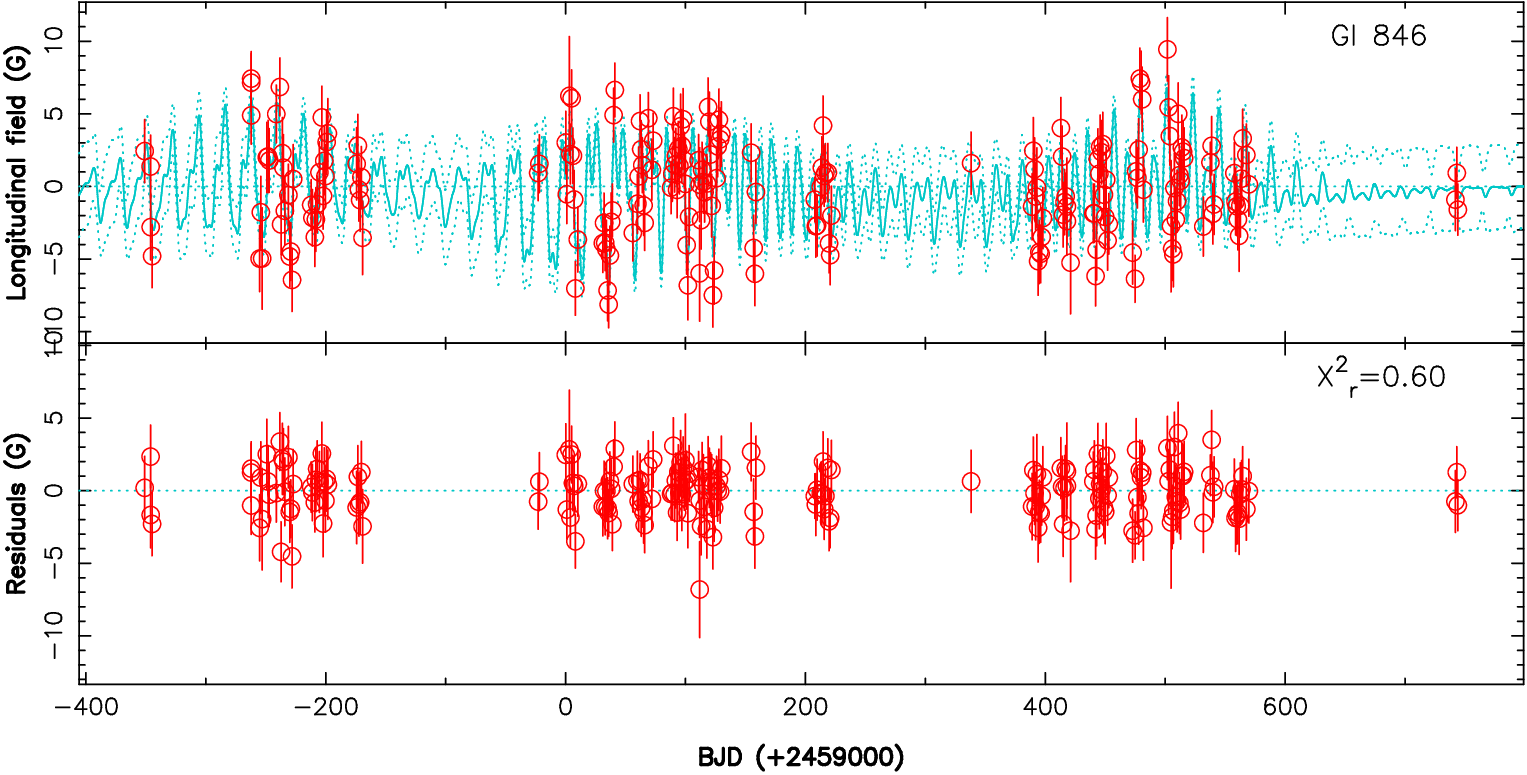}} 
\caption[]{Same as Fig.~\ref{fig:gjs} for Gl~338B (top), Gl~410 (middle) and Gl~846 (bottom). } 
\label{fig:a1}
\end{figure*}

\begin{figure*}
\centerline{\includegraphics[scale=0.58,angle=-90]{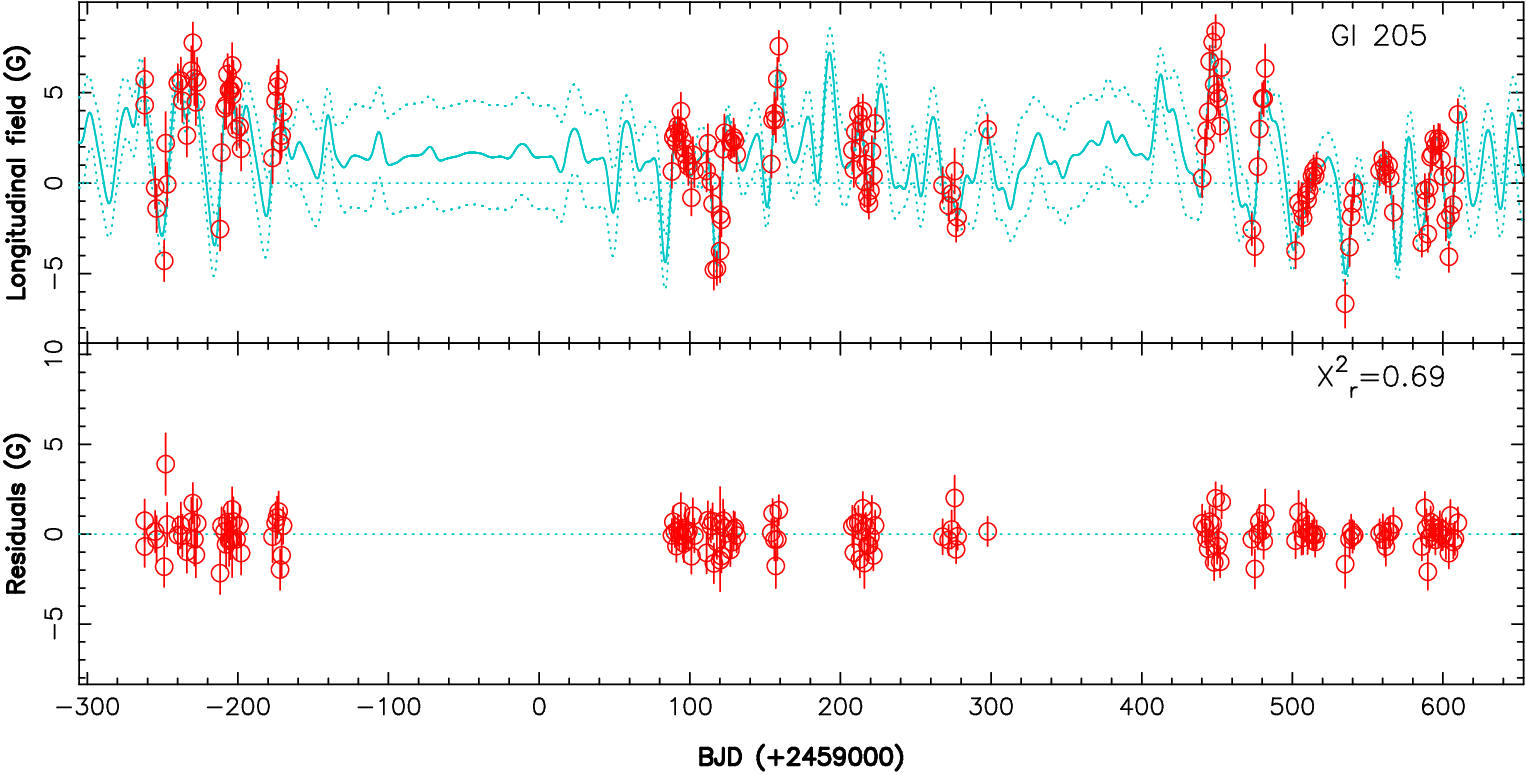} \vspace{2mm}} 
\centerline{\includegraphics[scale=0.58,angle=-90]{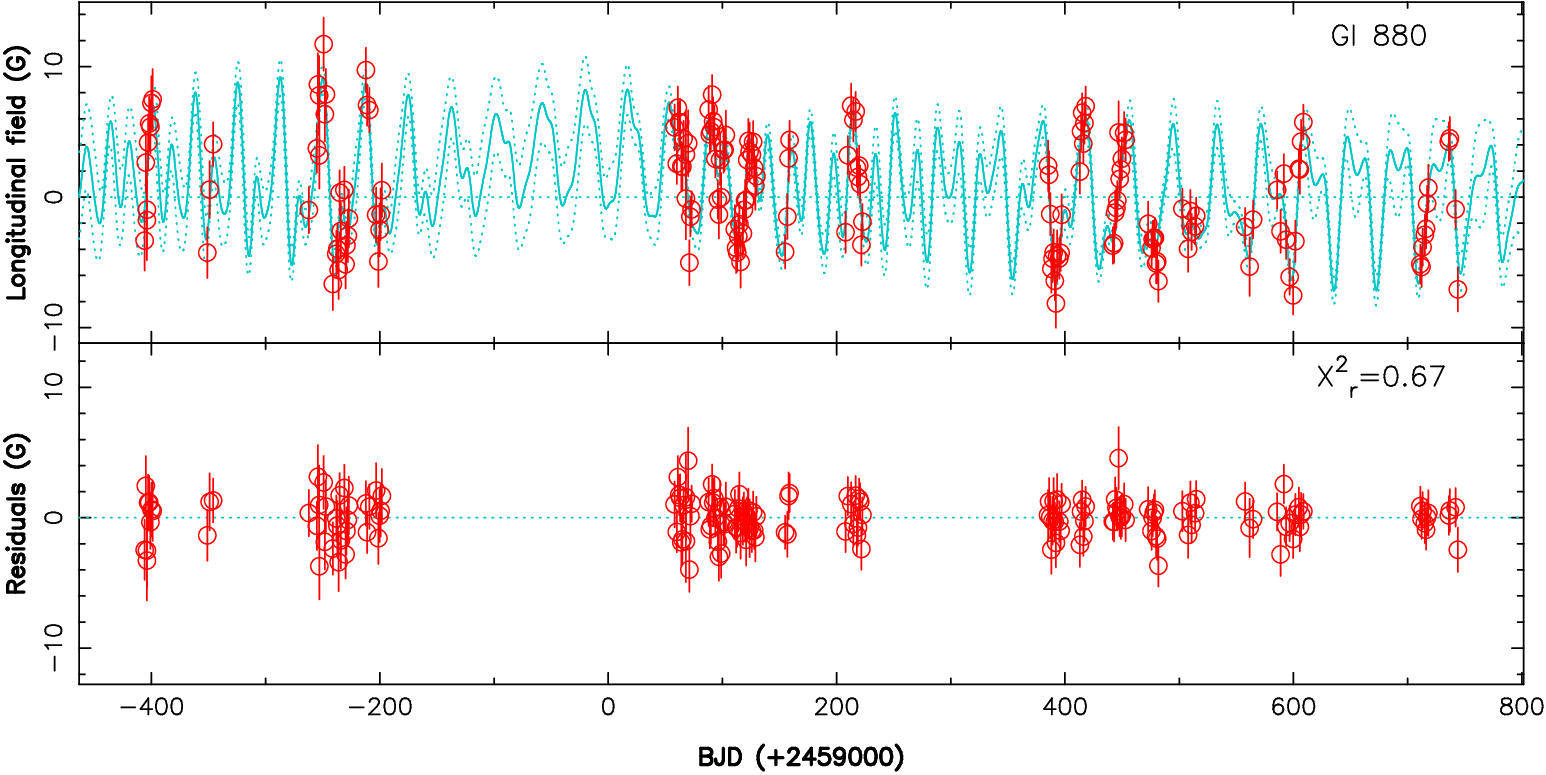} \vspace{2mm}} 
\centerline{\includegraphics[scale=0.58,angle=-90]{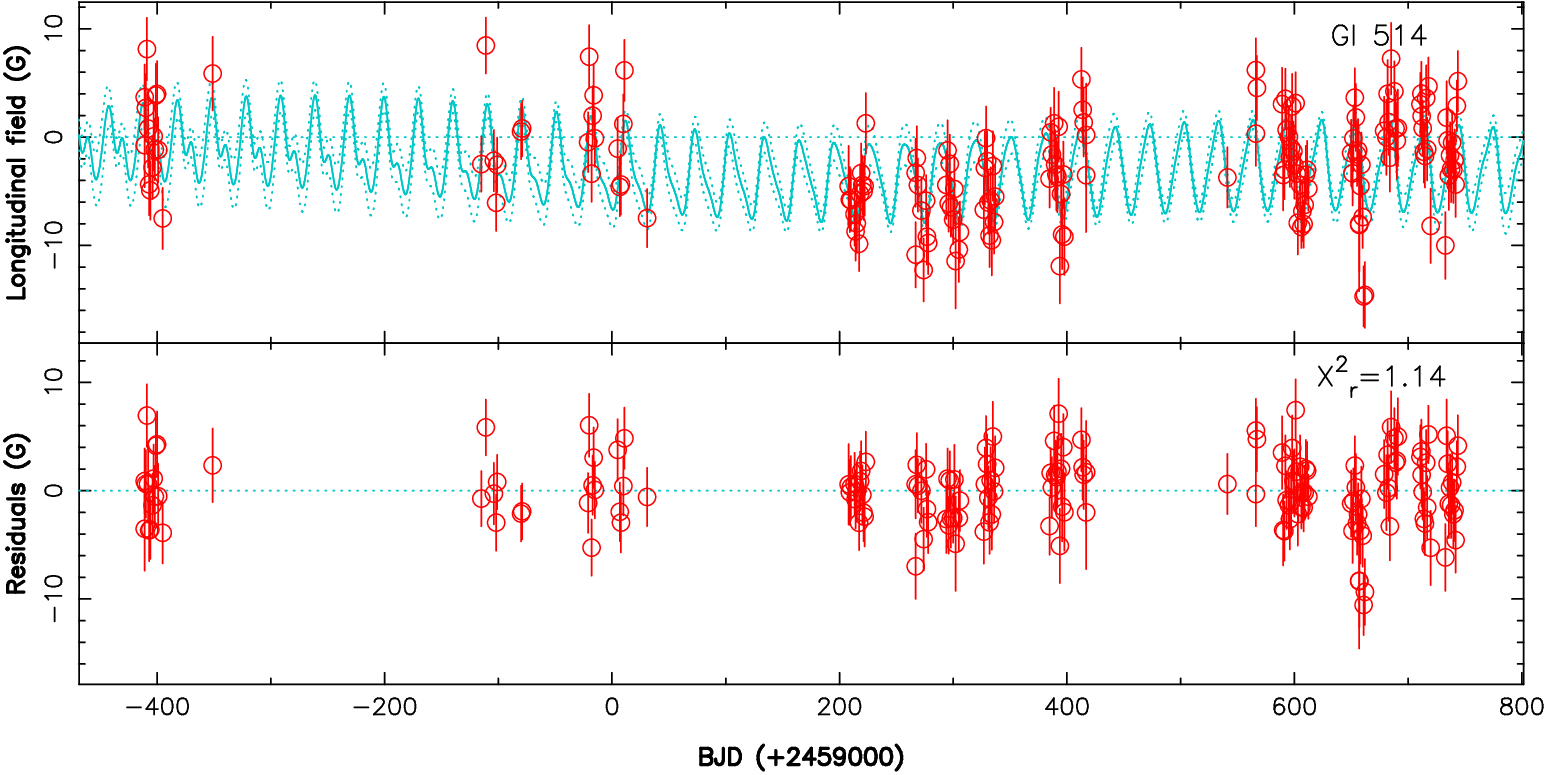}} 
\caption[]{Same as Fig.~\ref{fig:gjs} for Gl~205 (top), Gl~880 (middle) and Gl~514 (bottom). } 
\label{fig:a2}
\end{figure*}

\begin{figure*}
\centerline{\includegraphics[scale=0.58,angle=-90]{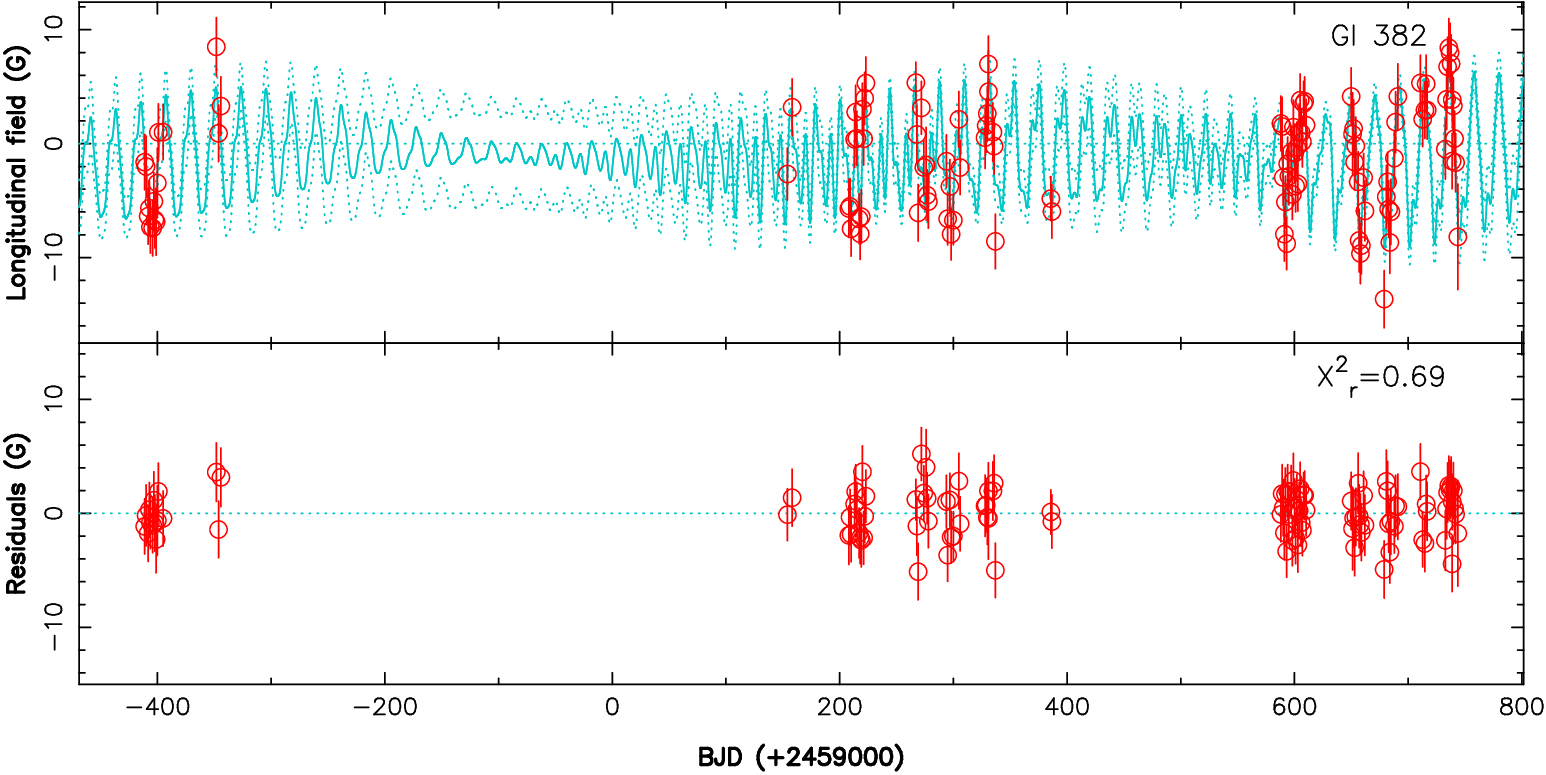} \vspace{2mm}} 
\centerline{\includegraphics[scale=0.58,angle=-90]{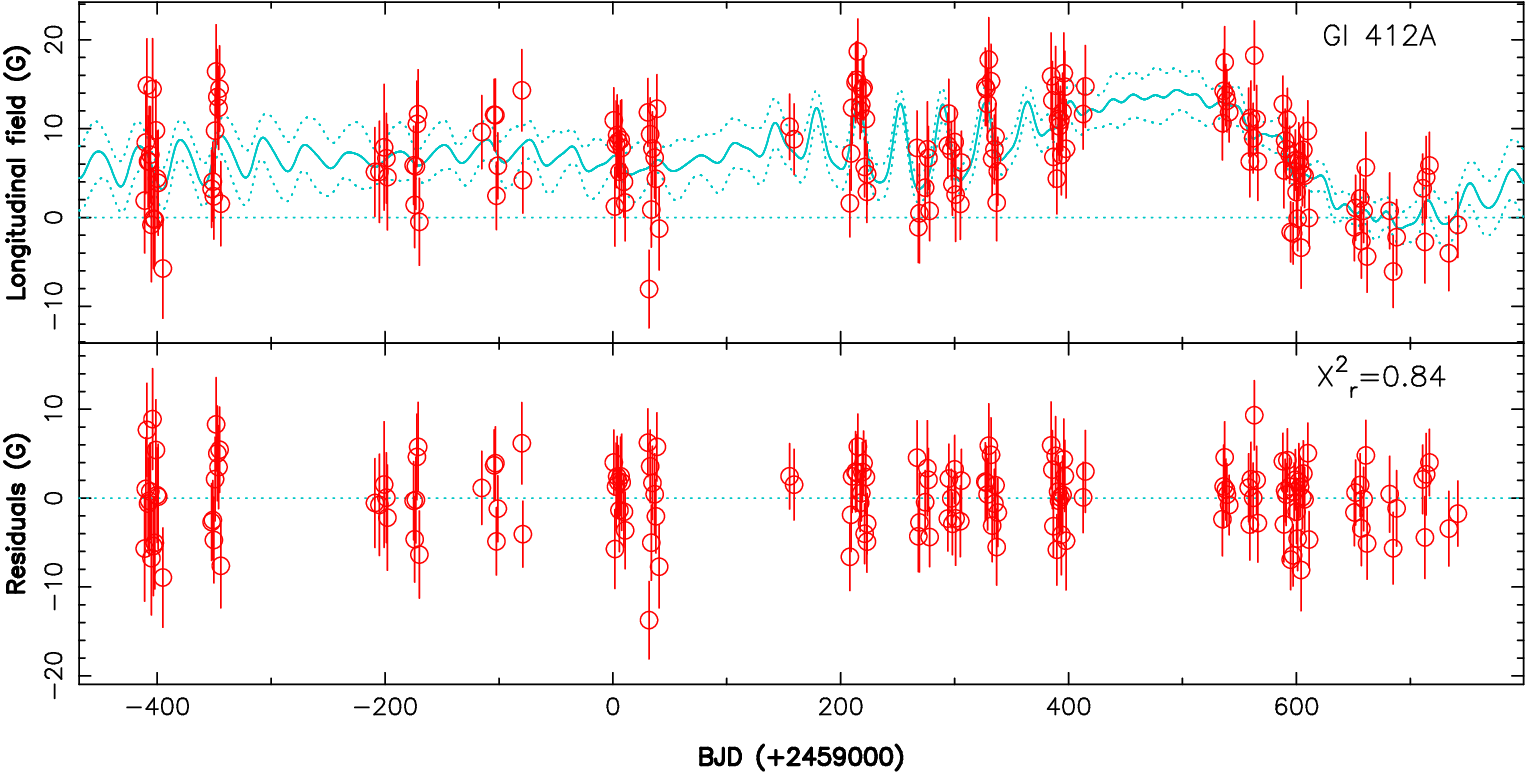} \vspace{2mm}} 
\centerline{\includegraphics[scale=0.58,angle=-90]{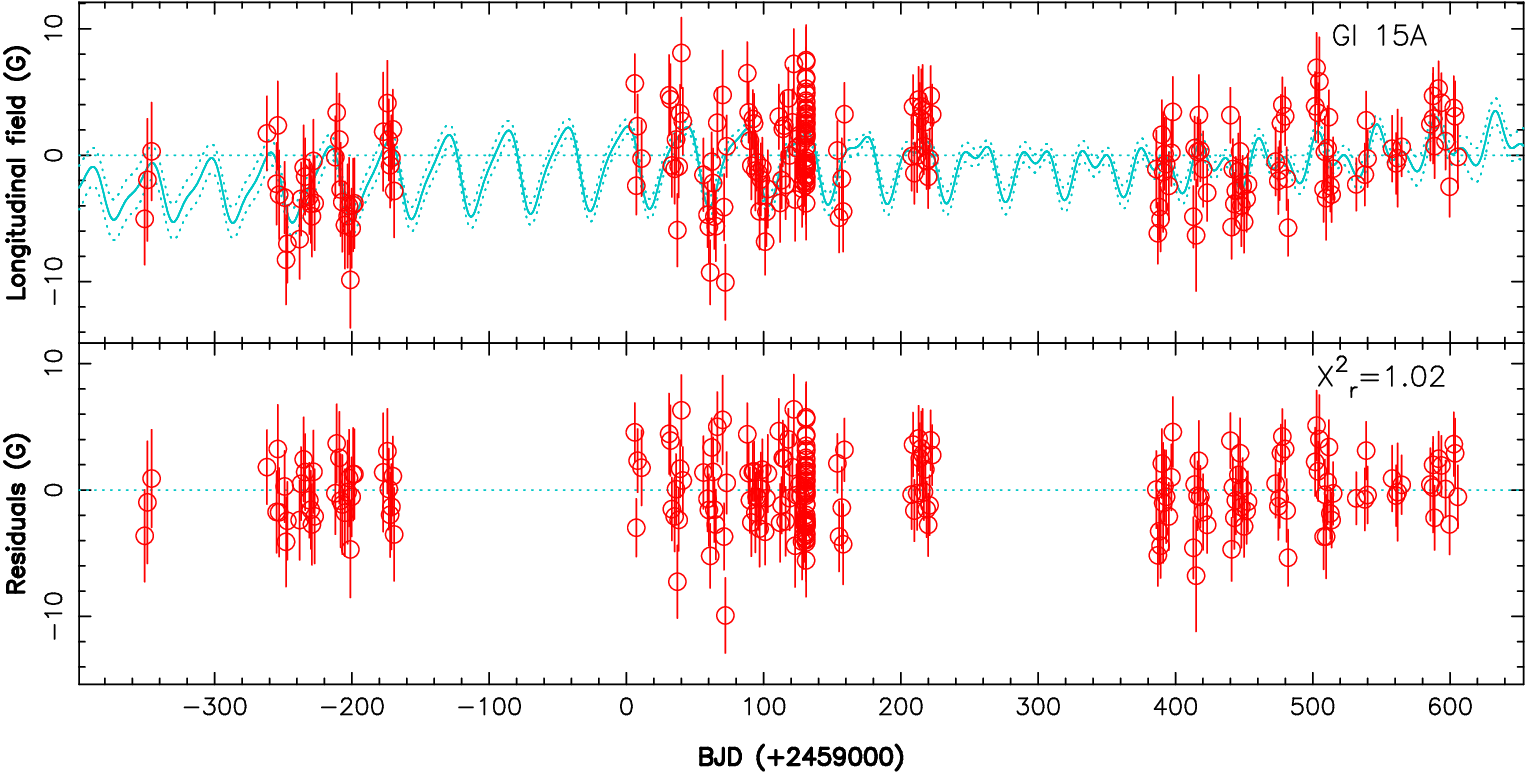}} 
\caption[]{Same as Fig.~\ref{fig:gjs} for Gl~382 (top), Gl~412A (middle) and Gl~15A (bottom). } 
\label{fig:a3}
\end{figure*}

\begin{figure*}
\centerline{\includegraphics[scale=0.58,angle=-90]{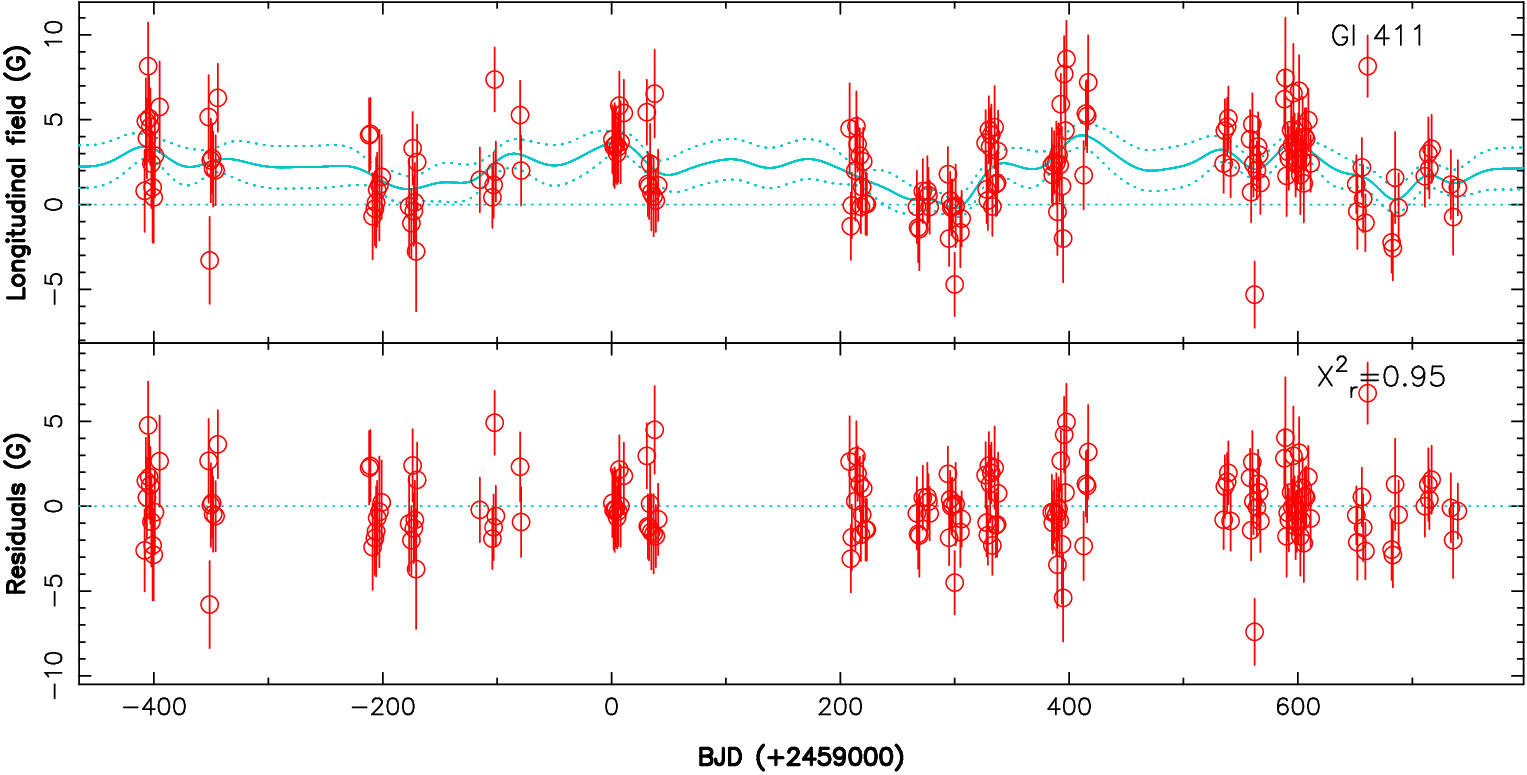} \vspace{2mm}} 
\centerline{\includegraphics[scale=0.58,angle=-90]{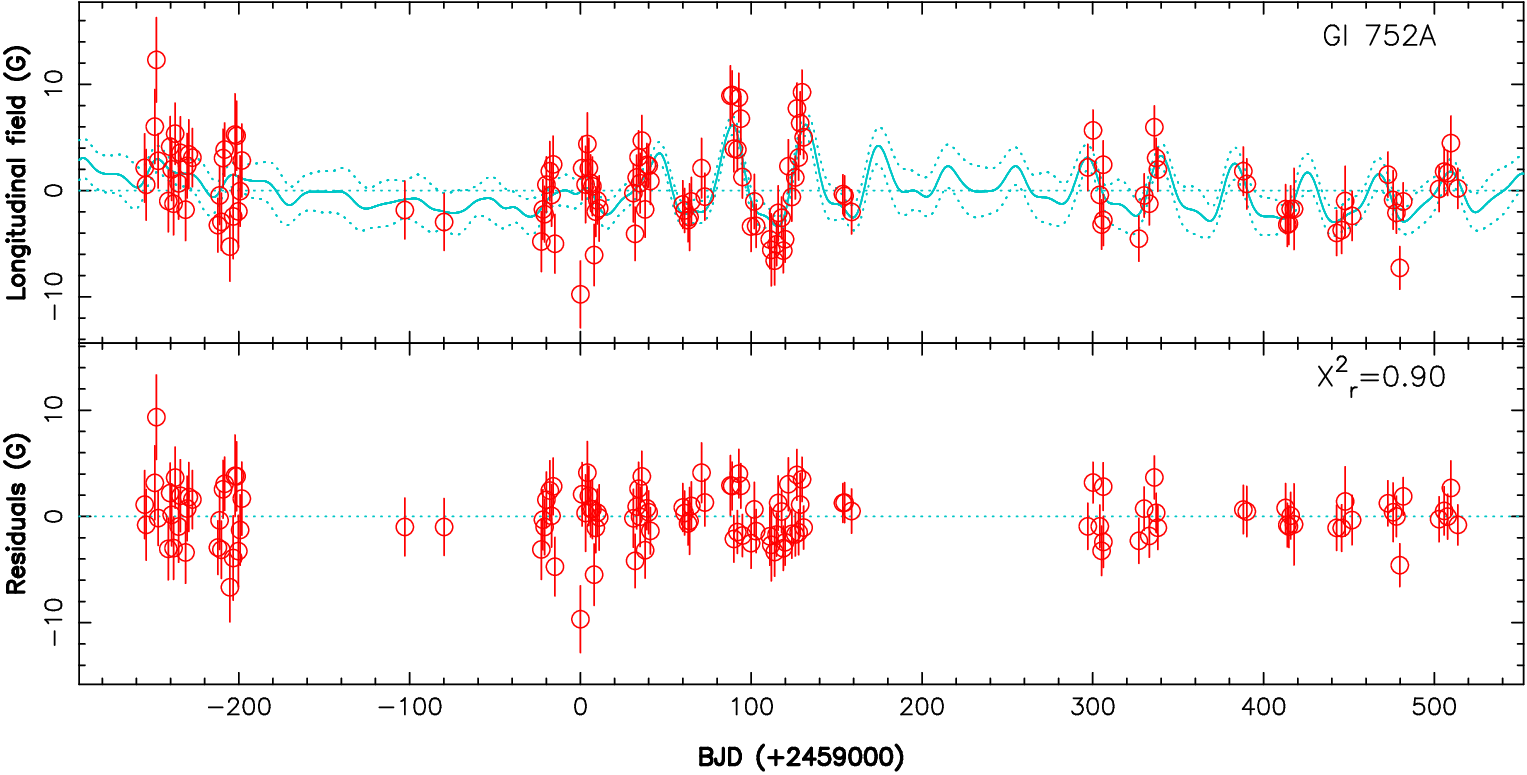} \vspace{2mm}} 
\centerline{\includegraphics[scale=0.58,angle=-90]{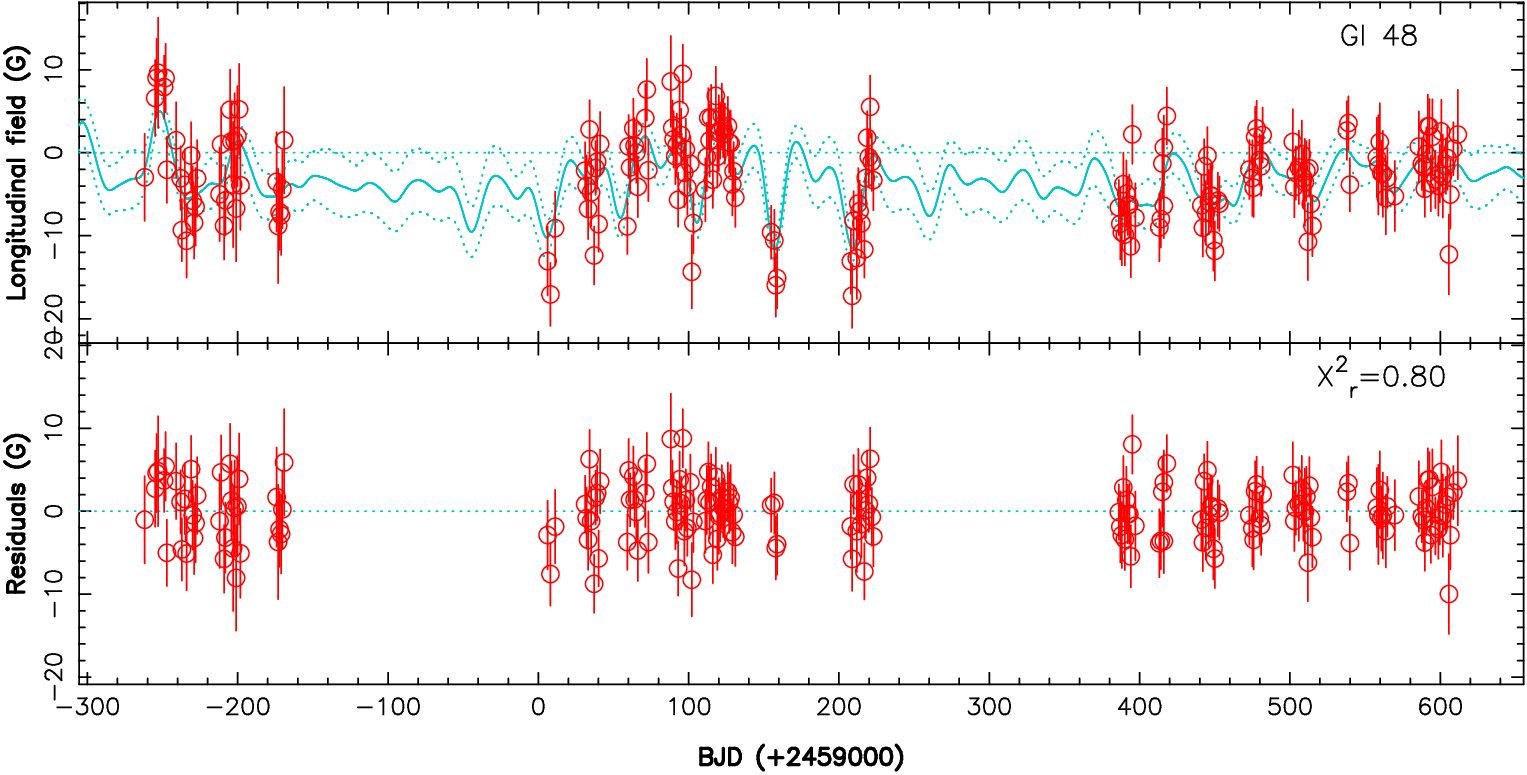}} 
\caption[]{Same as Fig.~\ref{fig:gjs} for Gl~411 (top), Gl~752A (middle) and Gl~48 (bottom). } 
\label{fig:a4}
\end{figure*}

\begin{figure*}
\centerline{\includegraphics[scale=0.58,angle=-90]{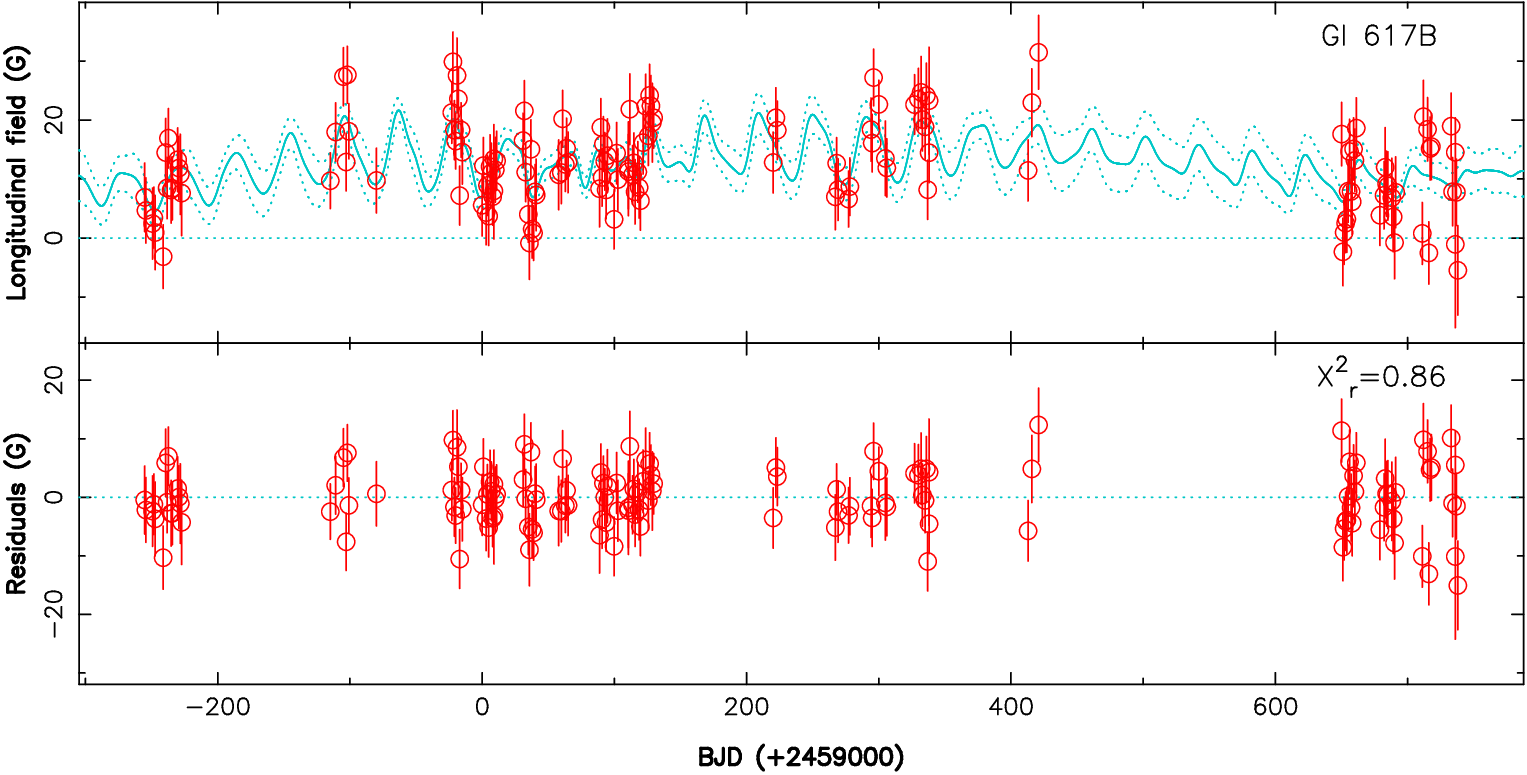} \vspace{2mm}} 
\centerline{\includegraphics[scale=0.58,angle=-90]{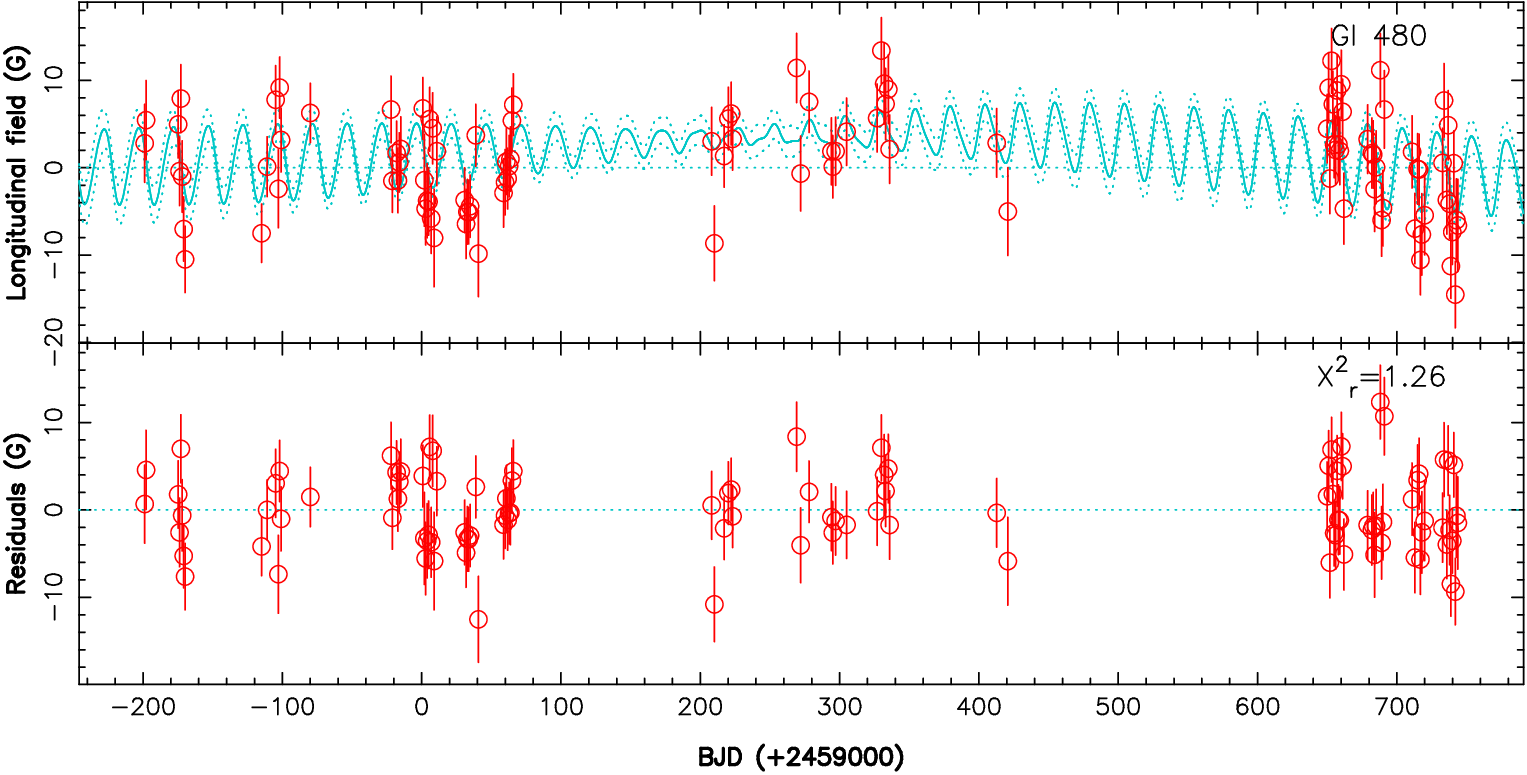} \vspace{2mm}} 
\centerline{\includegraphics[scale=0.58,angle=-90]{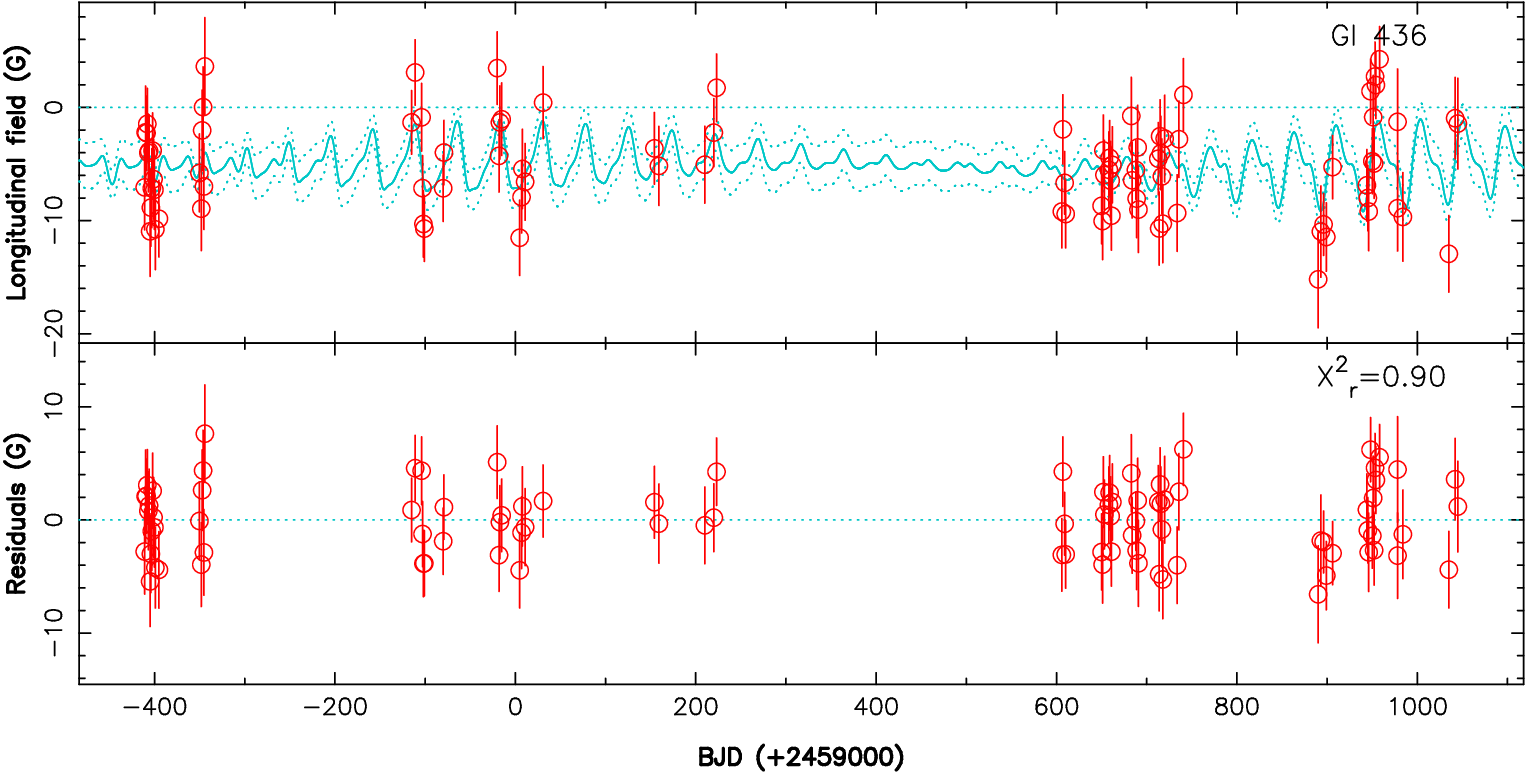}} 
\caption[]{Same as Fig.~\ref{fig:gjs} for Gl~617B (top), Gl~480 (middle) and Gl~436 (bottom). } 
\label{fig:a5}
\end{figure*}

\begin{figure*}
\centerline{\includegraphics[scale=0.58,angle=-90]{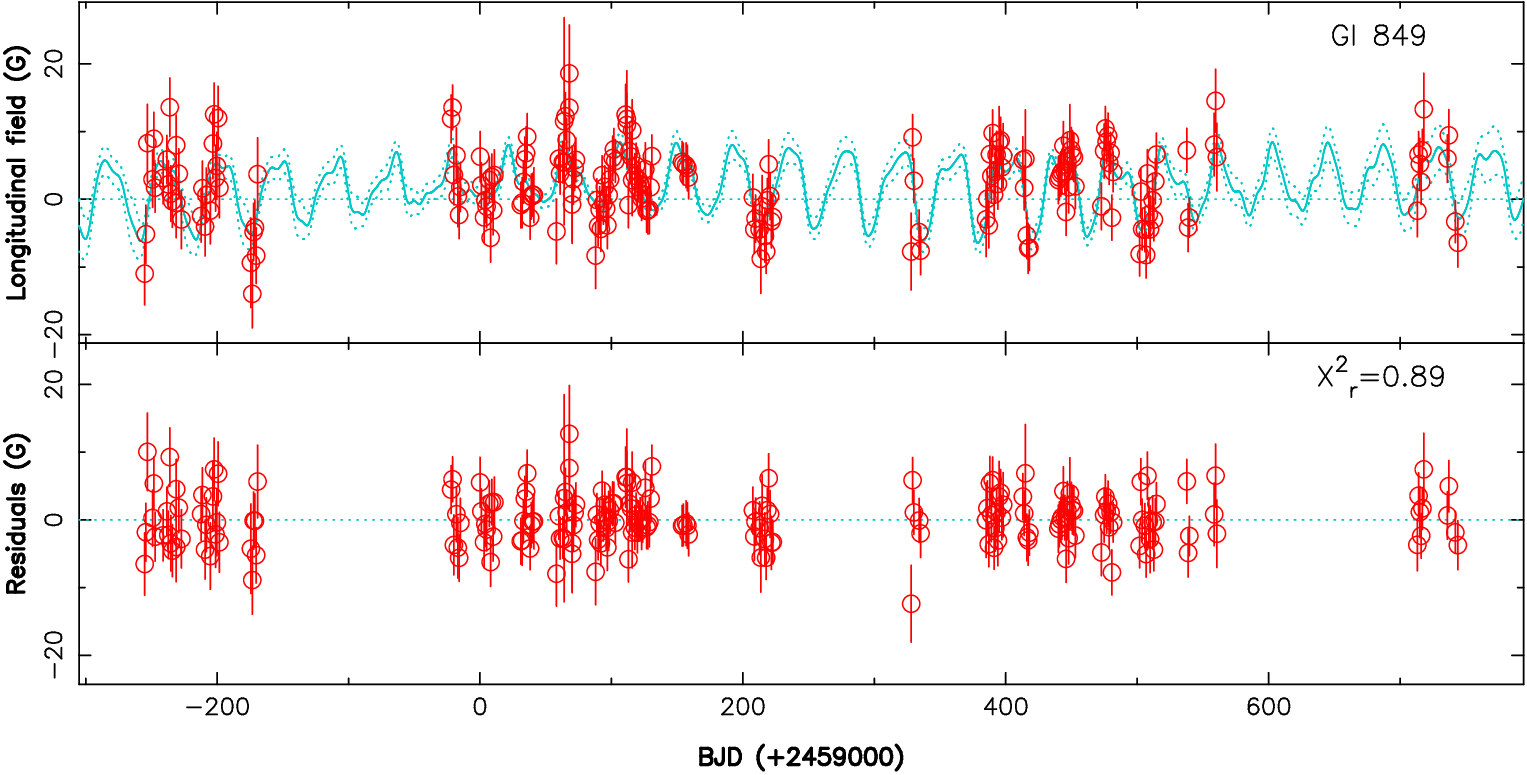} \vspace{2mm}} 
\centerline{\includegraphics[scale=0.58,angle=-90]{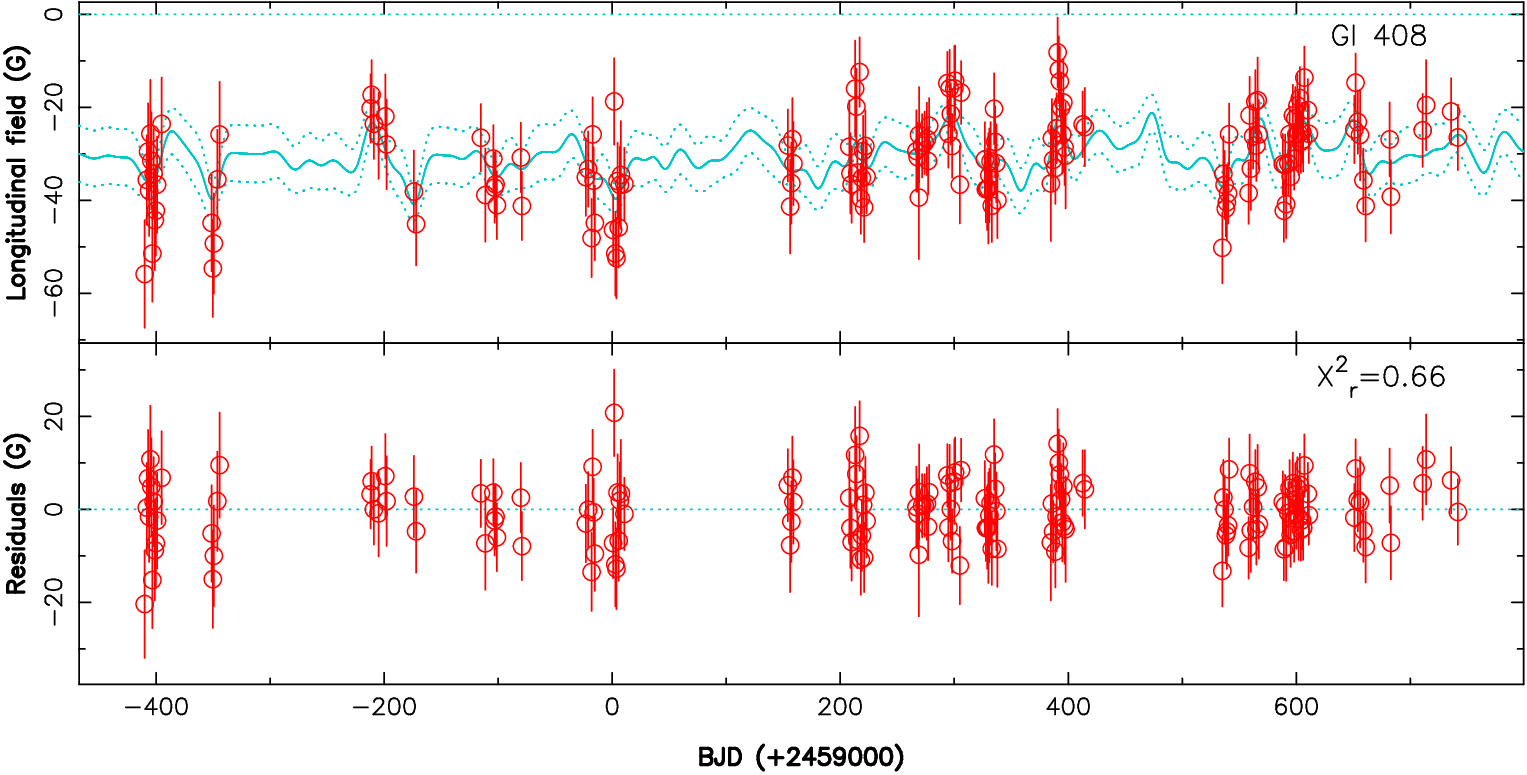} \vspace{2mm}} 
\centerline{\includegraphics[scale=0.58,angle=-90]{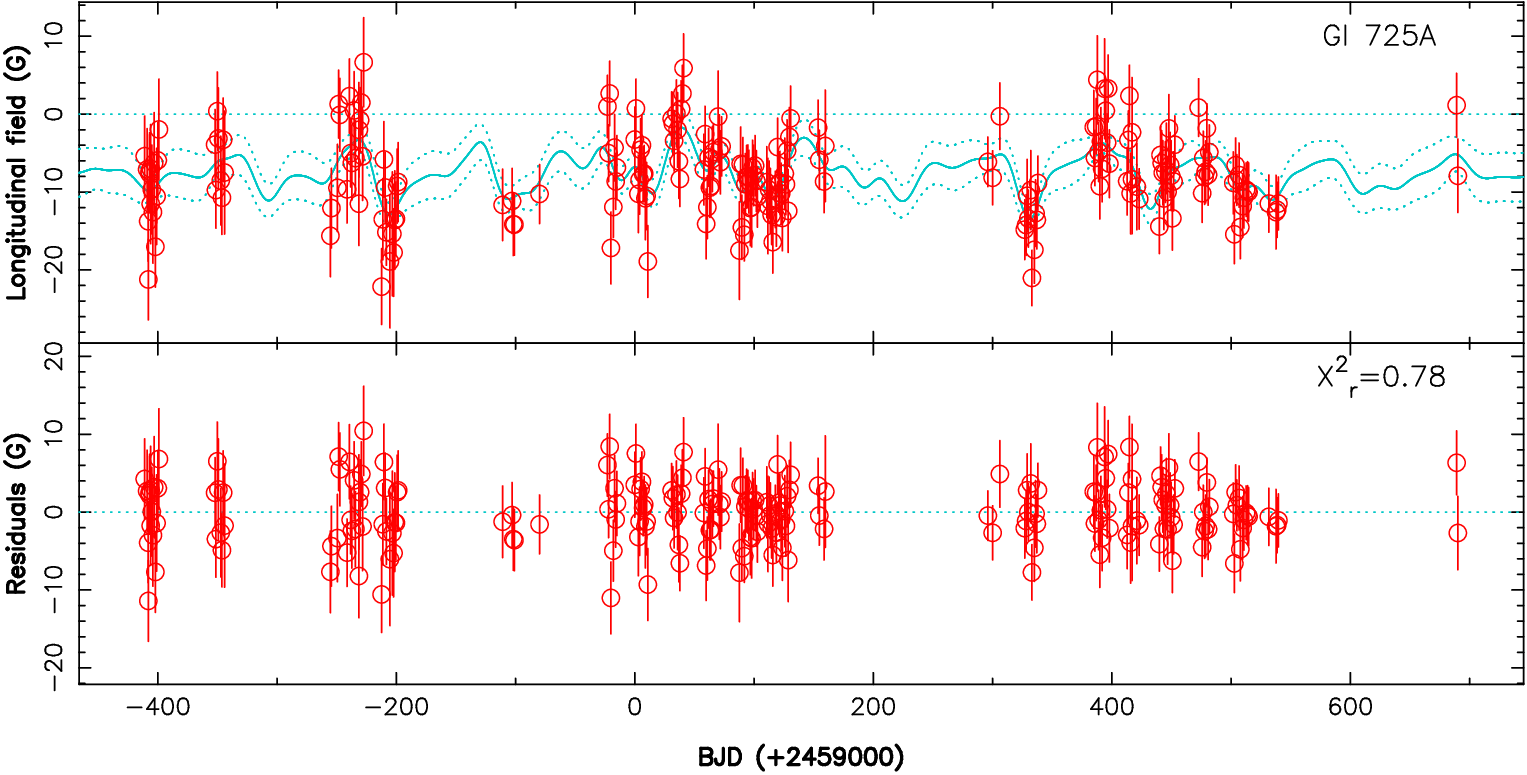}} 
\caption[]{Same as Fig.~\ref{fig:gjs} for Gl~849 (top), Gl~408 (middle) and Gl~725A (bottom). } 
\label{fig:a6}
\end{figure*}

\begin{figure*}
\centerline{\includegraphics[scale=0.58,angle=-90]{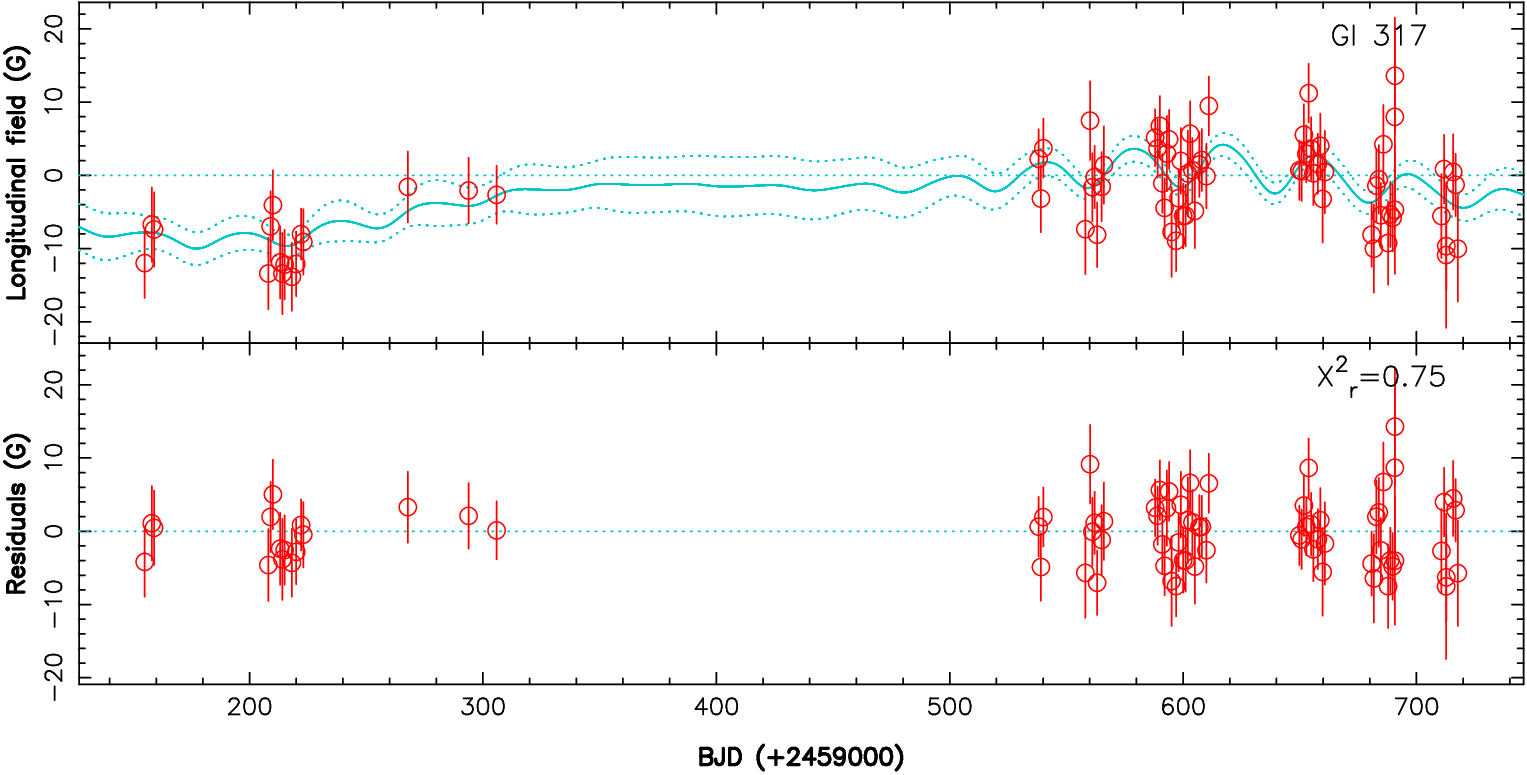} \vspace{2mm}} 
\centerline{\includegraphics[scale=0.58,angle=-90]{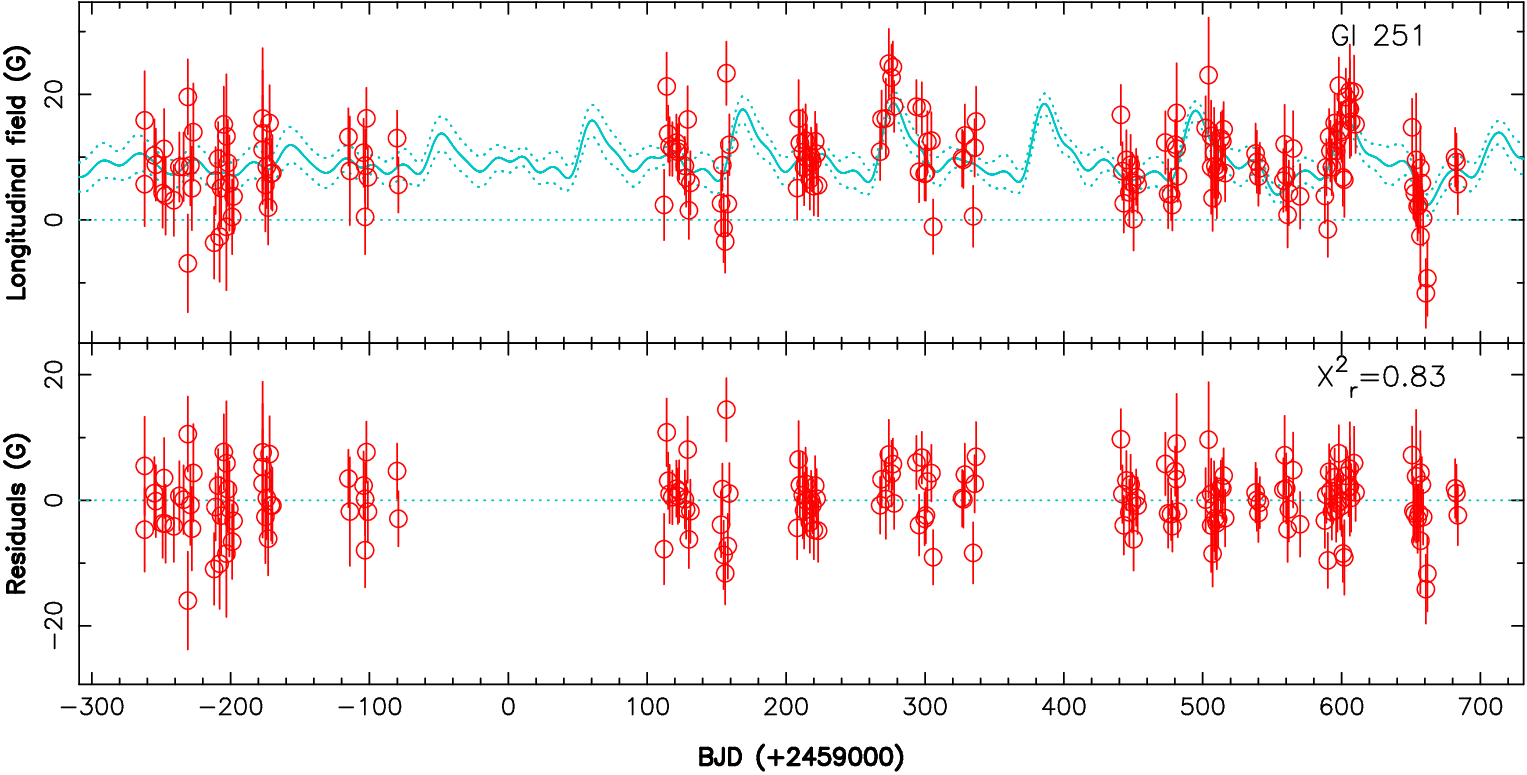} \vspace{2mm}} 
\centerline{\includegraphics[scale=0.58,angle=-90]{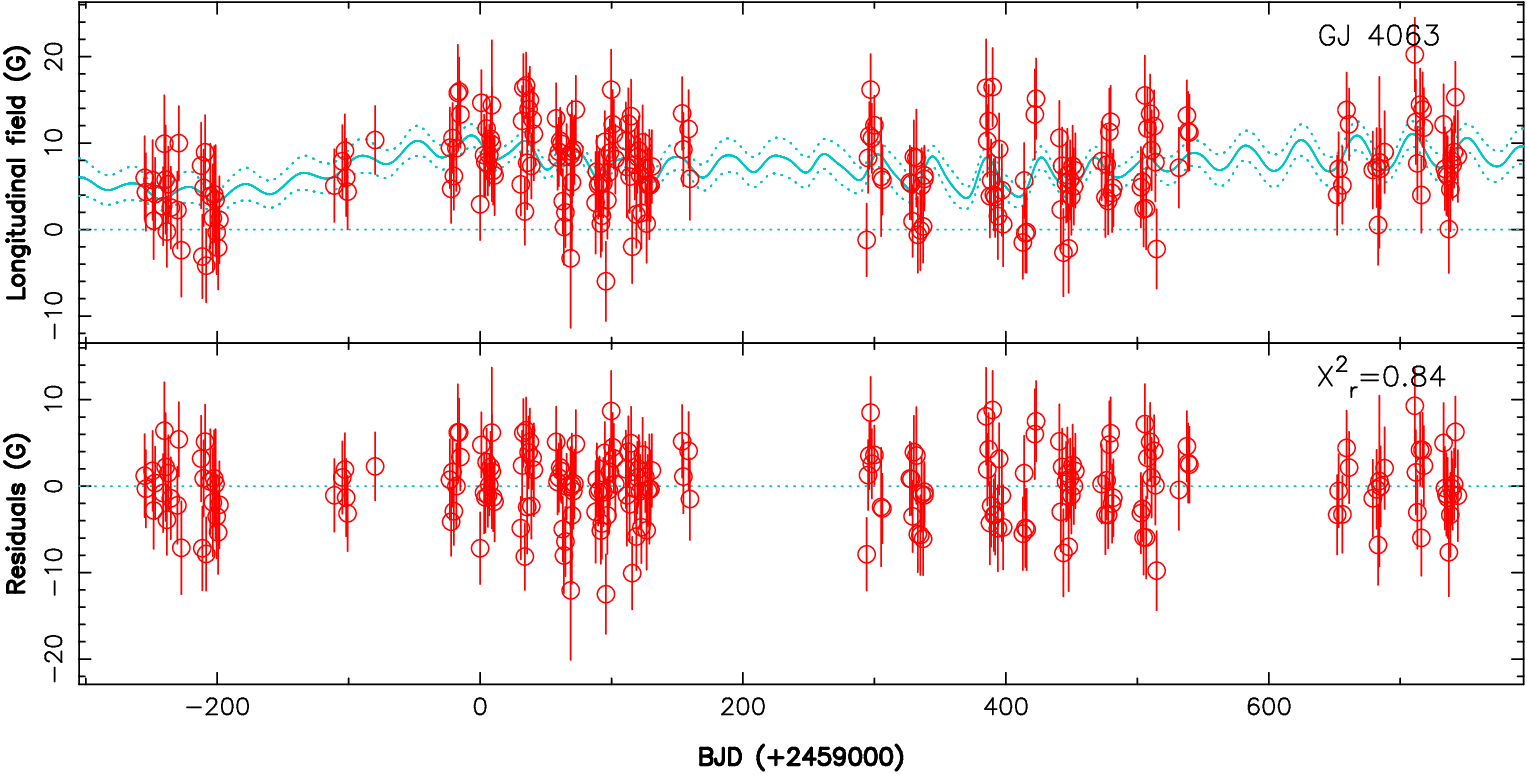}} 
\caption[]{Same as Fig.~\ref{fig:gjs} for Gl~317 (top), Gl~251 (middle) and GJ~4063 (bottom). } 
\label{fig:a7}
\end{figure*}

\begin{figure*}
\centerline{\includegraphics[scale=0.58,angle=-90]{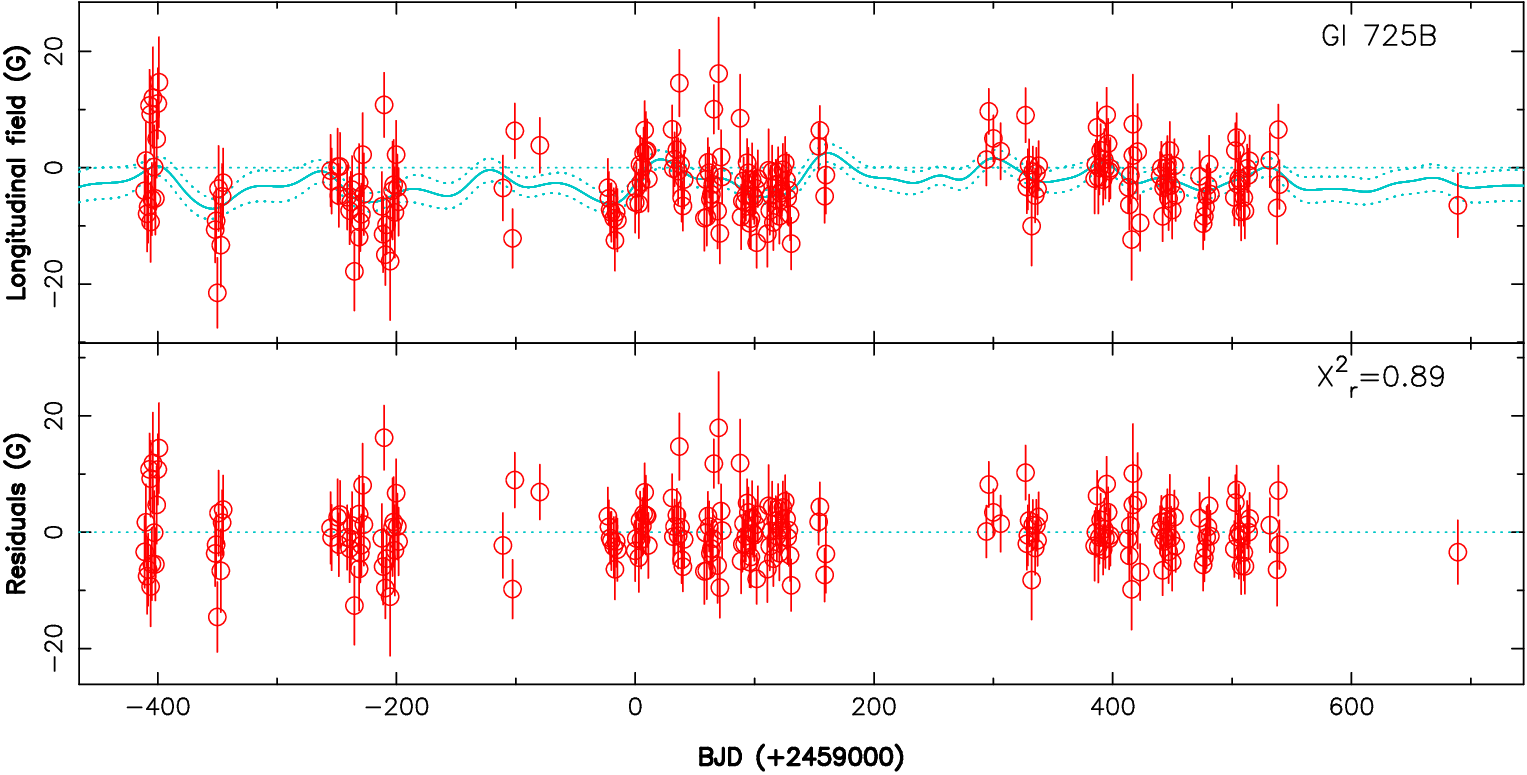} \vspace{2mm}} 
\centerline{\includegraphics[scale=0.58,angle=-90]{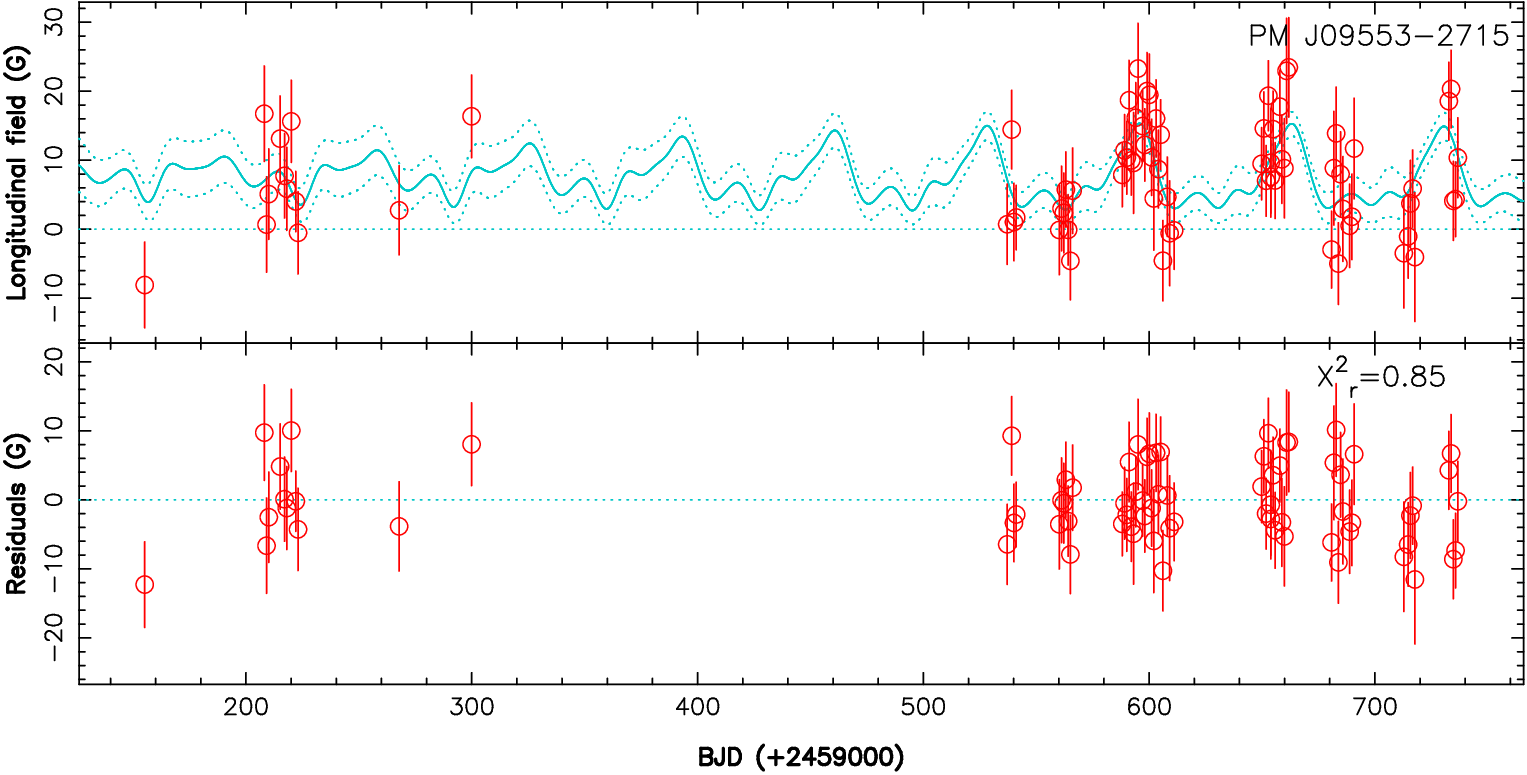} \vspace{2mm}} 
\centerline{\includegraphics[scale=0.58,angle=-90]{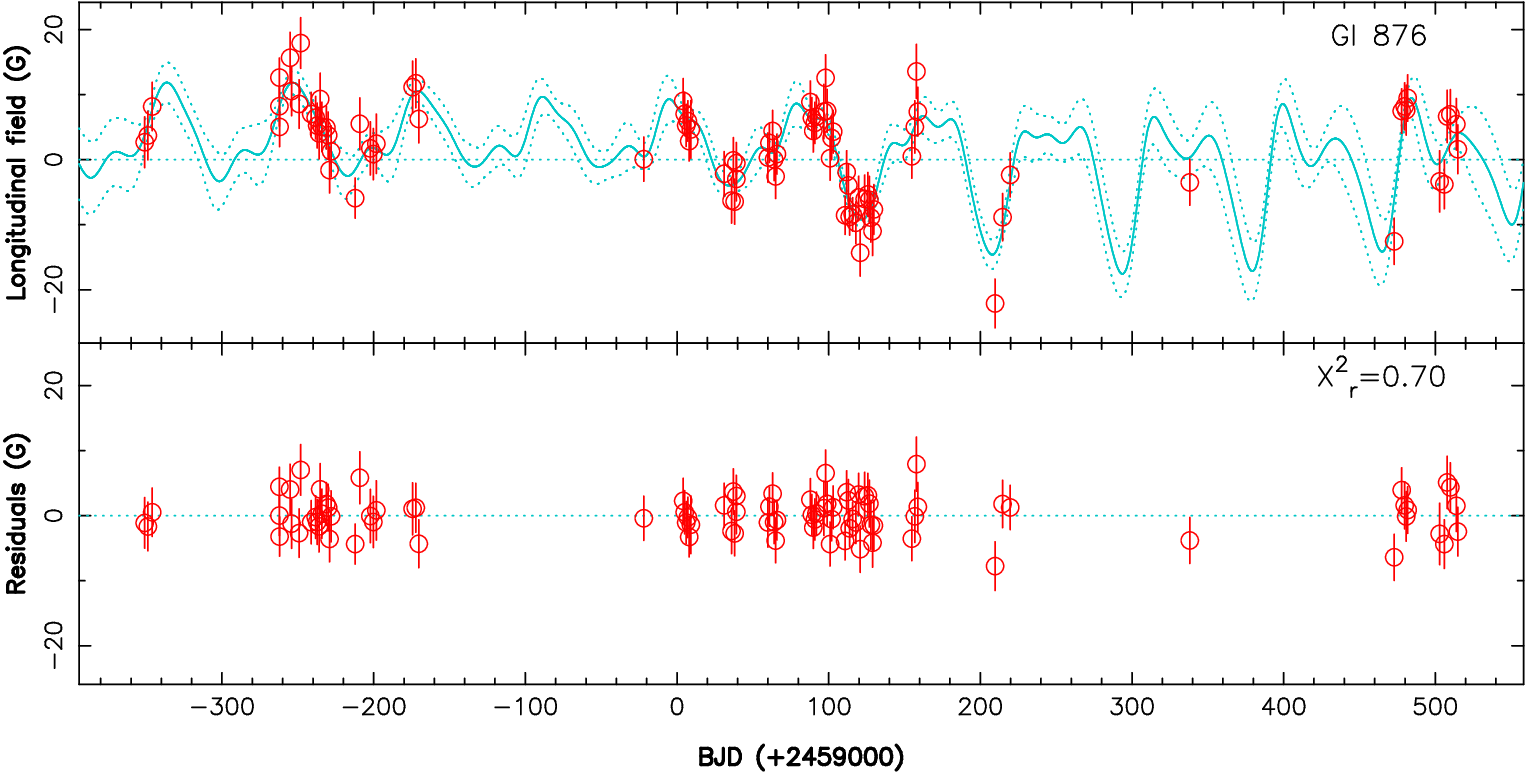}} 
\caption[]{Same as Fig.~\ref{fig:gjs} for Gl~725B (top), PM~J09553-2715 (middle) and Gl~876 (bottom). } 
\label{fig:a8}
\end{figure*}

\begin{figure*}
\centerline{\includegraphics[scale=0.58,angle=-90]{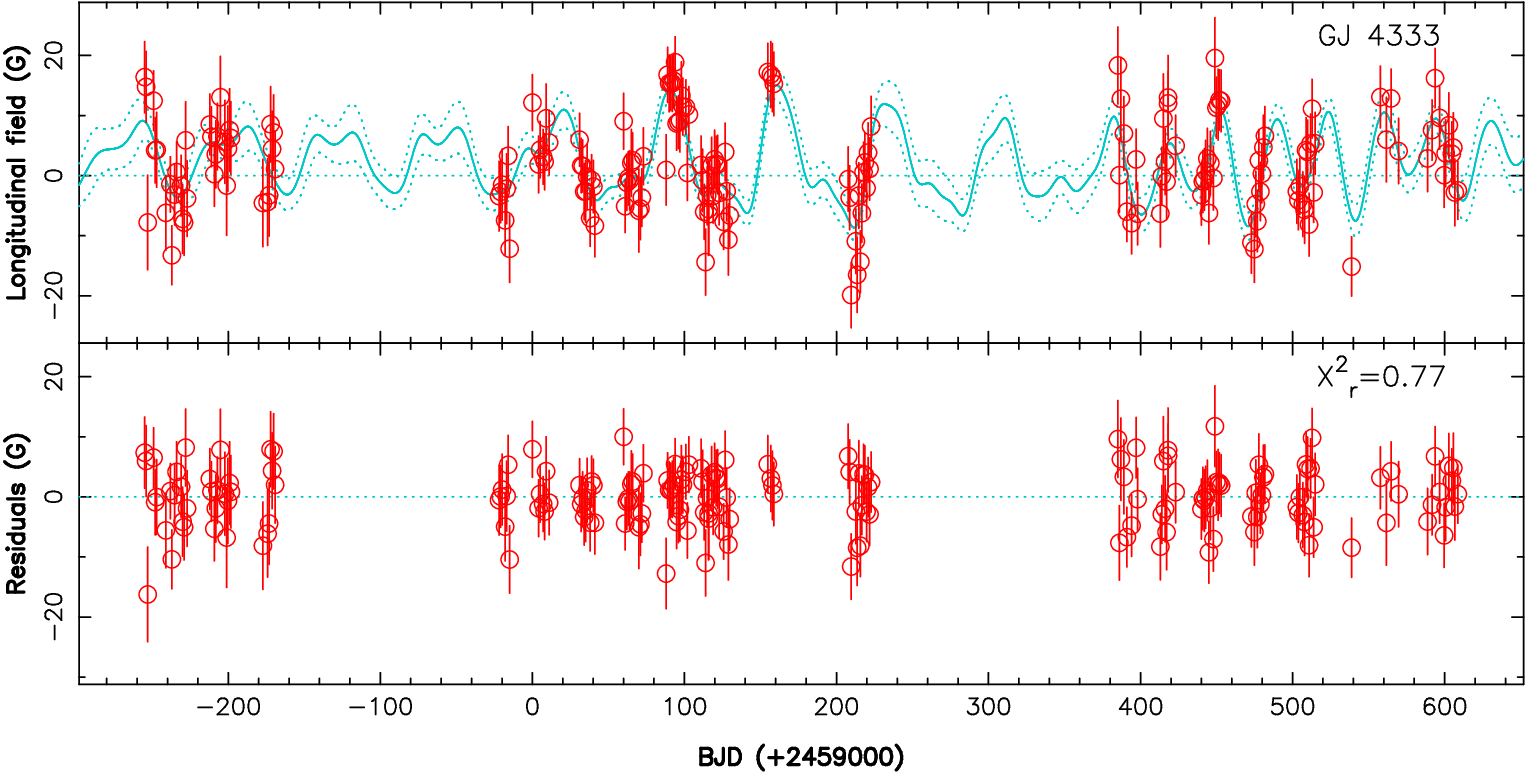} \vspace{2mm}} 
\centerline{\includegraphics[scale=0.58,angle=-90]{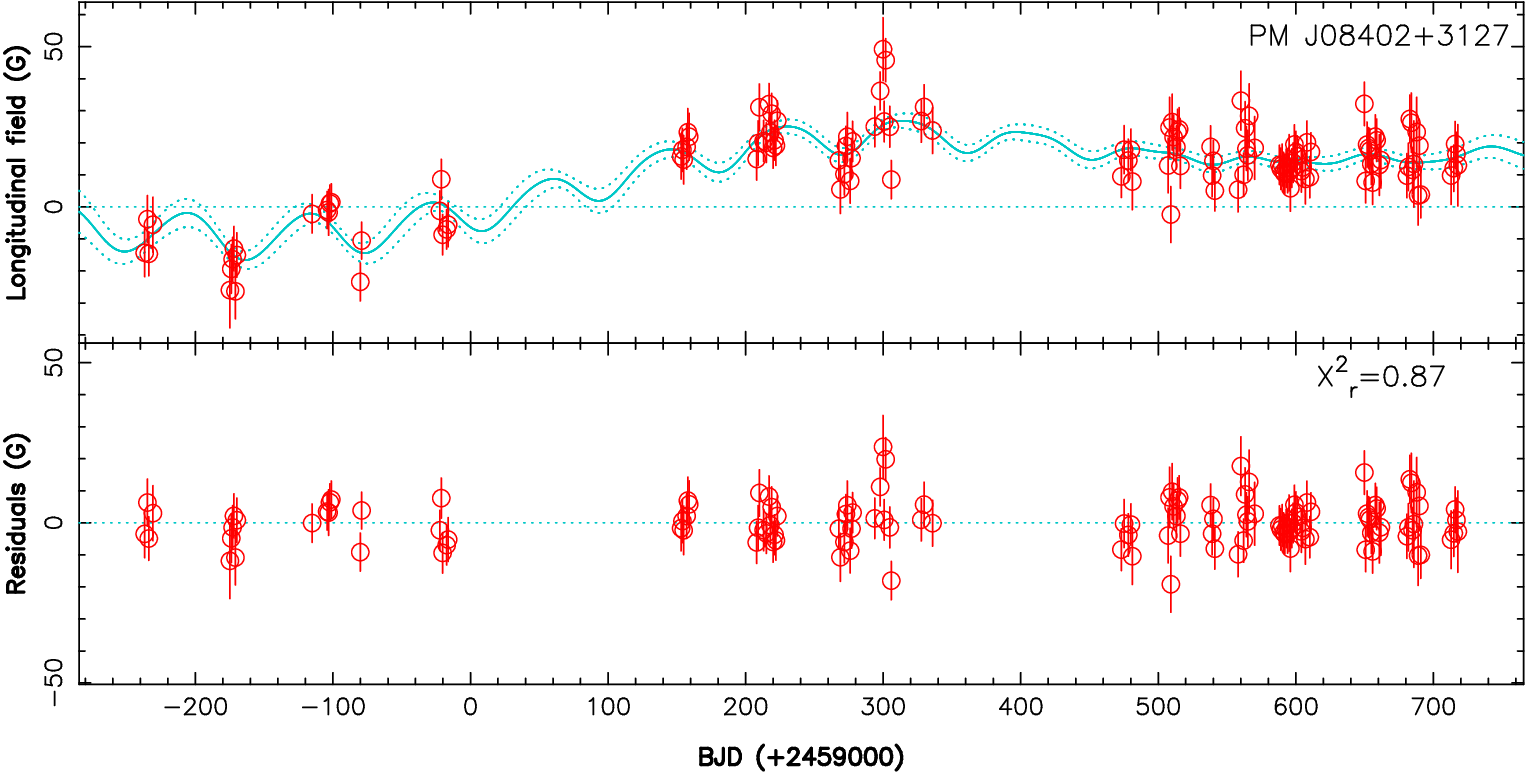} \vspace{2mm}} 
\centerline{\includegraphics[scale=0.58,angle=-90]{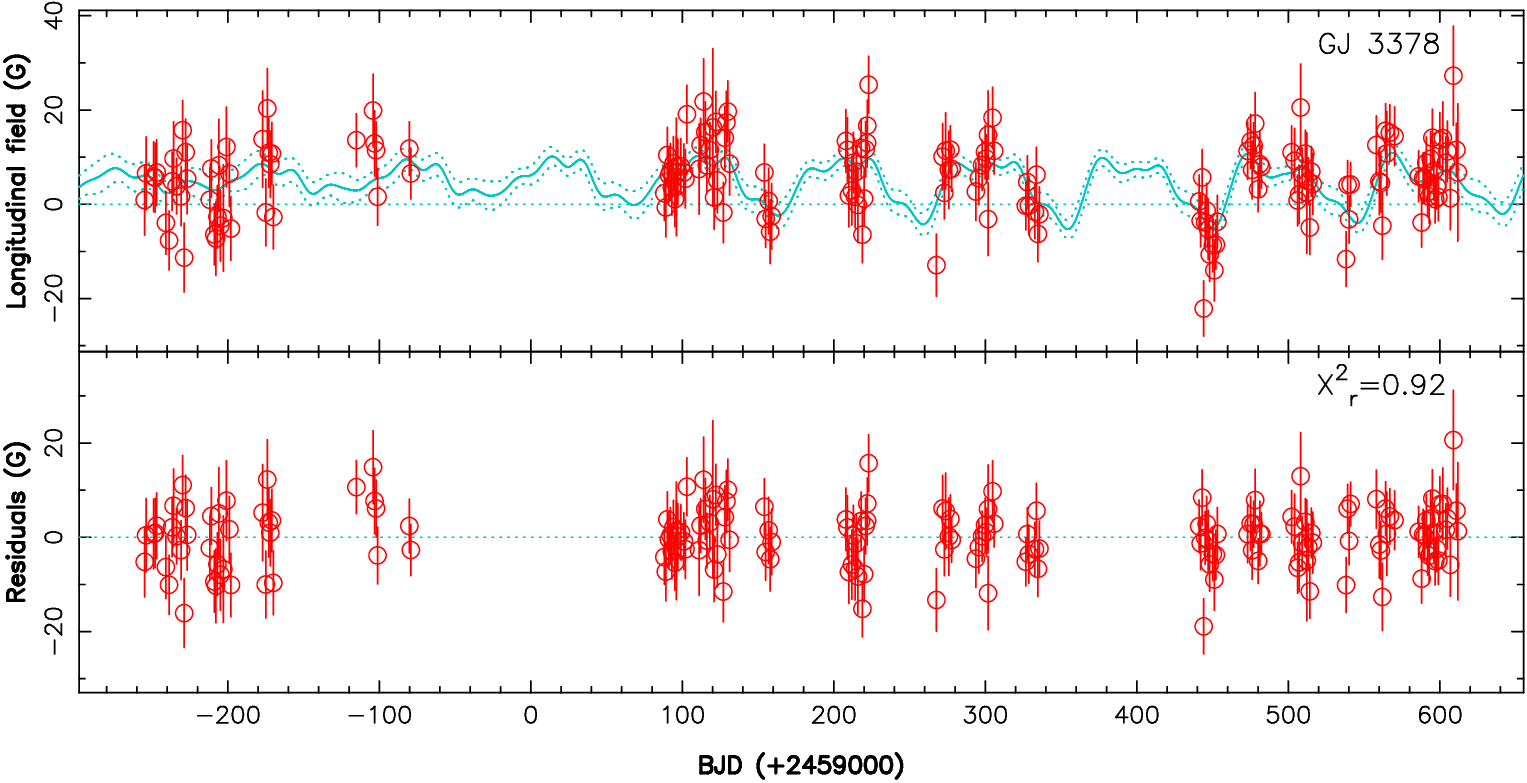}} 
\caption[]{Same as Fig.~\ref{fig:gjs} for GJ~4333 (top), PM~J08402+3127 (middle) and GJ~3378 (bottom). } 
\label{fig:a9}
\end{figure*}

\begin{figure*}
\centerline{\includegraphics[scale=0.58,angle=-90]{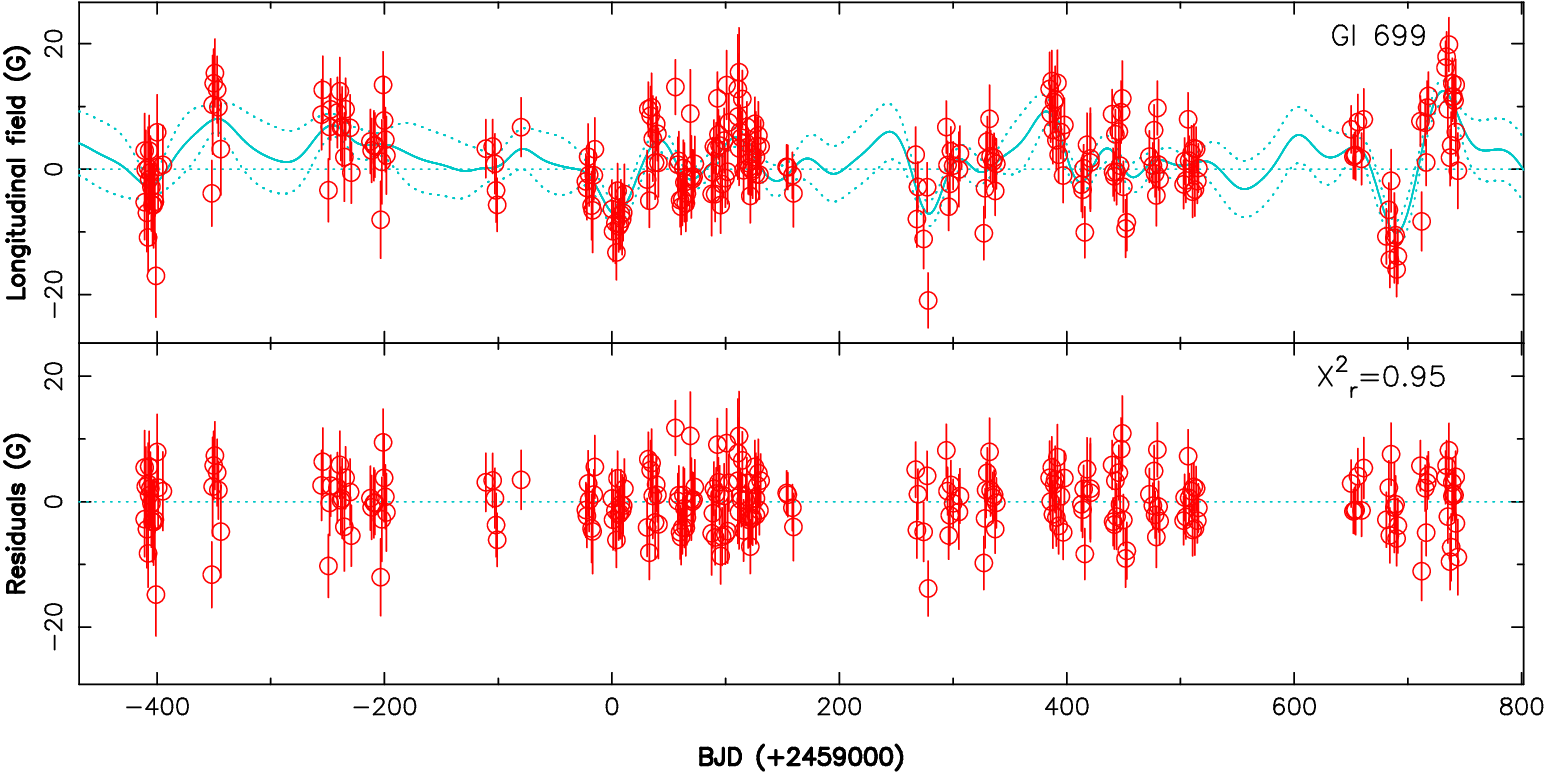} \vspace{2mm}} 
\centerline{\includegraphics[scale=0.58,angle=-90]{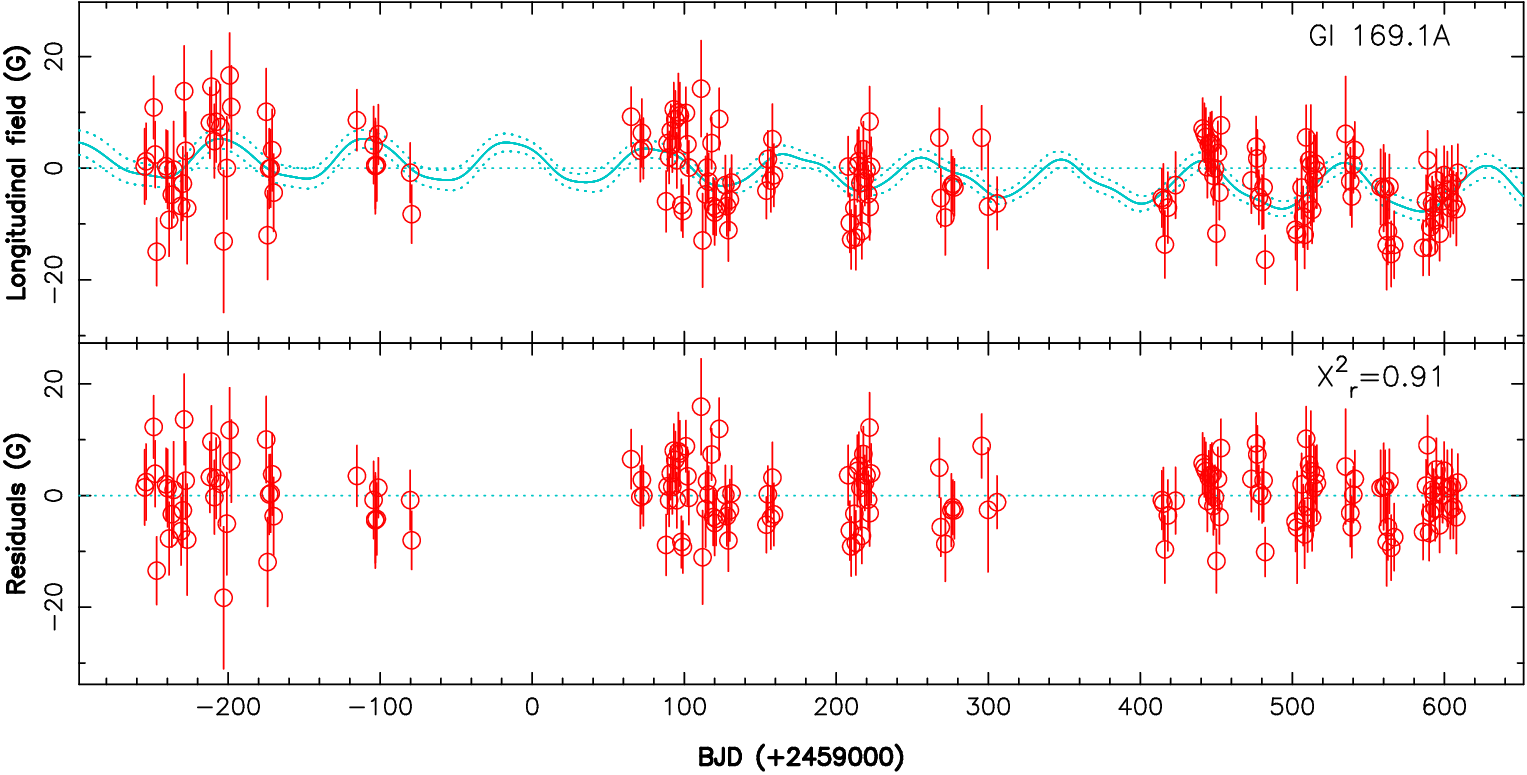} \vspace{2mm}} 
\centerline{\includegraphics[scale=0.58,angle=-90]{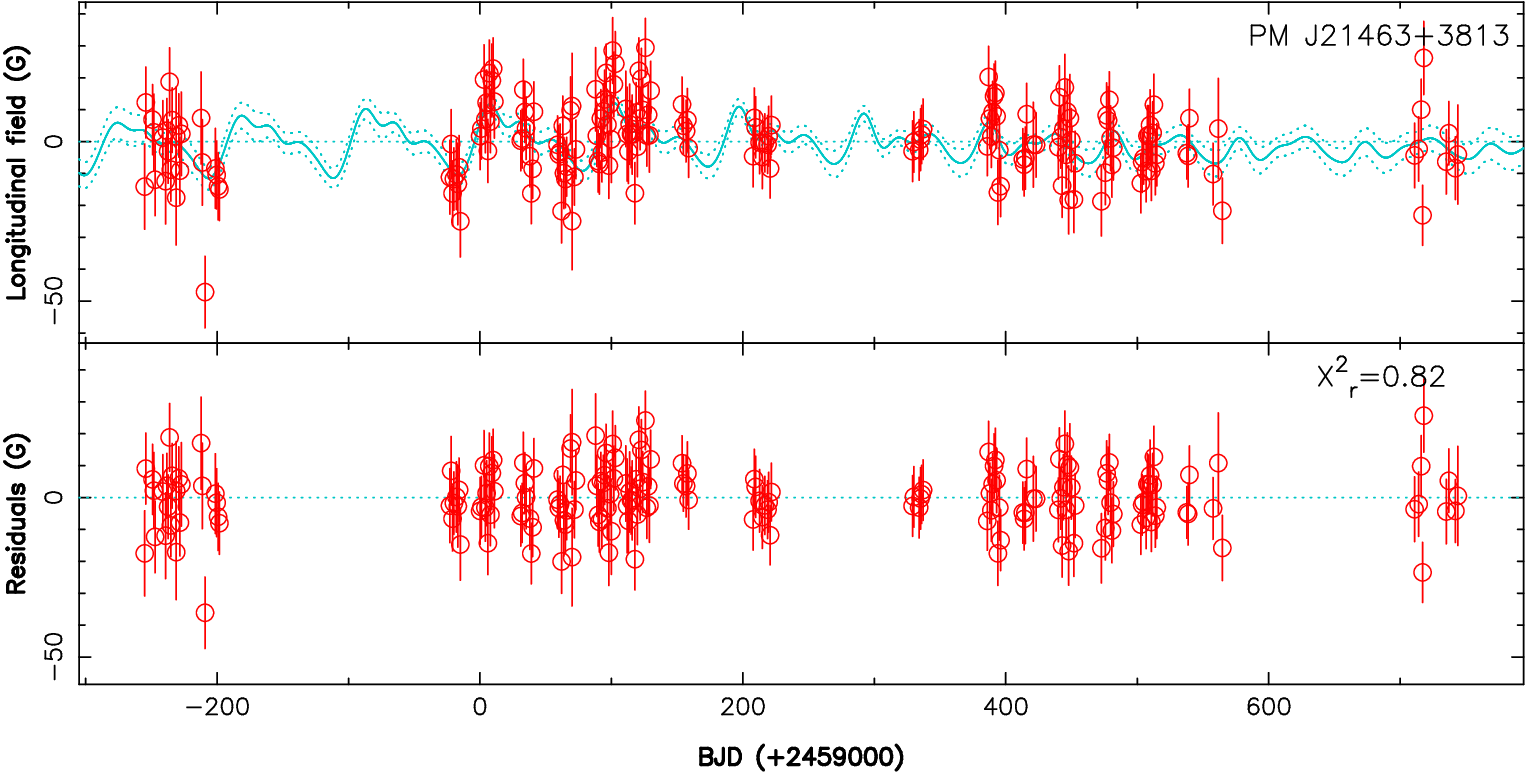}} 
\caption[]{Same as Fig.~\ref{fig:gjs} for Gl~699 (top), Gl~169.1A (middle), PM~J21463+3813 (bottom). } 
\label{fig:a10}
\end{figure*}

\begin{figure*}
\centerline{\includegraphics[scale=0.58,angle=-90]{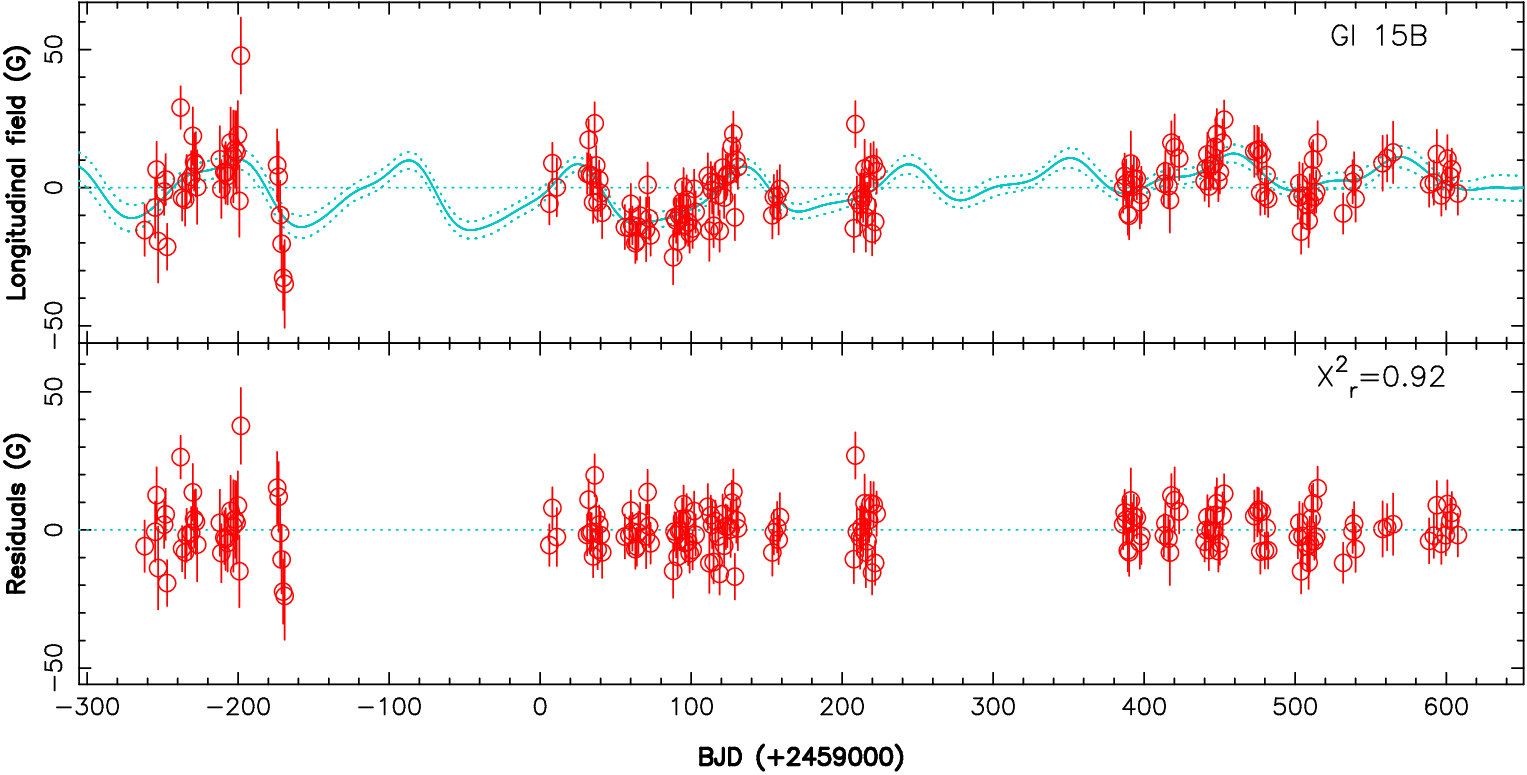} \vspace{2mm}} 
\centerline{\includegraphics[scale=0.58,angle=-90]{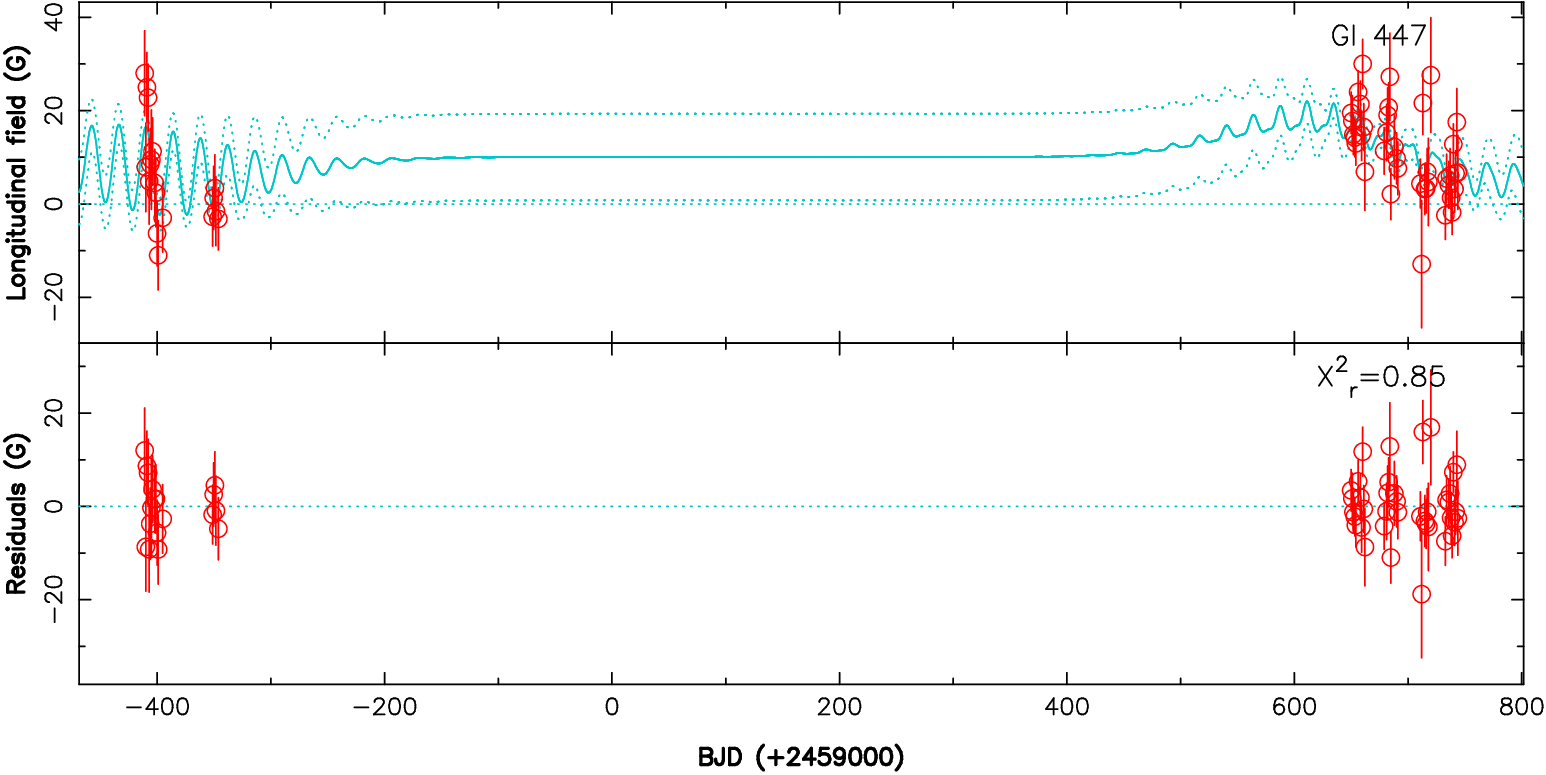} \vspace{2mm}} 
\centerline{\includegraphics[scale=0.58,angle=-90]{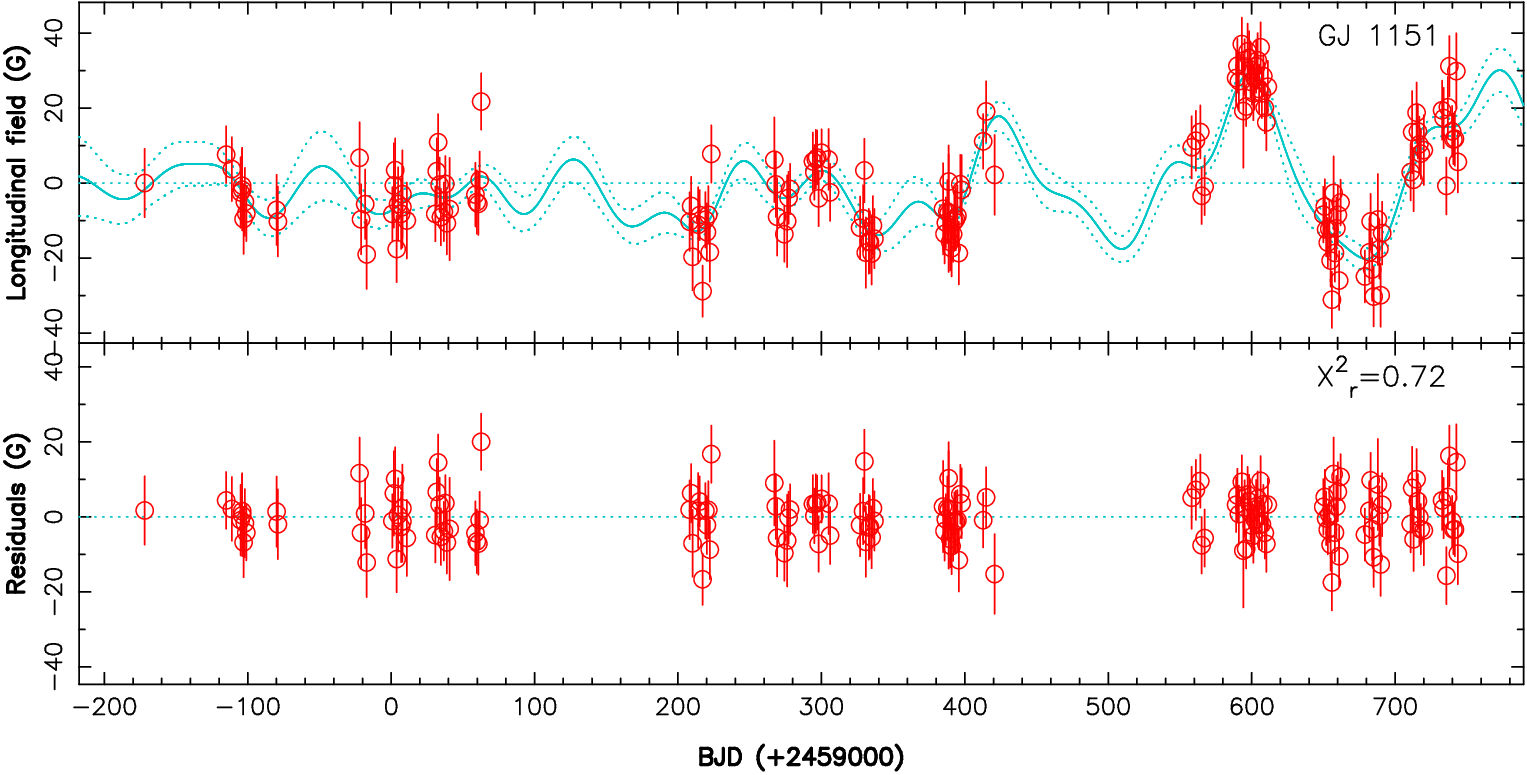}} 
\caption[]{Same as Fig.~\ref{fig:gjs} for Gl~15B (top), Gl~447 (middle) and GJ~1151 (bottom). } 
\label{fig:a11}
\end{figure*}

\begin{figure*}
\centerline{\includegraphics[scale=0.58,angle=-90]{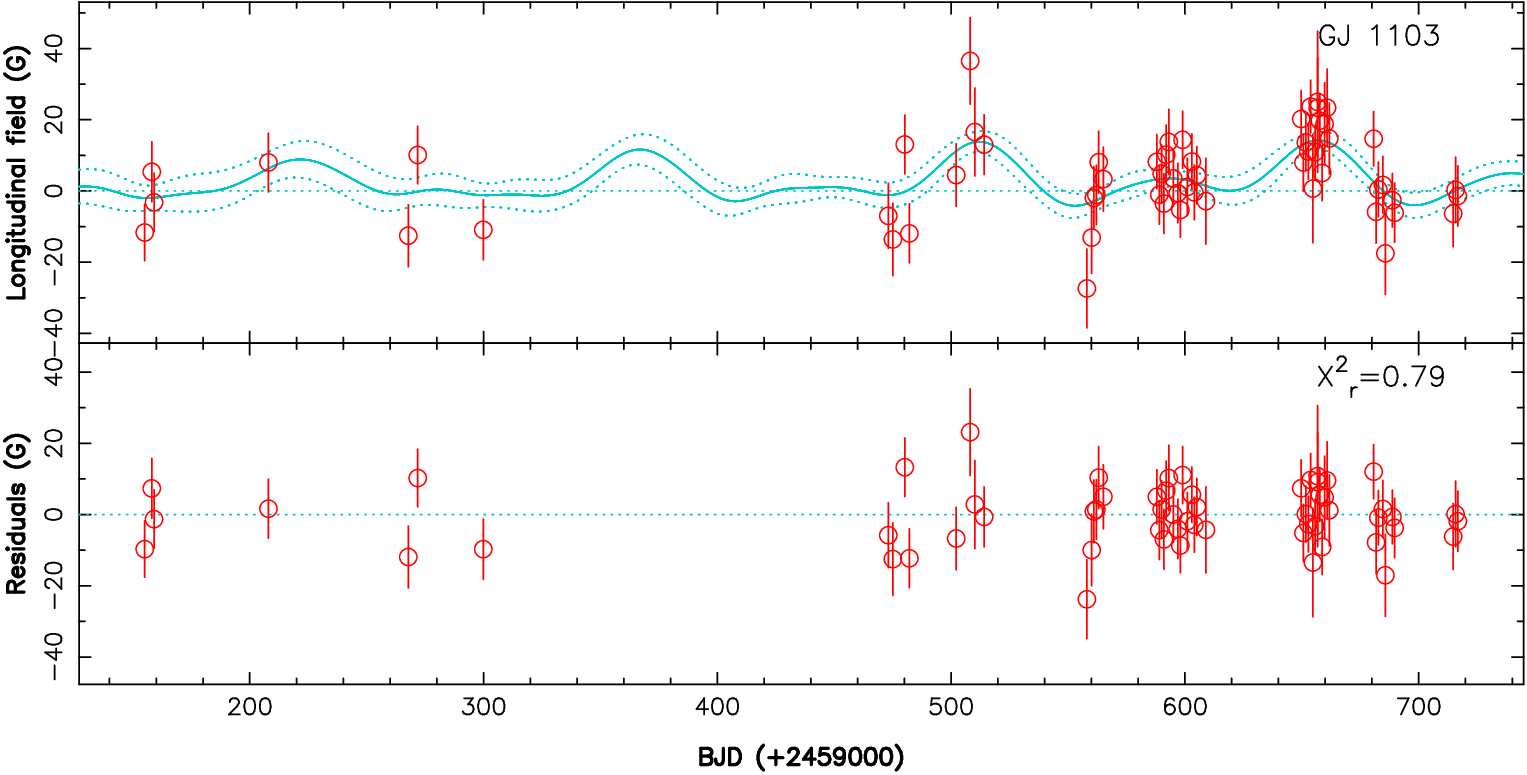} \vspace{2mm}} 
\centerline{\includegraphics[scale=0.58,angle=-90]{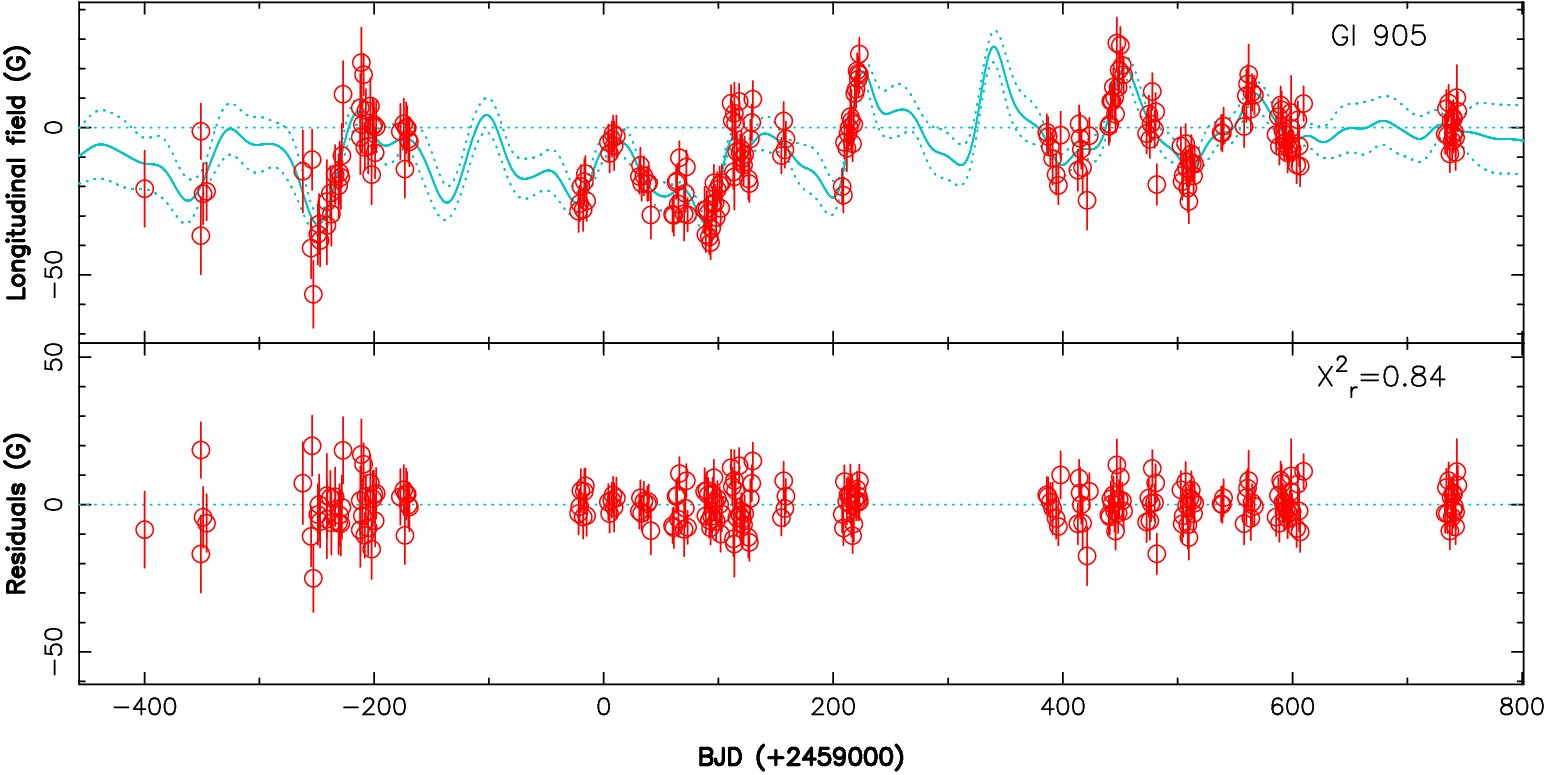} \vspace{2mm}} 
\centerline{\includegraphics[scale=0.58,angle=-90]{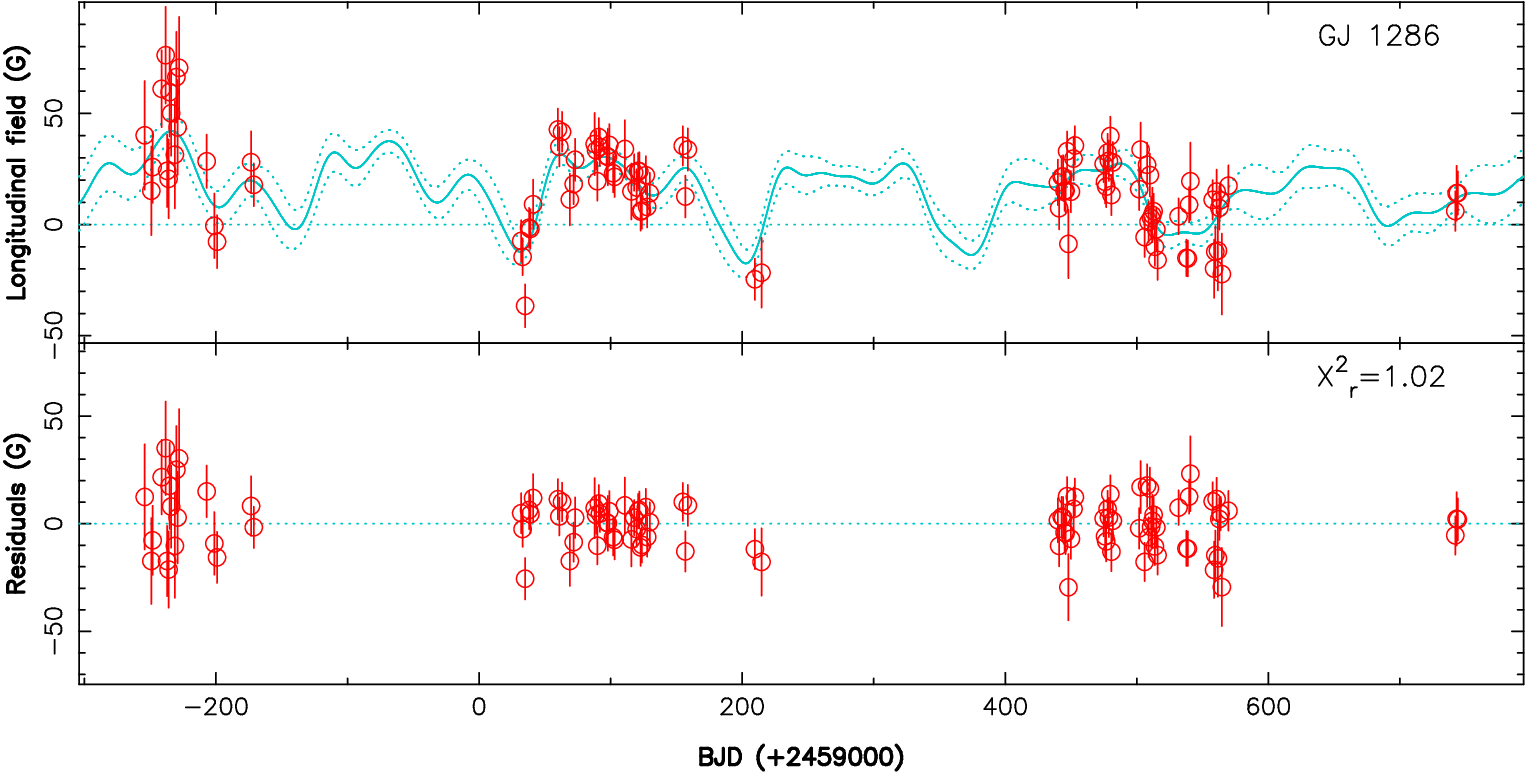}} 
\caption[]{Same as Fig.~\ref{fig:gjs} for GJ~1103 (top), Gl~905 (middle) and GJ~1286 (bottom). } 
\label{fig:a12}
\end{figure*}

\bsp	
\label{lastpage}
\end{document}